\documentclass[fleqn,usenatbib]{mnras}
\usepackage{float,graphicx,deluxetable,amssymb}
\usepackage[T1]{fontenc}
\usepackage{ae,aecompl}
\usepackage{graphicx}

\def\solar{\ifmmode _{\mathord\odot}\else $_{\mathord\odot}$\fi}
\def\msun{\ifmmode { M}_{\mathord\odot}\else $M_{\mathord\odot}$\fi}
\def\lsun{\ifmmode { L}_{\mathord\odot}\else $L_{\mathord\odot}$\fi}
\def\rsun{\ifmmode { R}_{\mathord\odot}\else $L_{\mathord\odot}$\fi}
\def\msol{\ifmmode { M}_{\mathord\odot}\else $M_{\mathord\odot}$\fi}
\def\lsol{\ifmmode { L}_{\mathord\odot}\else $L_{\mathord\odot}$\fi}
\def\rsol{\ifmmode { R}_{\mathord\odot}\else $L_{\mathord\odot}$\fi}
\def\sol{\ifmmode {\mathord\odot}\else ${\mathord\odot}$\fi}

\def\calm{{\cal M}}

\newcommand{\Alfven}{Alfv\`en }

\def\va{v_{\rm A}}
\def\ma{\calm_{\rm A}}
\def\mao{\calm_{\rm A,0}}

\def\maf{\calm_{\rm A,>0.8}}
\def\rms{{\rm  rms}}
\renewcommand{\vec}[1]{{\bf #1}}

\usepackage[section]{placeins}
\usepackage{color}

\newcommand{\altaffilmark}[1]{{$^{#1}$}}
\newcommand{\altaffiltext}[1]{{$^{#1}$}}

\title[Effects of Magnetic Fields on Cluster Formation]{The Effects of Magnetic Fields and Protostellar Feedback on Low-mass Cluster Formation}

\author[Cunningham et al.]{Andrew J. Cunningham\altaffilmark{1}, 
Mark R. Krumholz\altaffilmark{2}, 
Christopher F. McKee\altaffilmark{3,4}, 
\and 
Richard I. Klein\altaffilmark{1,3}
\\
\altaffiltext{1}{Lawrence Livermore National Laboratory, Livermore, CA 94550}\\
\altaffiltext{2}{Research School of Astronomy \& Astrophysics, Australian National University, Canberra, ACT, Australia 2611}\\
\altaffiltext{3}{Department of Astronomy, University of California Berkeley, Berkeley, CA 94720}\\
\altaffiltext{4}{Department of Physics, University of California Berkeley, Berkeley, CA 94720}\\
}

\begin{document}

\label{firstpage}
\pagerange{\pageref{firstpage}--\pageref{lastpage}}
\maketitle

\begin{abstract}
We present a large suite of simulations of the formation of low-mass star clusters. Our simulations include an extensive set of physical processes -- magnetohydrodynamics, radiative transfer, and protostellar outflows -- and span a wide range of virial parameters and magnetic field strengths. Comparing the outcomes of our simulations to observations, we find that simulations remaining close to virial balance throughout their history produce star formation efficiencies and initial mass function (IMF) peaks that are stable in time and in reasonable agreement with observations.  Our results indicate that small-scale dissipation effects near the protostellar surface provide a feedback loop for stabilizing the star formation efficiency. This is true regardless of whether the balance is maintained by input of energy from large scale forcing or by strong magnetic fields that inhibit collapse. In contrast, simulations that leave virial balance and undergo runaway collapse form stars too efficiently and produce an IMF that becomes increasingly top-heavy with time. In all cases we find that the competition between magnetic flux advection toward the protostar and outward advection due to magnetic interchange instabilities, and the competition between turbulent amplification and reconnection close to newly-formed protostars renders the local magnetic field structure insensitive to the strength of the large-scale field, ensuring that radiation is always more important than magnetic support in setting the fragmentation scale and thus the IMF peak mass. The statistics of multiple stellar systems are similarly insensitive to variations in the initial conditions and generally agree with observations within the range of statistical uncertainty.
\vspace{0.2in}
\end{abstract}

\begin{keywords}
ISM: clouds --- magnetohydrodynamics (MHD) --- radiative transfer --- stars: formation --- stars: luminosity function, mass function --- turbulence
\end{keywords}

\section{Introduction} \label{introduction}

The formation of stars through the collapse of prestellar molecular
clouds is a rich process, mediated by the coupled effects of gravity,
magnetic fields, turbulence, and feedback from protostellar evolution
in the form of outflows and radiation.  To date, numerical
models of star cluster formation have mostly probed the coupling of at most
two of these effects.  
While most contemporary protostellar cluster models include turbulence and self-gravity, the remaining effects have received a more piecemeal treatment.
Models that include magnetic effects and
protostellar outflow feedback have typically ignored radiative
transfer \citep{linak06,wang10,federrath15}, 
while simulations that
include radiative transfer with protostellar outflows have neglected
magnetic fields \citep{hansen,krumholz12}. Yet other works have
included magnetic fields and radiative transfer but not protostellar
outflows \citep{price08,price09,peters11}.  Two exceptions that includes all
three effects are the work of \citet{myers14} and \citet{li17}, who consider the
formation of star clusters consistent with observations of regions of
high-mass star formation with mean column densities $\Sigma = 1~{\rm g~
cm}^{-3}$. In this work, we perform an analogous study of regions
of low-mass star formation, characterised by a mean column density
$\Sigma = 0.1~{\rm g~ cm}^{-3}$. Under these conditions
the coupling of radiation from young stars to the infalling gas is weaker
\citep{krumholz08, krumholz10, cunningham11}, and
it is possible that magnetic and protostellar outflow effects are more
prominent.

A consistent and comprehensive theory of star formation must
simultaneously confront a number of observational constraints, as
summarised in the review by \citet{krumholzrev}.  The
relative strengths of the magnetic field and gravity in a clump of
molecular gas are measured by the mass-to-flux ratio. Statistical
surveys of Zeeman splitting measurements indicate that this ratio is
$2-3$ times larger than the critical value at which the magnetic field
is strong enough to prevent gravitational collapse
\citep{falgarone,crutcher}.  Recent
modeling that allows for the difference between the theoretical
mass-to-flux ratio, which is area weighted, and the observed one,
which is density weighted, show that the data summarized by
\citet{crutcher} are consistent with a value for this ratio that is
three times the critical value \citep{li15}.

Next, the collapse of a molecular cloud must give rise to a distribution of
stellar masses consistent with the observed stellar initial mass
function (IMF).  The observed IMF is well-represented by a power law
for stars more massive than $1\,\msol$ \citep{salpeter} and a log-normal
distribution peaked at $0.2\,\msol$ for lower mass sources
\citep{chabrier2005}.  Simulations suggest that magnetic fields influence
the process by which the IMF is established by adding support to
otherwise super-Jeans scale masses, thereby suppressing fragmentation
and influencing the distribution of stellar masses
\citep{padoan07,commercon,hennebelle11,federrath12,myers13}.
Radiative heating from protostellar sources similarly acts to increase
the local Jeans mass that characterizes the gas in the protostars'
vicinity, further reducing fragmentation and inhibiting the formation of lower mass
protostars \citep{bate09,offner09,myers11,krumholz11,krumholz12,krumholz16}.
Outflows also indirectly play a role in mediating the IMF by reducing the rate
of accretion onto stars, which in turn reduces the rate of radiative
heating.  In spite of the complexity of the coupling among
magnetic, radiative and outflow feedback, the shape of the
IMF is remarkably insensitive to the star-forming environment,
particularly for the majority of stars that are heavier than
brown-dwarfs but not so massive as to be exceptionally rare, where
observational statistics are best-constrained \citep{offner14}.

Not only must the complex physical processes associated with star
formation collude to fragment molecular clouds into a distribution of
masses that is consistent with the observed IMF, but the star
formation rate must also agree with observation.  Star formation is
remarkably inefficient.  Absent turbulent or magnetic support, a
molecular cloud should collapse entirely into stars on a free-fall
timescale.  In fact, the star-formation efficiency per free-fall time
$\epsilon_{\rm ff}$, defined as the fraction of gas mass that is
converted to stars per free-fall time, was found to lie between 0.001 and 0.09 by
\citet{krumholz07}; a much larger and more recent compilation of data
gives $\epsilon_{\rm ff}=0.015$ with a scatter of $\sim 0.5$ dex for both galactic and extraglaactic regions \citep{kdm12},
and subsequent authors have obtained similarly low values
\citep[e.g.,][]{garcia-burillo12, evans14, salim15, usero15, 
heyer16}.\footnote{In contrast to
these results, \citet{murray11} and \citet{lee16} argue
$\epsilon_{\rm ff}\sim 1$ in the most
luminous star clusters, based on their high ratios of ionizing luminosity
to molecular gas mass. This interpretation has been disputed by
\citet{feldmann11} and \citet{krumholzrev}, who suggest that these
ratios are high not because $\epsilon_{\rm ff}$ is large, but rather 
because very massive clusters have strong feedback that destroys
their parent clouds efficiently, thereby rendering invalid 
\citeauthor{murray11}'s and \citeauthor{lee16}'s assumption that
one can identify and measure the gas mass of these clusters'
progenitors. However, this dispute is immaterial
for the low-mass, low-density clusters with which we are concerned here,
for which there is a strong observational consensus that $\epsilon_{\rm ff}$
is low.}

Turbulence provides a significant source of support against rapid
collapse.  Gas velocity dispersions inferred from observed molecular
cloud line width-size relations \citep{larson81} are consistent with
the inertial range scaling of both supersonic hydrodynamic and
magnetohydrodynamical turbulence \citep{gammie96,ostriker99} and the
magnitude of the velocity dispersions is consistent with a near virial
equipartition between turbulent support and gravitational contraction
\citep{mckee10}. Numerical studies of magnetised turbulence show that
turbulence at virial levels, in conjunction with protostellar outflows,
can produce values of $\epsilon_{\rm ff}$ close to the observed, low
values \citep{linak06,wang10,padoan12,krumholz12,myers14,federrath15}.

Magnetohydrodynamical turbulence, however, undergoes
exponential decay via dissipation in shocks on a turbulent crossing time
\citep{stone98,maclow99}, which is comparable to the free fall
time in a virialized cloud.  Gravitational collapse can drive turbulence,
but only at the cost of increasing the density and thus lowering the
free-fall time \citep{robertson12, murray15}. As a result, in the absence
of an external or internal energy source to drive the turbulence,
hydrodynamical models fall into a state of rapid, near free-fall
collapse \citep{hansen,krumholz12}. A number of forcing mechanisms
applicable to low-mass star clusters have been proposed, including
accretion from larger scales \citep[e.g.,][]{kle10, gol11, lee16a,
lee16b}, internal forcing by protostellar outflows
\citep[e.g.,][]{linak06,matzner07,wang10,cunningham11,peters14,federrath15}, and
forcing by external H~\textsc{ii} regions 
\citep[e.g.,][]{matzner02, krumholz06, gol11} and supernova explosions
\citep[e.g.,][]{padoan16,padoan17,pan16}. 

In this paper we present a series of simulations of the formation of
low-mass stellar systems, which we compare to the various observational
constraints on the star formation process we have outlined above. Our
goal is to search for models that are capable of simultaneously
reproducing the observed IMF and star formation efficiency, determine
the relationship between these two constraints, and identify the
physical mechanisms that are responsible for satisfying them.
In order to accomplish this we carry out a suite of simulations
All our simulations use radiative transfer and protostellar radiation
feedback, since these are likely key to setting the IMF. In order
to explore the influence of outflows, we toggle these on and off.
We also vary the initial magnetisation of our regions, and
consider both models with turbulence that is continuously driven
to maintain virial equilibrium and those where no artificial
driving is included. We describe our numerical method and simulation setup in
\autoref{models}, we describe the results in \autoref{results}, we explore
the question of magnetic fields in more detail in \autoref{s4}, and we
summarise our findings in \autoref{conclusion}.

\section{Simulation Setup} \label{models}

\subsection{Initial Conditions} \label{ssec:initial_conditions}

We perform all our simulations using the \textsc{orion2} code. Our
setup shares several initial parameters with \citet{offner09} and
\citet{hansen}.  The initial state consists of a solar metallicity
molecular gas with a mean molecular weight $2.33 m_p$ and a
temperature $T_g=10~{\rm K}$, corresponding to an initial sound speed
$c_s=0.19~{\rm km~s}^{-1}$. The length of the cubical domain is
$L=0.65~{\rm pc}$ and the average density is
$\bar{\rho}=4.46\times10^{-20}~{\rm g~cm}^{-3}$ ($n_{\rm H_2} =
1.15\times 10^4$ cm$^{-3}$) corresponding to a total mass of $M = 185
\msol$.  The initial mean state corresponds to a minimum stable Jeans
length $\lambda_J=0.20~{\rm pc}$, and a Jeans mass $M_J=2.7\msol$.
Our models are carried forward with periodic boundary conditions for
the gravity and magnetohydrodynamics. We use Marshak boundary
conditions with an external radiation field corresponding to
a $10~{\rm K}$ isotropic blackbody for the radiation field.

The ability of a magnetized cloud to resist gravitational collapse is
measured by the magnetic critical mass 
\citep{mou76},
\begin{equation}
\label{eq:mphi}
M_\Phi = c_\Phi\frac{\Phi}{\sqrt{G}}, \label{mcrit}
\end{equation}
where $\Phi$ is the total magnetic flux threading the cloud and
$c_\Phi$ is a dimensionless constant that depends weakly on
the equilibrium shape of the cloud in the absence of gravity.  A cloud
of mass less than $M_\Phi$ cannot collapse, whereas a cloud with a
greater mass cannot be supported against collapse by magnetic fields
alone.  Because the value of $c_\Phi$ is only weakly sensitive to the cloud
geometry \citep{mckee10}, we adopt the value appropriate for a slab
threaded by a uniform, perpendicular field, $c_\Phi=1/2\pi$, consistent
with the convention chosen in several contemporary theoretical
\citep{li15} and observational works \citep{crutcher}.  We then define
the normalized mass-to-flux ratio
\begin{equation}
\mu_\Phi = \frac{M}{M_\Phi} = 2\pi\sqrt{G}\left(\frac{M}{\Phi}\right)=2\pi\sqrt{G}\left(\frac{\Sigma}{\bar B}\right),
\end{equation}
where $\bar B\equiv \Phi/S$ is the magnitude of the mean magnetic
field, $\Sigma$ is the mass surface density and $S$ the cloud's
cross-sectional area in the plane orthogonal to the magnetic field direction.
Since mass and magnetic flux are conserved for the entire simulation
box, the value of $\mu_\Phi$ for the box is constant and can be used
to specify the initial mean magnetic field.  We define initial fields
large enough to render the system close to magnetically critical,
$\mu_\Phi \approx 1$, ``strong,'' initial fields that make the system
moderately supercritical, $\mu_\Phi\approx 2-3$ ``moderate,''
and initial fields corresponding to $\mu_\Phi\gg 1$ ``weak.''

It is convenient to express the strength of the magnetic
field relative to the kinetic energy in terms of the \Alfven Mach
number,
\begin{equation}
\ma=\frac{v_\rms}{\va},
\end{equation}
where $\va^2\equiv B_\rms^2/(4\pi\bar\rho)$ is the \Alfven velocity,
$B_\rms$ is the root-mean-square (rms) volume-weighted field strength,
and $v_\rms$ is the rms
 mass-weighted, 3D
velocity dispersion.  Relative to thermal pressure, the strength of the magnetic field can be expressed in terms of
the plasma-$\beta$ parameter
\begin{equation}
\beta = \frac{8\pi\bar{\rho}c^2_s}{B_\rms^2} = 2\left(\frac{\ma}{\calm}\right)^2.
\end{equation}

In this work we carry out simulations with several different mean magnetic
field strengths. Our models begin with an initially uniform magnetic
field $B_0$ in the $\hat{z}$ direction.  We consider cases with
mass-to-flux ratios $\mu_\Phi=1.56\textrm{, }2.17\textrm{,
}23.1\textrm{, and } \infty$.  The corresponding initial \Alfven Mach numbers
are $\mao=1.0\textrm{, }1.4\textrm{, }15\textrm{, and } \infty$,
and the initial plasma-$\beta$'s are $\beta_0=0.046\textrm{, }0.089\textrm{,
}10\textrm{, and } \infty$.  We note that the recent models by
\citet{federrath15} consider a somewhat more weakly magnetized model
($\mu_\Phi=3.8$) than our two most strongly magnetized cases, and
the model by \citet{wang10} has an initial cloud with
$\mu_\Phi=1.4$, comparable to our strongest field model.

Turbulent conditions for starting the cluster simulation are obtained
by applying the driving recipe of \citet{maclow99}.  The procedure
consists of applying a force consisting of a purely solenoidal,
uniform spectrum of modes spanning wave numbers $\vec{k}$ such
that $1 \le |\vec{k}|L/2\pi
\le 2$ to a uniform initial gas.  This turbulent driving force is
intended to mock up the effect of sources of mechanical energy on
scales larger than the simulation box, via one of the numerous
possible mechanisms discusses in \autoref{introduction}.

The observed velocity dispersion to size relation can be met by models
of compressive turbulence that span a range of relative contributions
from solenoidal and compressive forcing \citep{federrath09}.  Recent
simulations of supernova-driven turbulence \citep{padoan16, pan16} show that
even when the initial driving is purely compressive, the turbulence
becomes primarily solenoidal in the inhomogeneous ISM; indeed, these
authors conclude that external supernova-driven turbulence is best
characterized as $84\%$ solenoidal after mediation through the ISM to
adjacent star-forming regions.  The work of \citet{federrath10}
indicates that the statistical properties of turbulent flow are not
sensitive to $\la 20\%$ compressive contributions to compressive
driving.  This suggests that our purely solenoidal forcing is a
plausible approximation for the driving of turbulence in a molecular
cloud.

The turbulent driving force is scaled to maintain a roughly constant
mass-weighted rms thermal Mach number of $\calm = v_\rms/c_s \equiv
\sqrt{3}\sigma_v/c_s= 6.6$, where $\sigma_v$ is the 1D velocity
dispersion.  The virial parameter,
\begin{equation}
\alpha_{\rm vir}\equiv\frac{5\sigma_v^2 R}{GM},
\end{equation}
measures the balance between kinetic energy and gravitational energy; a spherical cloud of uniform density
has $\alpha_{\rm vir}=1$ in virial equilibrium. 
With $R=L/2$ and our chosen value of $\calm$,
the computational box has $\alpha_{\rm vir}=1.05.$

The gas is initially evolved under the action of the driving force without
gravity according to the equations of ideal, isothermal
magnetohydrodynamics for two crossing times,
\begin{equation}
t_{\rm cross}=\frac{L}{v_\rms}=0.51~\textrm{Myr}, 
\end{equation}
on a uniform mesh with $512^3$ zones with periodic boundary
conditions.  This resolution has been shown by \citet{licode} to be
sufficient to capture a well-resolved inertial cascade down to scales
of $~L/30 = 0.02~\textrm{pc}$, an order of magnitude smaller than the
initial Jeans scale of the turbulent cloud.  Consequently, the
turbulent flows on the scales that give rise to the initial
fragmentation of the simulated cluster are well resolved.
Henceforth,
we denote the state of the turbulent gas after evolving for two
crossing times as the time $t=0$.  
\footnote{We note that the time required to saturate the magnetic energy depends on the initial field strength.  For models initialized with magnetic fields much weaker than those considered in this paper, significantly more than two dynamical times are required to fully saturate the magnetic energy of the system due to small-scale dynamos \citep{federrath11,tricco}. However, the models considered in this work are too strongly magnetized for small-scale eddies to generate significant field amplification, and the  magnetic power spectra have been shown to saturate within a few crossing times in these regimes \citep{licode}.}
We take these conditions as the
initial state of the turbulent cluster simulation, which includes a
more comprehensive set of physics that we describe in \autoref{ssec:physics}.

\subsection{Defining The Free-Fall Time and Star Formation Rate} \label{ssec:free-fall}

Because our simulations take place in a periodic box that represents
only a portion of a cloud, we must give some care to how we go
about defining the free-fall time and the efficiency of star
formation. Absent magnetic or thermal support to oppose self-gravity, the
free-fall time of gas structures of characteristic density $\rho$ is
\begin{equation}
t_{\rm ff} = \sqrt{\frac{3\pi}{32G \rho}}.
\end{equation}
The coefficient $\sqrt{3\pi/32}$ applies to spherical geometries,
but by convention we retain it for all geometries. The mean cloud density
$\bar{\rho}$ corresponds to a free-fall time $t_{\rm ff,IC} =
0.315~\textrm{Myr}$, where we use the subscript IC to refer to
quantities defined for the uniform-density state before driving the
initial turbulence to steady state.  We can measure star formation
rates relative to this fee-fall time via the usual dimensionless star
formation rate parameter
\begin{equation}
\epsilon_{\rm ff,IC} = \frac{\dot{M}_* t_{\rm ff,IC}}{M}.
\end{equation}

While $\epsilon_{\rm ff,IC}$ provides one measure of the star
formation rate, it can be somewhat misleading. As we drive the
turbulence to statistical steady-state, the gas density distribution
becomes non-uniform, and a majority of the mass ends up residing at
densities much higher than $\bar{\rho}$. This leads to two biases.
First, because most of the mass is at higher density, we might
expect $\epsilon_{\rm ff,IC}$ to end up relatively large even if
star formation in the dense gas is inefficient, simply because the
dense gas has a much shorter free-fall time. Second, because we are
using a periodic box to handle gravity, much of the low-density gas
is not self-gravitating at all, and thus would not collapse even
with efficient star formation. Including this mass in a calculation
of $\epsilon_{\rm ff}$ would lead to an artificially low value.

To counteract these biases, we define a new reference density 
that provides a better description of the gas after the turbulence
is fully developed. 
Let $\rho_{\rm med}$ be the mass-weighted median density--i.e., the density at which half the mass is denser and half less dense--at the end of the turbulent driving phase ($t=0$). Then we define $\rho_{\rm >med}$ as the average density of the regions with $\rho>\rho_{\rm med}$ at that time. 
In effect,
$\rho_{\rm >med}$ gives 
an estimate of the mean density of the gas that is actually
self-gravitating prior to any star formation. 
We denote the associated free-fall time as
$t_{\rm ff,>med}$. The dimensionless star formation rate is defined
as
\begin{equation}
\epsilon_{\rm ff,>med} = \frac{\dot{M}_* t_{\rm ff,>med}}{M_{\rm >med}}.
\end{equation}
Since $M = 2M_{\rm >med}$, this can also be expressed as
\begin{equation}
\epsilon_{\rm ff,>med} = \frac{2 \dot{M}_* t_{\rm ff,>med}}{M}.
\end{equation}

While $\epsilon_{\rm ff,>med}$ is a natural physical measure of
star formation efficiency, it cannot readily be compared to observations.
This is both because our simulation at $t=0$  represents a
somewhat artificial state of driven turbulence but no gravitational
effects, and because observations cannot usually measure
gas densities directly in any event. Instead, volume densities must
be inferred indirectly from molecular line critical densities, or,
more commonly for Galactic observations, using column densities that
can be measured almost directly. Following \cite{kdm12}, we define
the observationally-inferred characteristic density by taking
projected column densities and extracting those regions denser
than $0.048~\textrm{g cm}^{-2} = 230~\msun~\textrm{pc}^{-2}$,
corresponding to those regions with K-band extinction
$A_K>0.8~\textrm{mag}$, a threshold
used by a number of previous authors \citep[e.g.,][]{lada10, lada12, lada13}.
We compute the projected area of these regions $S_{\rm >0.8}$,
and from this determine the projected mass enclose $M_{\rm > 0.8}$ and
the effective radius $R_{\rm >0.8} = \sqrt{S_{\rm >0.8}/\pi}$. 
We then define the characteristic observed density of the projection 
as $\rho_{\rm obs, proj} = 3 M_{\rm >0.8} / (4 \pi R_{\rm >0.8}^3)$, and take the mean
value from projections in the three cardinal directions to be the characteristic
density $\rho_{\rm >0.8}$ for each simulation.
We compute this quantity at the
final time in each simulation
(in contrast to $\rho_{\rm > med}$, which is calculated at $t=0$, when gravity is turned on),
and we denote the corresponding 
free-fall time as $t_{\rm ff,>0.8}$. The star formation rate per
$t_{\rm ff,>0.8}$ is then
\begin{equation}
\epsilon_{\rm ff,>0.8} = \frac{\dot{M}_* t_{\rm ff,>0.8}}{M_{\rm >0.8}},
\end{equation}
and this is the quantity that is most natural to compare to observed star formation
rates.

In \autoref{t0}, we summarize the two characteristic densities
($\rho_{\rm >med}$, and $\rho_{\rm >0.8}$) that we have defined for
all our simulations.  The former is typically $7~\textrm{to}~9$ times
greater than the mean density $\bar{\rho}=4.46 \times 10^{-20} {\rm
  g~cm}^{-3}$, due to turbulent compression. The observationally-defined
$\rho_{\rm 0.8}$ ranges
from $2~\textrm{to}~7 \times \bar{\rho}$ due to the influence of the
combined effects of turbulent compression, gravitational collapse and
stellar feedback.  We also report the mean values of the \Alfven
Mach number and plasma $\beta$ at the end of the driving phase and at
the final time in the simulation, to make it clear how these evolve
with time as well.  This evolution is a result of field amplification
by turbulence and gravitational collapse.
For consistency, we
use the same notation for these other quantities as for the density
and free-fall time, i.e., quantities measured in the initial, uniform
state have a subscript ``IC'' (for initial condition), quantities
averaged over gas denser than the median density at the end of the
driving phase have subscript ``>med'', and those defined based on
averages within the $0.8$ mag contour are defined by the subscript
``$>0.8$''.

\begin{table*}
\begin{tabular}{cccccccccccc}
\hline
& \multicolumn{2}{c}{Physics included} & \multicolumn{2}{c}{IC} & \multicolumn{3}{c}{>med} & \multicolumn{3}{c}{>0.8} \\
$\mu_{\Phi}$ & Outflows & Driving & $\beta_{\rm IC}$  & $\mathcal{M}_{A,\rm IC}$ & $\rho_{\rm >med}~[{\rm g~cm}^{-3}]$ & $\beta_{\rm >med}$  & $\mathcal{M}_{\rm >med}$ & $\rho_{\rm >0.8}~[{\rm g~cm}^{-3}]$ & $\beta_{\rm >0.8}$  & $\maf$ \\ 
\hline
1.56 &      Y & N & 0.046    & 1.0      & $2.36\times10^{-19}$ & 0.041    & 0.95     & $3.21\times10^{-19}$ & 0.039    & 1.09 \\ 
1.56 &      N & N & 0.046    & 1.0      & $2.36\times10^{-19}$ & 0.041    & 0.95     & $1.45\times10^{-19}$ & 0.038    & 1.14 \\ 
1.56 &      Y & Y & 0.046    & 1.0      & $2.36\times10^{-19}$ & 0.041    & 0.95     & $9.76\times10^{-20}$ & 0.037    & 1.05 \\ 
\\
 2.17 &     Y & N & 0.089    & 1.39     & $2.18\times10^{-19}$ & 0.057    & 1.14     & $1.53\times10^{-19}$ & 0.074    & 1.92 \\ 
 2.17 &     Y & Y & 0.089    & 1.39     & $2.18\times10^{-19}$ & 0.057    & 1.14     & $2.35\times10^{-19}$ & 0.067    & 1.70 \\ 
 \\
 23.1 &     Y & N & 10.0     & 14.8     & $1.90\times10^{-19}$ & 0.46     & 3.14     & $1.62\times10^{-19}$ & 0.41     & 6.67 \\ 
 23.1 &     Y & Y & 10.0     & 14.8     & $1.90\times10^{-19}$ & 0.46     & 3.14     & $1.07\times10^{-19}$ & 0.75     & 4.45 \\ 
 \\
 $\infty$ & Y & N & $\infty$ & $\infty$ & $2.66\times10^{-19}$ & $\infty$ & $\infty$ & $1.33\times10^{-19}$ & $\infty$ & $\infty$ \\ 
 $\infty$ & Y & Y & $\infty$ & $\infty$ & $2.66\times10^{-19}$ & $\infty$ & $\infty$ & $9.24\times10^{-20}$ & $\infty$ & $\infty$ \\ \hline
\end{tabular}
\caption{
\label{t0}
Model parameters and initial conditions. The outflows and driving
columns indicate whether we included protostellar outflows, and
whether we continued driving the turbulence after gravity was turned
on. For the remaining parameters, the subscript ``IC'' denotes the
state in the uniform initial condition, the subscript ``>med''
denotes quantities averaged over gas denser than the median density
at the instant when self-gravity is switched on, and the subscript ``>0.8''
denotes averages over gas with a K-band extinction
$A_K>0.8~\textrm{mag}$ at the termination of the model.  See the text
for the details of how the the averaging regions are defined. 
The mean density in all cases is $4.46 \times 10^{-20}$ g cm$^{-3}$.
}
\end{table*}

\subsection{Physics Included and Numerical Method} \label{ssec:physics}

Our protostellar cluster simulations are performed using the
\textsc{orion2} adaptive mesh refinement (AMR) code \citep{licode}
beginning from the initial state generated following the procedure
described in the previous section.  The code solves the equations of
ideal magnetohydrodynamics (MHD) using the scheme of \citet{mignone},
along with coupled self-gravity \citep{truelove98,klein99} 
and radiation
transfer \citep{krumholzRad} in the two-temperature, mixed-frame,
grey, flux-limited diffusion approximation. The exact set of equations
solved are the same as those given in \citet{myers14}. \textsc{orion2}
contains many options for the numerical advection scheme, but we have
found the best stability with the HLLD Riemann solver \citep{miyoshi}
coupled to the constrained transport magnetic flux advection of
\citet{londrillo}, particularly for magnetically dominated flows.  Our
radiative transfer calculations use the frequency-integrated grey dust
opacities from the iron-normal, composite aggregates model of
\citet{semenov}.  For each magnetic field strength parameter
considered in this study, $\mu_\Phi=1.56$, 2.17, 23.1, and $\infty$,
we present two models -- one with decaying turbulence where the
large-scale driving forces described in the previous section are
turned off, and one where the turbulent forcing continues with a
constant rate of energy injection that balances the rate of turbulent
decay as estimated for a given mean magnetic field strength and
turbulent Mach number by \citet{maclow99}.  In the case without
turbulent forcing, self-gravity is switched on and turbulent forcing
is switched off simultaneously.

Our cluster evolution models neglect non-ideal MHD effects.  These effects have been shown to impact turbulent density profiles on small scales \cite{liad2} and therefore could influence the fragmentation dynamics of strongly magnetized cores.  Recent studies on the impact of non-ideal MHD effects on the evolution and fragmentation of self-gravitating cores indicate that the evolution of super-critical cores is most strongly influenced by their initial conditions and that the impact of non-ideal effects on their subsequent evolution and fragmentation are small by comparison \cite{wurster}.  Consequently, our models should reliably capture the gross impacts of varying large-scale magnetic field strength on the cloud fragmentation and the statistics of its collapse.

There are two limitations of
our radiative transfer method to which we should call attention. First, we
assume that the gas and dust temperatures are the same.
In the dense regions simulated by \citet{myers14} this is a very good
assumption, because the dust and gas temperatures become nearly
identical at densities above $\sim 10^4 - 10^5$ cm$^{-3}$ 
\citep[e.g.,][]{goldsmith01}. The assumption of strong coupling
is still valid for most of the gas in our simulation domain that
is actively participating in star formation 
(c.f.~$\rho_{>,\rm med}$ in \autoref{t0}), but it is not valid for
the lower density, non-self-gravitating regions. In these parts
of the flow we are somewhat overestimating the dust cooling rate
for the gas. 
The second approximation we make is to assume
that the absorption opacity is the same as the emission opacity, which
is equivalent to assuming that the dust and radiation temperatures are the same.
This approximation is clearly invalid very close to the star,
where dust is exposed to direct stellar radiation
and reprocesses the radiation into the infrared. However, the dust
opacity to direct stellar radiation is so high that all this
reprocessing occurs in a thin zone at the dust destruction front,
which for low mass star formation is confined to $\lesssim 1$ AU
from the star (except perhaps over the small range in solid
angle evacuated by outflows). Such small structures are
unresolved in our simulations.
Our approximation would only become problematic for stars larger
than $\sim 20$ $M_\odot$, massive enough that radiation pressure
can inflate bubbles that push the dust destruction front out to
thousands of AU \citep[e.g.,][]{rosen16}. While \textsc{orion2}
does include a hybrid radiative transfer method to handle this
situation \citep{rosen17}, we do not use it here because our
simulations do not form any stars massive enough to require it.
Once the radiation field is dominated by
reprocessed radiation, the dust and radiation temperatures
will remain close so long as the dust is opaque. Outside the
dust photosphere, the emission opacity declines while the absorption opacity
remains constant,
but this affects only the low-frequency part of the emission spectrum,
which has only a small fraction of the energy \citep{chak05}.\footnote{The 
choice of the temperature at which one
should evaluate the mean opacity for a gray method such as ours
is in fact very subtle, even in
calculations that separately track the dust, gas, and radiation
temperatures. This is because tabulated opacity tables, such as the
ones from \citet{semenov} that we use, generally assume that all
three temperatures are equal, and different temperatures matter for
different physical effects -- for example, condensation of ice
mantles on grains depends on the gas temperature, evaporation of
grains depends on the dust temperature, and the radiation spectrum
depends on the radiation temperature. Absent tabulated opacities
that consider out-of-equilibrium dust, gas, and radiation fields,
there is no single choice of temperature at which the
opacity can be computed that properly captures all these divergent
effects. 
However, greater accuracy could be obtained by distinguishing the
absorption opacity, which is primarily affected by the radiation
temperature, and the emission opacity, which is primarily determined by the 
dust temperature.} 

Our models include several source terms to capture the effects of
protostellar feedback that originate on smaller length scales than
those resolved in the model.  We initialize a sink particle in any zone
on the finest AMR level, and only on the finest level, that becomes
dense enough to reach a local Jeans number 
\begin{equation}
J=\sqrt{\frac{G \rho \Delta x^2}{\pi c_s^2}} > \frac{1}{4}. \label{jeansref}
\end{equation}
Sink particles
then evolve and interact with the cluster through gravity according
to the methodology of \citet{krumholz04}, updated to include the
effects of magnetic fields on the rate of gas accretion onto sink
particles (see the appendix of \citealp{lee14}).  The key assumption
in our treatment is that the point mass accretes mass, but not
flux. This assumption is motivated by observations that show that the
magnetic flux in young stellar objects is orders of magnitude less
than that in the gas that formed these objects, implying that flux
accretion is very inefficient \citep{mckeeostriker}.

The sink particles couple to their surrounding gas gravitationally and 
through accretion and feedback source terms that operate within a radius of $4 \Delta x$ on the finest AMR mesh level around each sink particle.
The sink particles emit radiation according to the protostellar
evolution model of \citet{offner09}.  
Our models include the atomic
line-cooling sources described in \citet{cunningham11} to treat
strongly heated regions behind outflow shocks that are sufficiently
fast $\sim 30~\textrm{km s}^{-1}$ to dissociate molecules.

Feedback due to protostellar
outflows is also included via momentum sources around each sink
particle following the procedure described in \citet{cunningham11}:
the outflow mass ejection rate is set by the fraction of accreted gas
that is ejected into the outflow $f_w$ and the outflow ejection speed
$v_w$.  In this work we make the same wind model parameter choices as
\citet{hansen} by setting $f_w=0.3$ and 
$v_{\rm Kep} = \min (v_{\rm Kep}, \,
60~\textrm{km s}^{-1})$, where $v_{\rm Kep}$ is the Kepler speed at
the surface of the protostar.  We note that these parameters differ
from the parameters used in \citet{cunningham11} and \citet{li17}
that set $v_w = v_{\rm Kep}/3$.  Consequently, the models presented
here impart relatively more outflow feedback during the earlier phases
of prestellar evolution relative to \citet{cunningham11} and
\citet{li17}.  Computational expediency motivates the cap on the
wind velocity so that isolated massive protostars do not dominate the
time-step of the cluster model. 

The AMR hierarchy is initialized on a $256^3$ base grid that we denote
as level $\mathcal{L}=0$. Our initial state is taken from turbulent
initial conditions that were established on a $512^3$ grid with no
adaptivity. After $t=0$, regions of the flow within 2 base-grid zones
of an interface with a density jump $\Delta \rho/\rho > 0.5$ on the
base level are refined to $\mathcal{L}=1$ so that the $512^3$ resolution
of the initial state is retained in these regions.  On all AMR levels,
refinement is triggered in 
regions in which the local Jeans scale is resolved by fewer than 8
zones, $J>1/8$ \citep{truelove97}.  Note that this refinement criterion
triggers refinement to the finest level, $\mathcal{L}_{\rm MAX}$, at a density one fourth that
required to trigger insertion of a sink particle.  This ensures that
strongly collapsing regions are refined to the finest level.
Refinement is also triggered when there is a large jump in radiation
energy density, $|\nabla E_r|\Delta x/E_r > 0.125$. 

A resolution sensitivity test was conducted for the case with
$\mu_\Phi=2.17$ by first performing this run with $\mathcal{L}_{\rm
  MAX}=4$ and then repeating it without the density gradient
refinement triggering on level $\mathcal{L}_0$ and with
$\mathcal{L}_{\rm MAX}=3$.  We found that the total mass accreted onto
protostars out to $t=1.35 t_{\rm ff}$ in the two models agreed to
within $3\%$ and that the mass distribution of protostars had a
similar level of agreement.

Our cluster models with decaying turbulence begin with
$\mathcal{L}_{\rm MAX}=4$.  The effective resolution of all collapsing
and/or heated regions is therefore $L/(256 \times 2^{\mathcal{L}_{\rm
    MAX}}) = 32$~AU.  Given the good agreement between the runs with
$\mathcal{L}_{\rm max}=3$ and $\mathcal{L}_{\rm max}=4$, the
$\mu_\Phi=1.56$ and $2.17$ models with decaying turbulence were
switched to $\mathcal{L}_{\rm MAX}=3$ with $65$~AU effective
resolution at $t=1.7 t_{\rm ff}$ in order to reduce the run time.  The
models with turbulent forcing were also performed with
$\mathcal{L}_{\rm MAX}=3$ and $65$~AU effective resolution, except for
the $\mu_\Phi=23.1$ case which retained $\mathcal{L}_{\rm MAX}=4$
throughout the evolution of the model.  

We note that the limited resolution of these models results in a gain
in computational expediency but comes at some cost to the fidelity of
the model in treating the fragmentation of low mass cores.  The
density threshold triggering sink formation at our background $10$ K
temperature is given by inverting \autoref{jeansref} as
$1.1\times10^{-15}$ g cm$^{-3}$ to $4.5\times10^{-15}$ g cm$^{-3}$ for
$65$~AU and $32$~AU effective resolution.  The isothermal-to-adiabatic
transition where compressional heating balances cooling occurs at a
substantially higher density of $\sim 10^{-14} (T/10{\,\rm K})^6$ g
cm$^{-3}$ (see chapter 16 of \citet{krumholzbook}), and thus our
resolution would not be sufficient to capture fragmentation in a
simulation without radiative transfer. However, once stars form,
radiative heating powered by accretion luminosity heats material at
much lower densities than one would infer simply by balancing
adiabatic compression against cooling. This substantially eases the
resolution required to capture fragmentation, and we show below that
our resolution is sufficient that in practice we essentially always
resolve the region where gas transitions from the background
temperature to elevated temperatures around most protostars in the
simulations. Nonetheless, we caution that our limited resolution may
still cause us to miss fragmentation in rare regions with little
radiative heating.

\section{Results} \label{results}

In this section we examine the results of the simulations summarised
in \autoref{t0}. In the discussion and figure legends that follow we
denote the various models by their initial mass-to-flux ratio; cases
that include turbulent driving are labeled with the designation
``Driven''. We have one model with outflows turned off, to which we
refer as ``no wind''; this model also is not driven.

\subsection{Protostellar Mass Accretion} \label{s3.1}

\begin{figure*}
  \begin{center}
    \includegraphics[clip=true,width=0.45\textwidth]{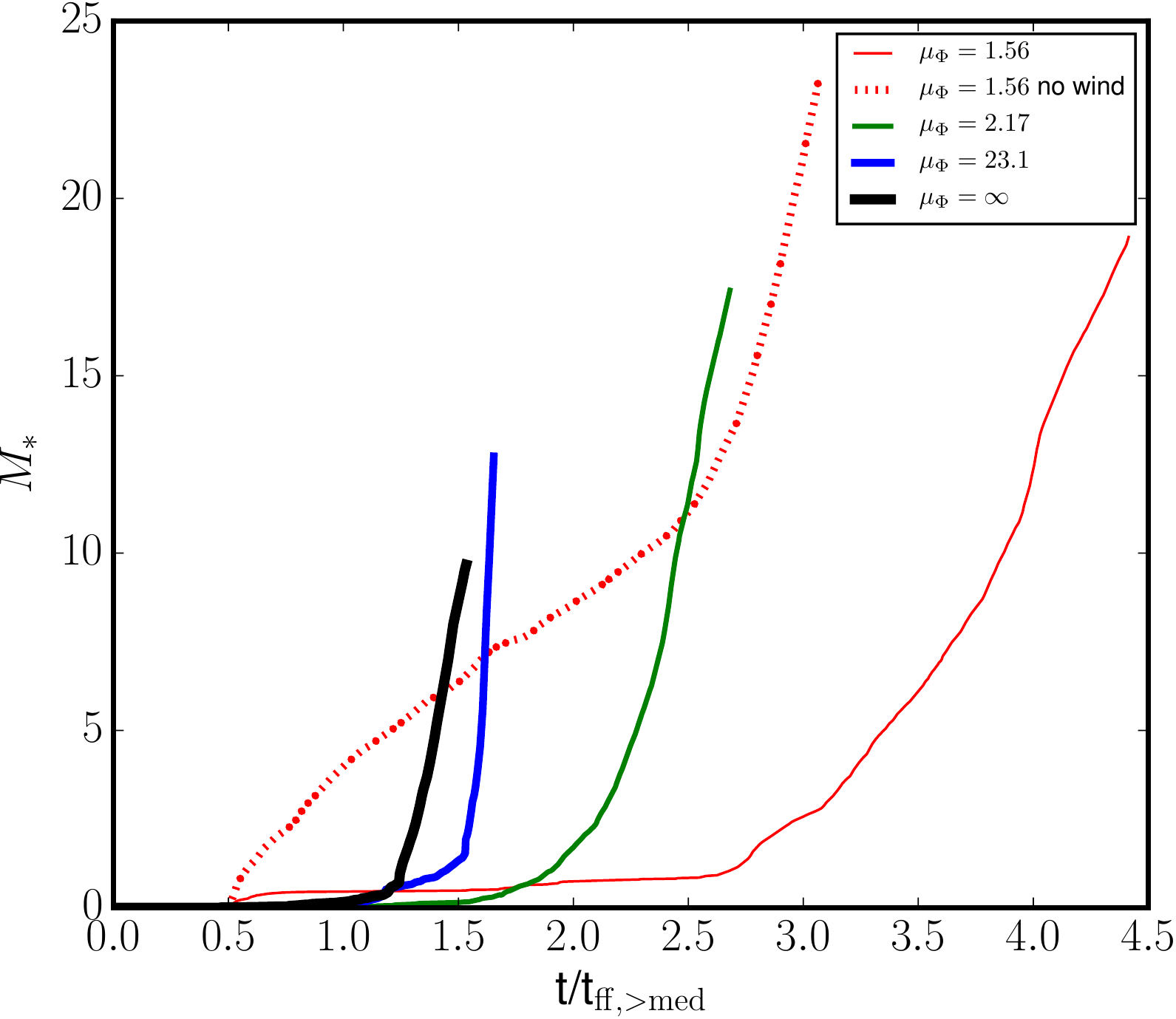}
    \includegraphics[clip=true,width=0.45\textwidth]{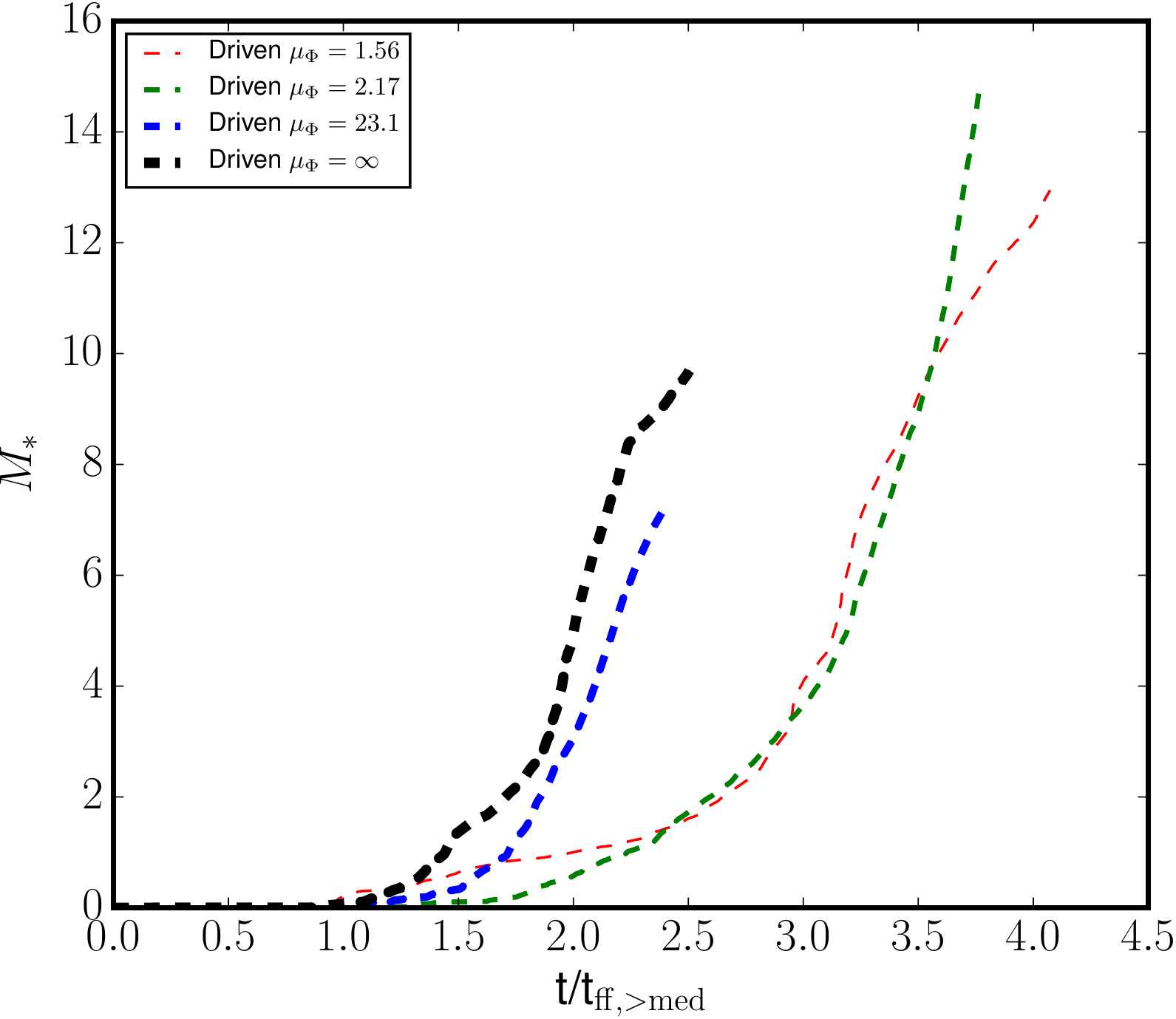} \\
  \end{center}
  \caption{Total mass accreted for the models without (left) and with (right) turbulent forcing. \label{f1}}
\end{figure*}

\begin{figure*}
  \begin{center}
    \includegraphics[clip=true,width=0.45\textwidth]{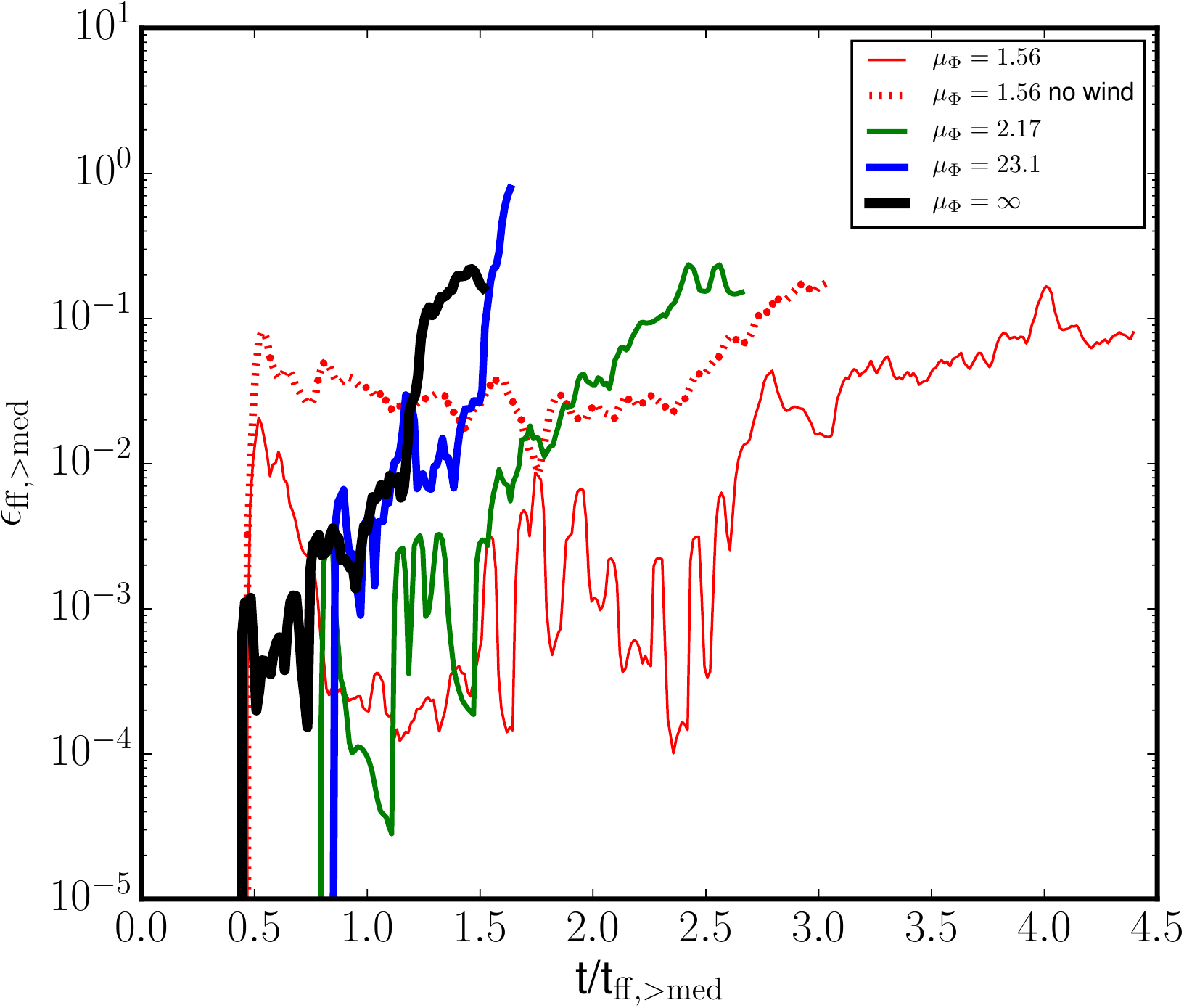}
    \includegraphics[clip=true,width=0.45\textwidth]{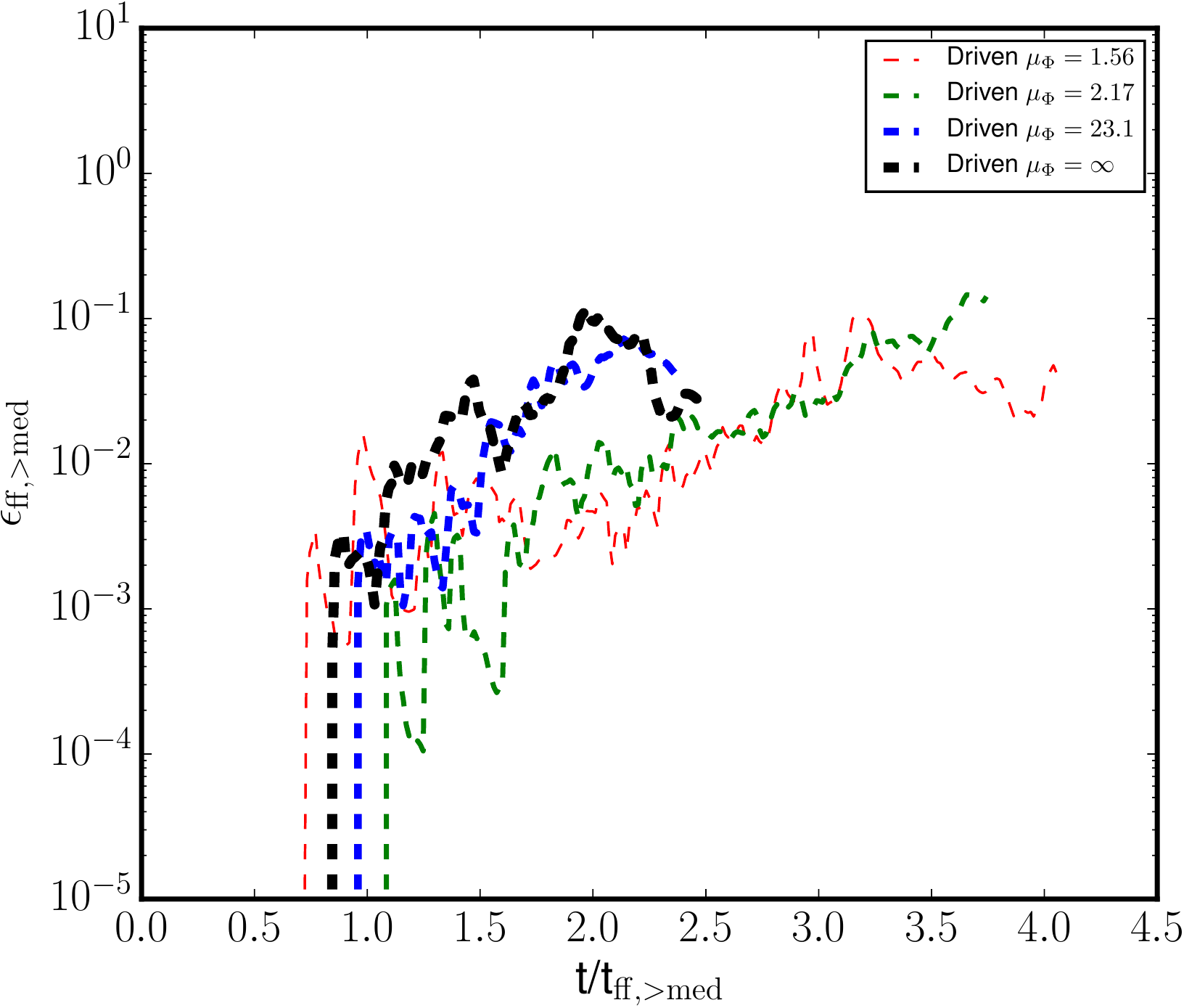} \\
  \end{center}
  \caption{Instantaneous value of the dimensionless star formation rate
  $\epsilon_{\rm ff,>med}$ for the models without (left) and with (right)
  turbulent forcing (right). The data have been boxcar-smoothed over a
  timescale of $0.05~t_{\rm ff,>med}$ to remove short-timescale variability.
  \label{f2}}
\end{figure*}

\begin{figure*}
  \begin{center}
    \includegraphics[clip=true,width=0.45\textwidth]{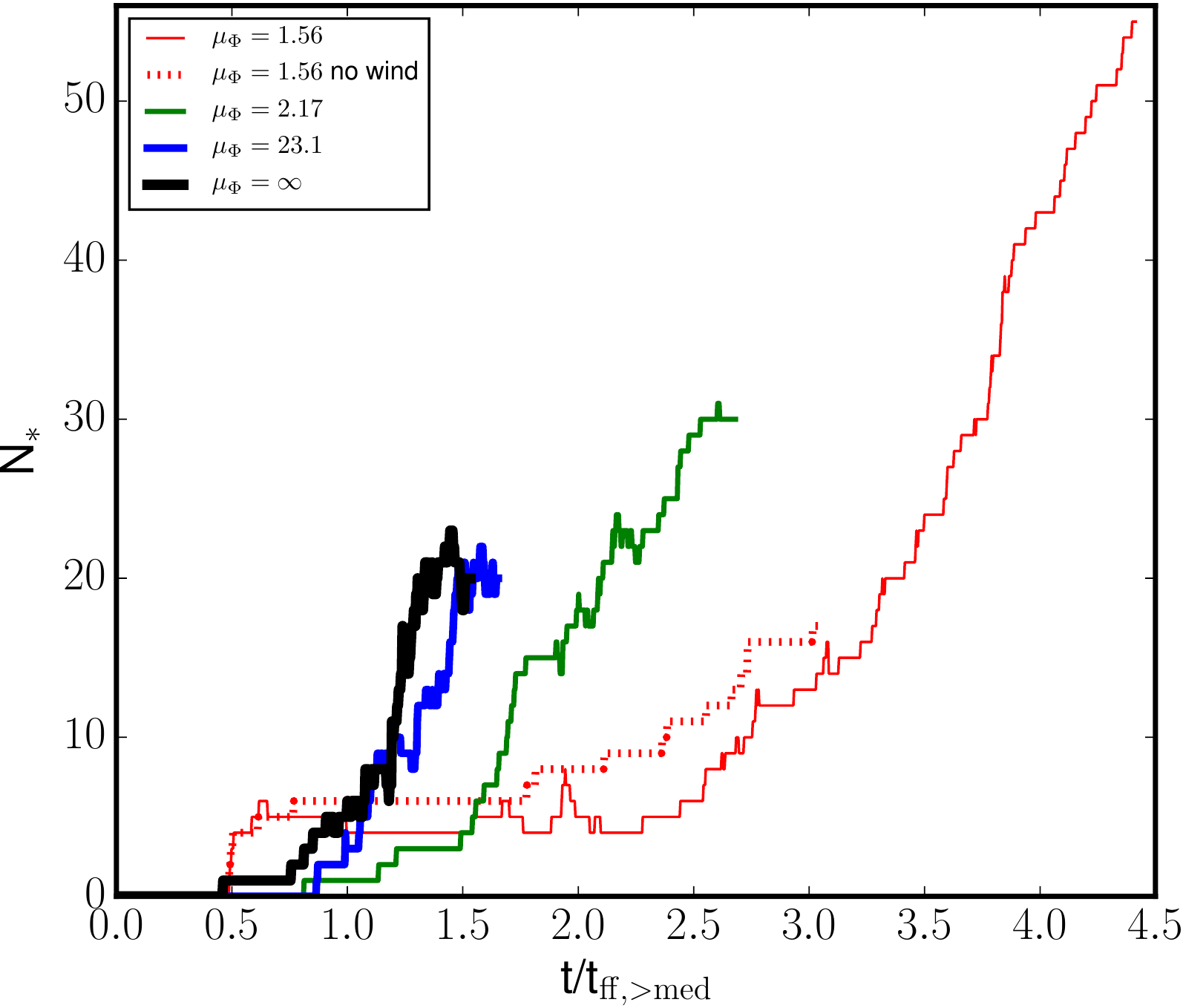}
    \includegraphics[clip=true,width=0.45\textwidth]{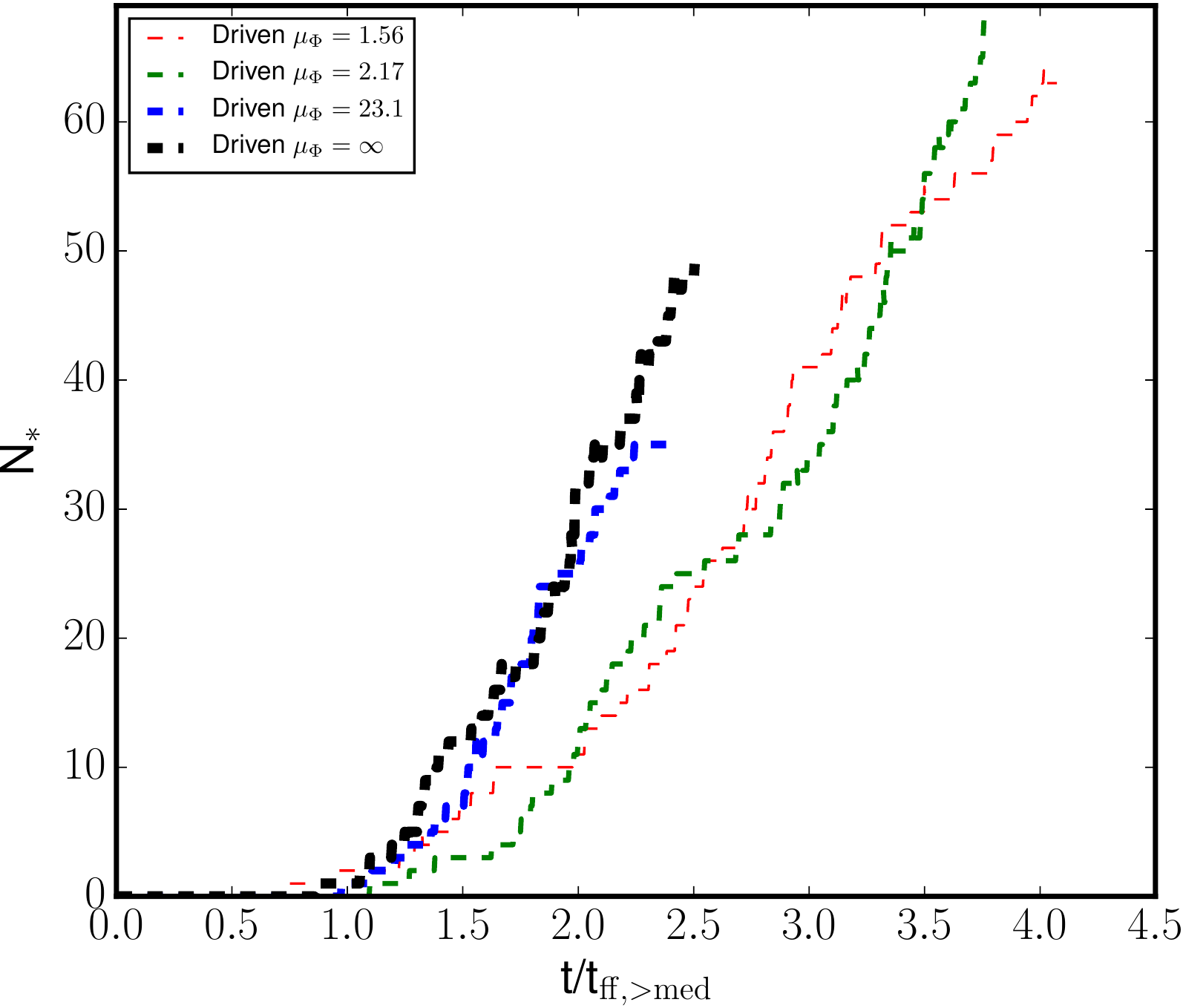} \\
  \end{center}
  \caption{Total number of protostars formed for the models without (left)
    and with (right) turbulent forcing. \label{f3}}
\end{figure*}

\autoref{f1} shows the temporal evolution of the total mass $M_*$
assembled into protostars for each model as a function of time.  The
left panel shows the simulations with decaying turbulence and the
right panel shows the simulations with large-scale turbulent driving,
a convention we shall follow throughout this paper.  \autoref{f2}
shows the star formation efficiency per initial free fall time
$\epsilon_{\rm ff,>med}$ (see \autoref{ssec:free-fall} for the precise
definition) as a function of time. \autoref{f3} shows the number of
protostars formed in each model as a function of time.

We also report, in \autoref{t1}, both $\epsilon_{\rm ff,>med}$
and the observed value of the star formation efficiency
per free-fall time, $\epsilon_{\rm ff,>0.8}$ (again see \autoref{ssec:initial_conditions}).
The quantities reported in the Table are  time-weighted averages over the
last $0.5 t_{\rm ff,>med}$ of the simulation. For most models we run the simulation
until the value of $\epsilon_{\rm ff,>med}$ stabilises, and thus the values reported
in \autoref{t1} are the steady-state ones. There are two exceptions to this
statement. First, the two weakest field models with
decaying turbulence ($\mu_\Phi=23.1$ and $\infty$) do not appear to have
well-converged values of $\epsilon_{\rm ff}$, and instead appear to approach
near-free fall collapse. We run these models for less time than the others, as
such efficient collapse is generally inconsistent with the observational
constraints. For them the values of $ \epsilon_{\rm ff}$ reported in \autoref{t1}
are only lower limits, and we flag them as such in the Table. The second
exception are the models with $\mu_\Phi=2.17$. We are forced to halt
these runs due to computational constraints, as the number of sink
particles and highly refined zones eventually becomes such
that exploring these models further is prohibitively expensive. 
While we would, absent consideration of computation
constraints, prefer to continue the $\mu_\Phi=2.17$ runs further, we note
that these models have begun to exhibit stabilized $\epsilon_{\rm ff.0}$
by the termination time. Thus we do not report these as upper limits in
\autoref{t1}, but warn readers here that they are less secure than the
values reported for the other runs.

For comparison, \citet{kdm12} combine data from galactic
clouds, nearby galaxies, and galaxies at high redshift, and find that all
are consistent with a star formation efficiency $\epsilon_{\rm  ff,>0.8}
=0.015$, with a scatter of $\sim 0.5$ dex. As noted in \autoref{introduction},
numerous other studies have confirmed this basic result.
Comparing to the values shown in \autoref{f2} and
reported in \autoref{t1}, we see that the models with $\mu_\Phi = 1.56$ fall within
the observational envelope if we include either driving or outflows, as do all the
more weakly magnetized models with driven turbulence, although these tend
to lie toward the top end of the observationally-permitted range.  We therefore
find that the star formation rate can be kept low enough to be compatible with
observations either if there are external mechanical energy inputs to maintain
the turbulence or if there are no such influences, but the magnetic field lies 
close to the critical value.

\begin{table*}
\begin{tabular}{ccccccc}
  \hline 
  $\mu_{\Phi}$ & Outflows & Driving & $\epsilon_{\rm ff, IC}$ & $\epsilon_{\rm ff,>med}$ & $\epsilon_{\rm ff,>0.8}$ \\
  \hline 
  $\mu_\Phi = 1.56$   & Y & N & 12\%            & 8.2\%           & 6.0\% \\ 
  $\mu_\Phi = 1.56$   & N & N & 23\%            & 15.8\%          & 19\% \\
  $\mu_\Phi = 1.56$   & Y & Y & 4.8\%           & 3.2\%           & 4.2\% \\ \\
  $\mu_\Phi = 2.17$   & Y & N & 26\%            & 18.8\%          & 19\% \\
  $\mu_\Phi = 2.17$   & Y & Y & 15\%            & 11.0\%          & 8.7\% \\ \\
  $\mu_\Phi = 23.1$   & Y & N & $\gtrsim 30\%$  & $\gtrsim 24\%$  & $\gtrsim 22\%$ \\ 
  $\mu_\Phi = 23.1$   & Y & Y & 8.2             & 6.4\%           & 6.6\% \\ \\
  $\mu_\Phi = \infty$ & Y & N & $\gtrsim 50\%$  & $\gtrsim 34\%$  & $\gtrsim 37\%$ \\ 
  $\mu_\Phi = \infty$ & Y & Y & 6.0\%           & 4.0\%           & 5.2\% \\ 
\hline
\end{tabular}
\caption{
\label{t1}
Star formation efficiency per free fall time in each run. The various values of $\epsilon_{\rm ff}$ are computed for different definitions of the free-fall time, as discussed in \autoref{ssec:free-fall}. The quantities reported are averaged over the last $0.5 t_{\rm ff,>med}$ of each simulation. A value preceded by $\gtrsim$ indicates that $\epsilon_{\rm ff}$ was still rising strongly and the end of the simulation, and the values listed are therefore lower limits.
}
\end{table*}


\subsection{Cluster Morphology} \label{s3.2}

\begin{figure*}
  \begin{center}
    \includegraphics[width=0.32\textwidth]{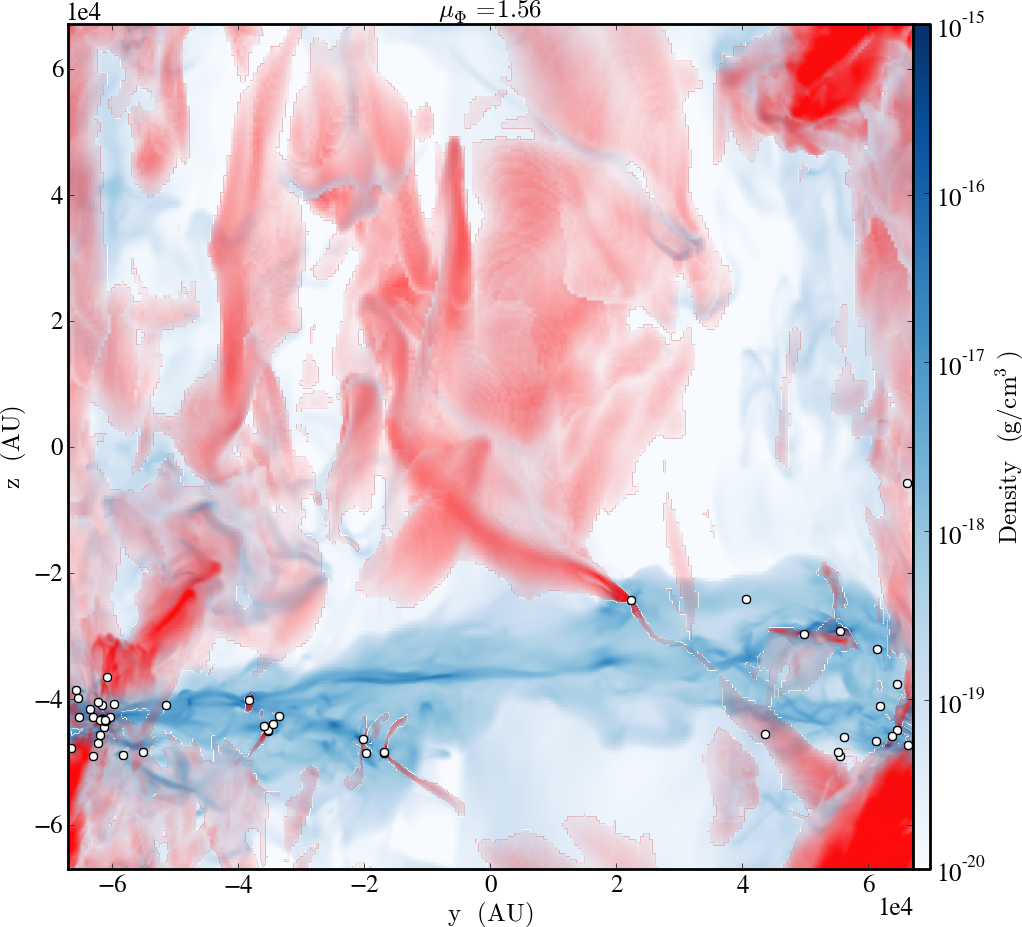}
    \includegraphics[width=0.32\textwidth]{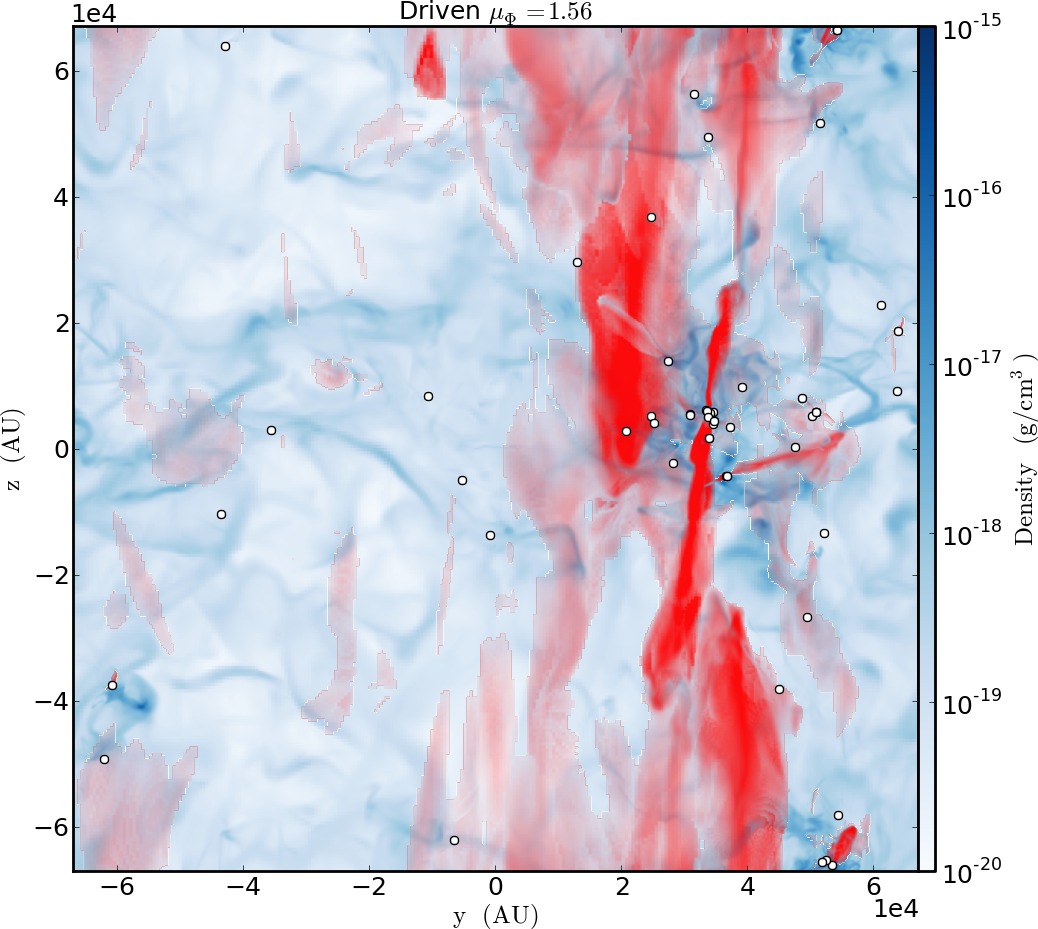} \\
    \includegraphics[width=0.32\textwidth]{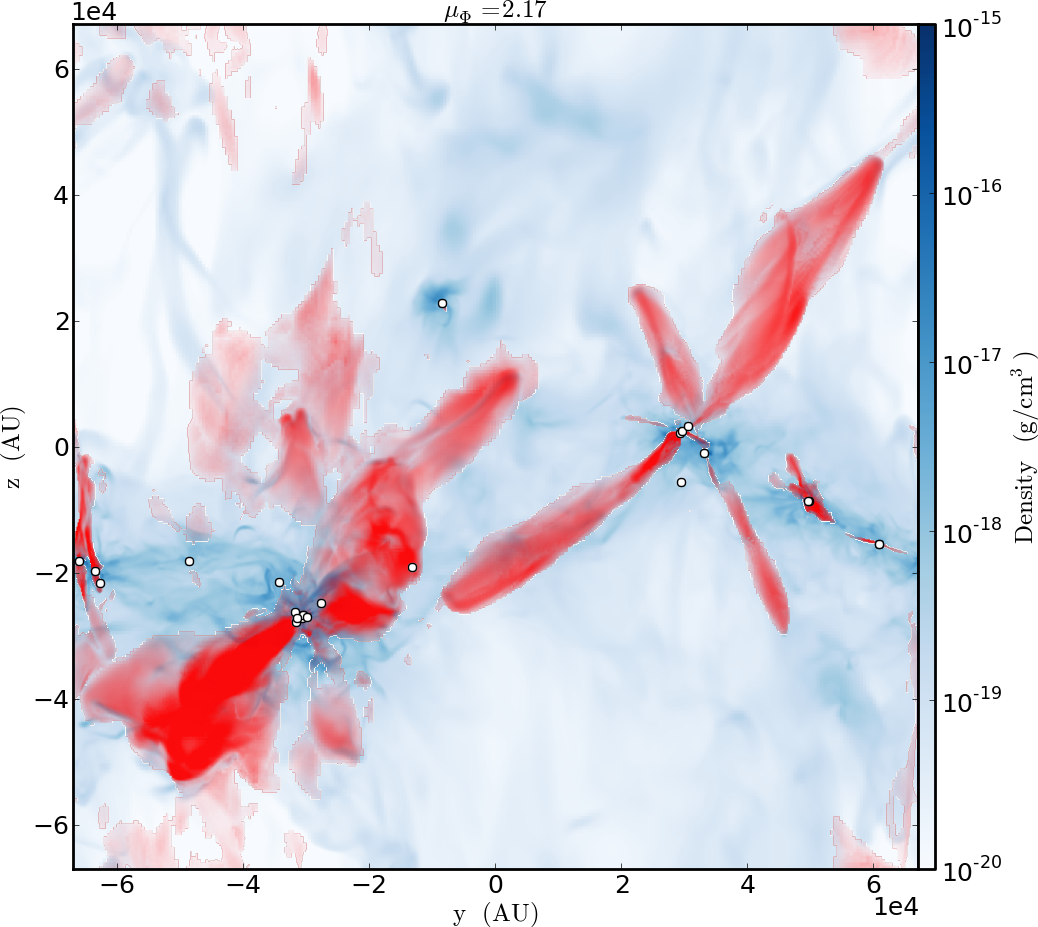}
    \includegraphics[width=0.32\textwidth]{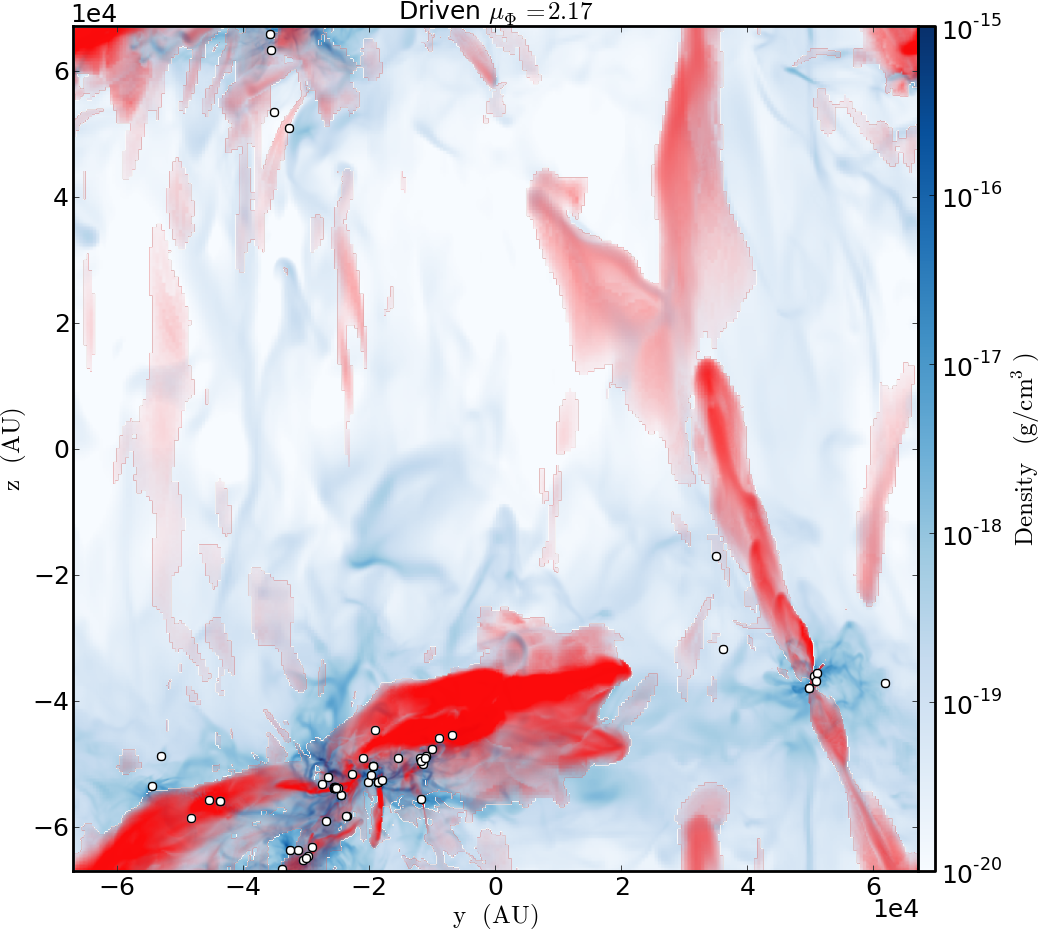} \\
    \includegraphics[width=0.32\textwidth]{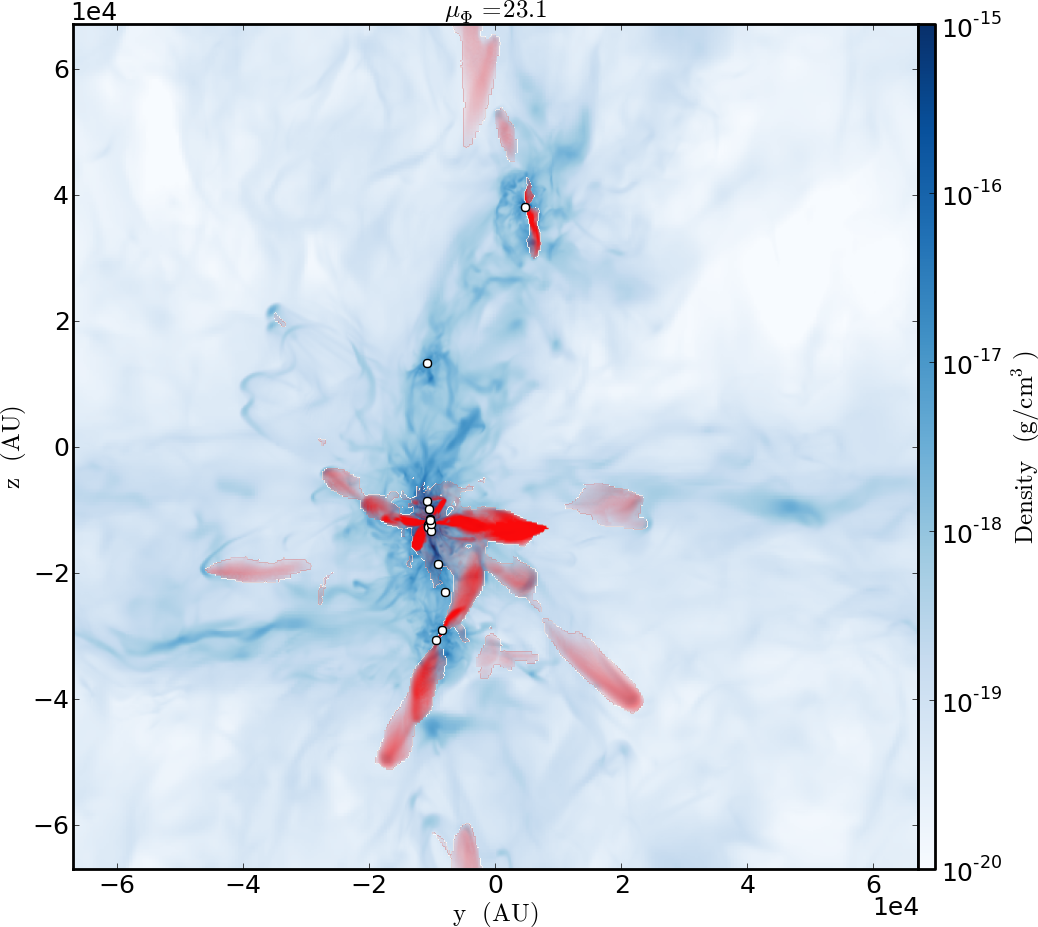}
    \includegraphics[width=0.32\textwidth]{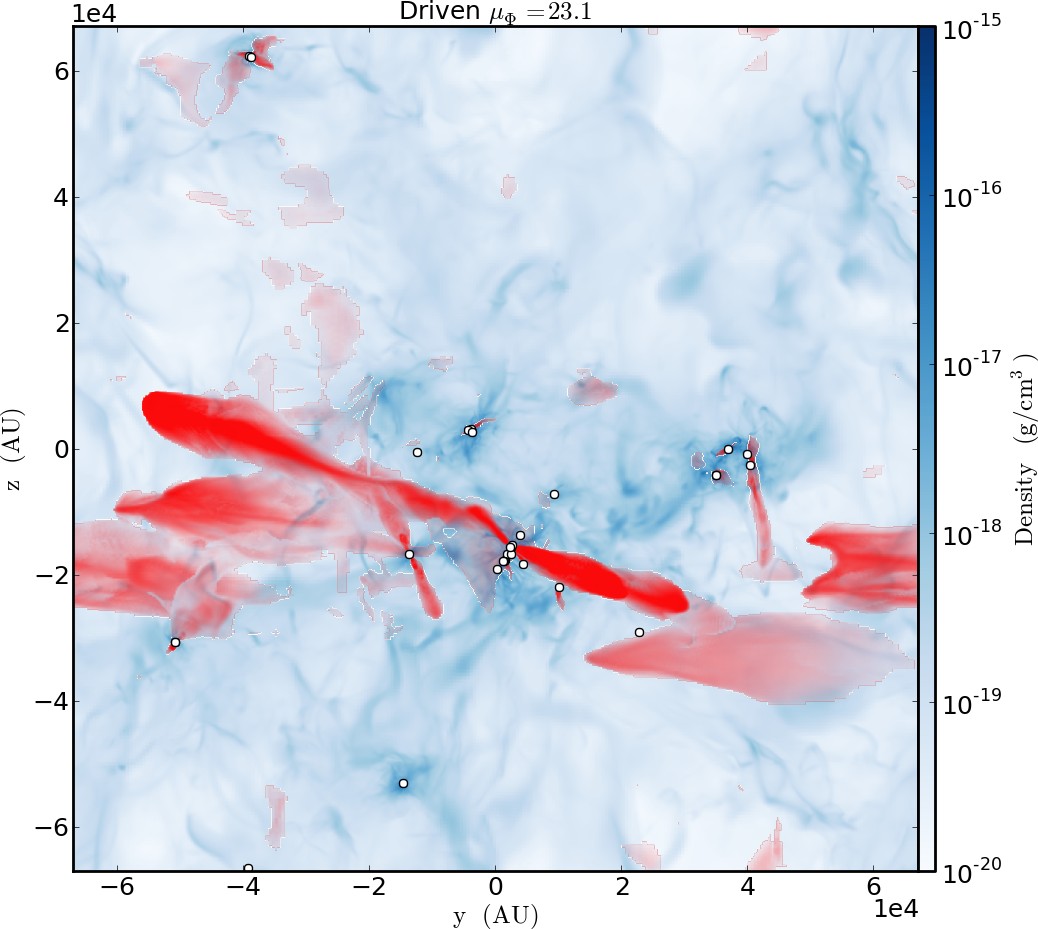} \\
    \includegraphics[width=0.32\textwidth]{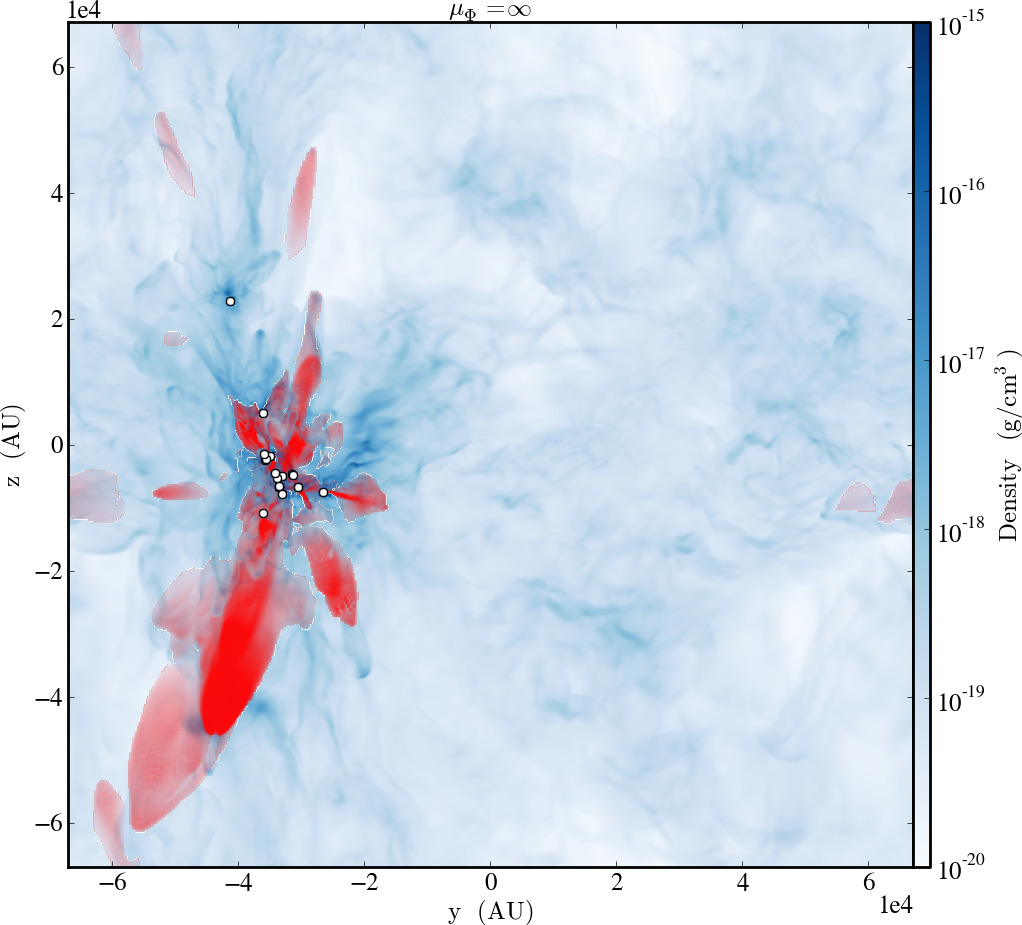}
    \includegraphics[width=0.32\textwidth]{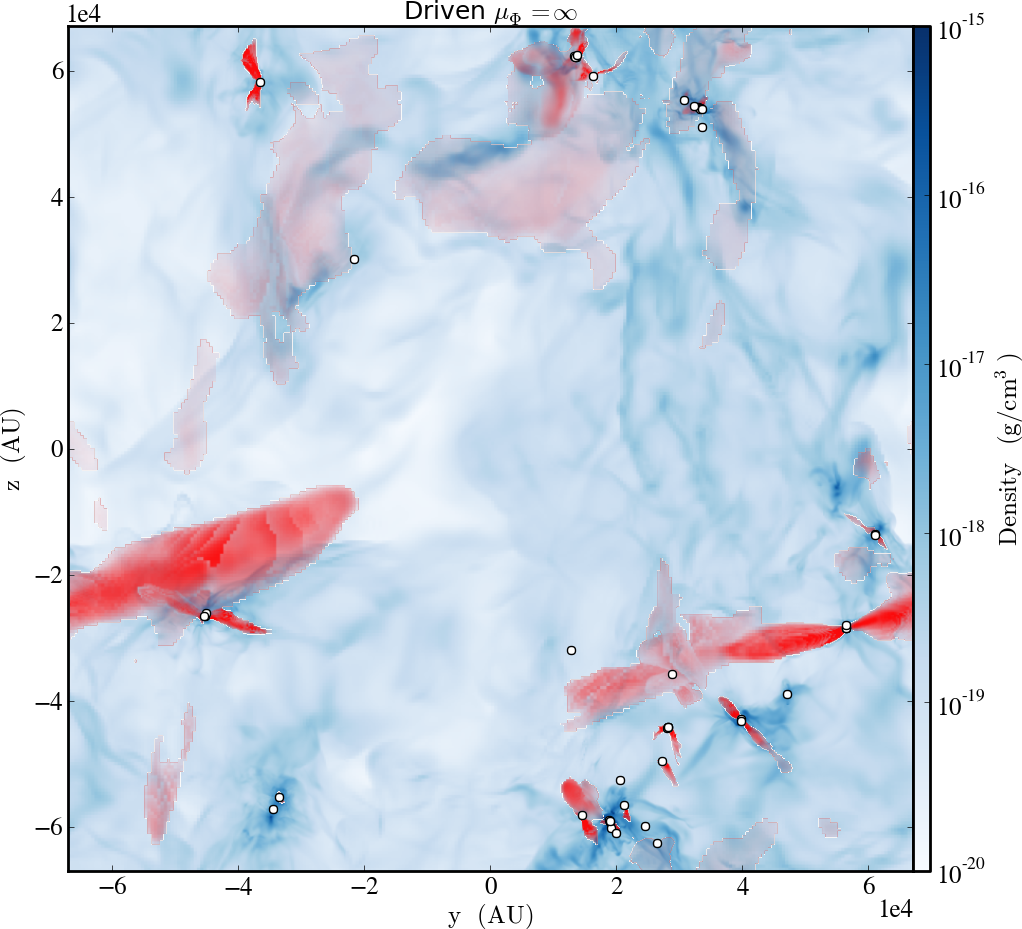} \\
    \hspace*{0.32\textwidth}
  \end{center}
  \caption{Volume-weighted mean densities projected along the $\hat{\vec{x}}$
    direction for the models with
    turbulent forcing (right) and with decaying turbulence (left).  The
    column density can be recovered by multiplying the mean density shown by
    the computational domain length $0.65~{\rm pc}$; for reference, this means
    that a density of $10^{-18}$ g cm$^{-3}$ corresponds to a column density
    of $2.0$ g cm$^{-2}$.  The four rows
    show the $\mu_\Phi=1.56$, $\mu_\Phi=2.17$, $\mu_\Phi=23.1$, and
    $\mu_\Phi=\infty$ runs, from top to bottom.  Regions with gas velocity
    greater than $2v_\rms$ are shaded in red with the opacity of the
    red regions increasing to a maximum for velocities exceeding
    $5v_{\rm RMS}$. White circles indicate the projected position of
    the protostellar sink particles.\label{f4}}
\end{figure*}

\autoref{f4} shows the 
mean density along lines of sight in the $\hat{\vec{x}}$ direction
at the termination
time of each model, overlaid with a semi-transparent rendering of
regions with $|v|>2~v_\rms$. 
Recall that the initial
magnetic field is in the $\hat{\vec{z}}$ direction, and is therefore
upwardly oriented on the page in \autoref{f4}.  We
note that the 
mean density along the line of sight, which is proportional to the surface density,
becomes increasingly
filamentary as the strength of the magnetic field increases.  In the
cases where the magnetic field is the most dynamically significant,
the outflows are oriented predominately perpendicularly to the
orientation of the densest structures. The $\mu_\Phi=1.56$ case with
decaying turbulence indicates a final cluster geometry that is most
dominated by the magnetic field.  In this case, the cluster is
arranged mostly in a planar geometry normal to the $\hat{\vec{z}}$ direction.
The magnetic fields in this case are strong enough to constrain large-scale collapse to be
predominantly in the direction of the initial magnetic field, with less
harassment from large-scale turbulent forcing than the equivalent
model with turbulent forcing.

\subsection{Protostellar Mass Functions} \label{s3.3}

\begin{figure*}
  \begin{center}
    \includegraphics[clip=true,width=0.4\textwidth]{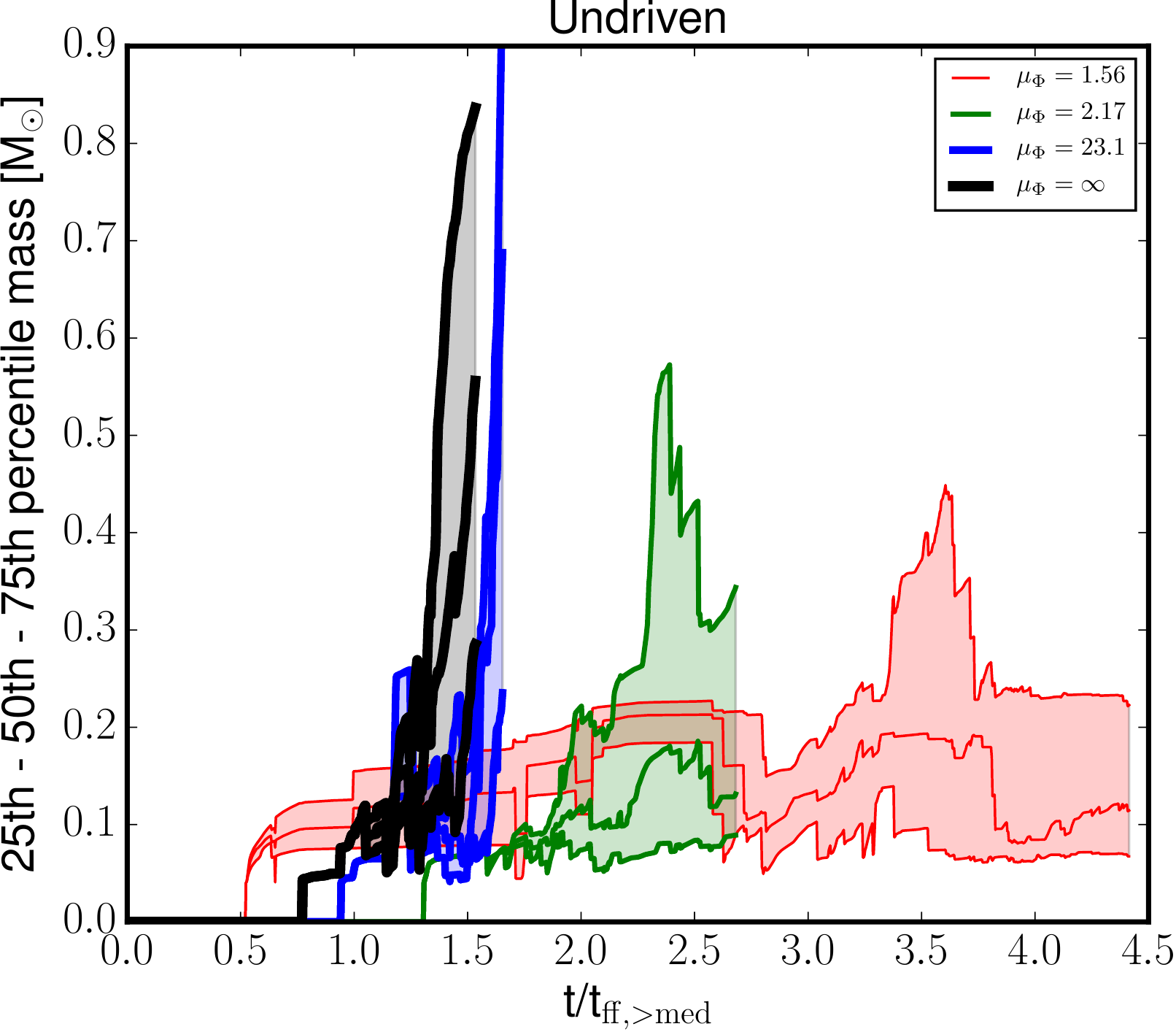}
    \includegraphics[clip=true,width=0.4\textwidth]{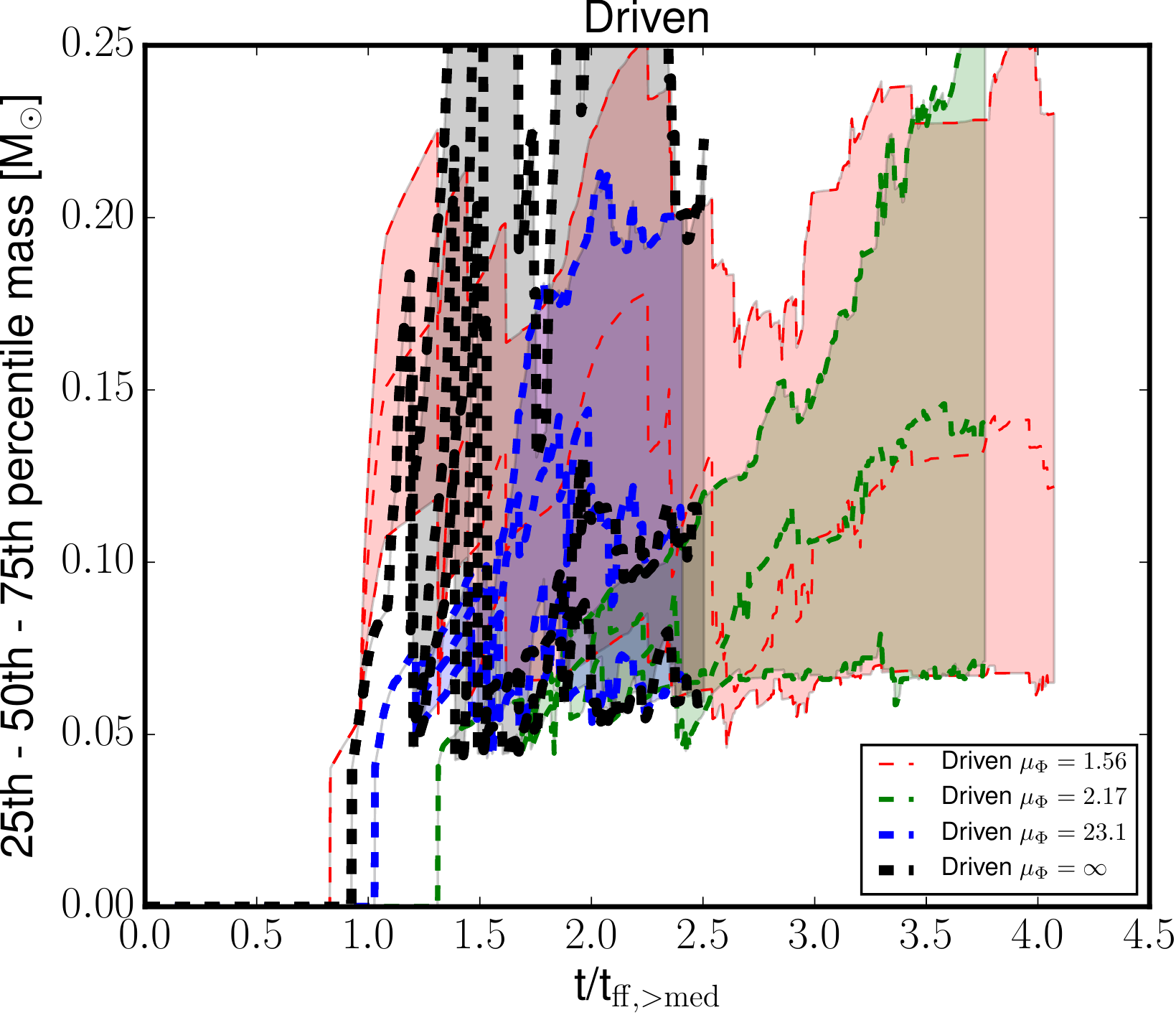}
  \end{center}
  \caption{The evolution of the stellar mass distribution for models with turbulent
    forcing (right) and with decaying turbulence (left). In these figures, the
    central lines show the 50$^{\rm th}$ percentile mass, while the shaded regions
    indicate the $25^{\rm th}-75^{\rm th}$ percentile range. Colours correspond to
    different values of $\mu_\Phi$, as indicated in the legend. \label{f5}}
\end{figure*}

We next examine the mass functions for the stars formed in our
simulations.  As a first step to this, we verify that these
distributions have come close to being stable in time. To demonstrate
this, in \autoref{f5} we plot the temporal evolution of the $25^{\rm
  th}$ percentile, median and $75^{\rm th}$ percentile of the
distribution of protostellar masses in each model as a function of
time.  Similar to the temporal evolution of $\epsilon_{\rm ff}$, we
see that the protostellar mass percentiles have mostly stabilised by
the end of the runs, except for the cases of a weak initial field
($\mu_\Phi=23.1$ and $\infty$) with decaying turbulence.
Consequently, we can interpret the protostellar mass distributions of
the moderate and strong initial field models with decaying turbulence,
and all of the models with driven turbulence, as being only weakly
time-dependent. We do warn, however, that this stabilisation is a
result of an equilibrium between new stars forming at low mass, and
existing stars growing. As noted in \citet{krumholz16}, if some
process were to halt the formation of new stars but the existing stars
were to continue accreting their envelopes, the mass distribution
would shift upward.

\begin{figure*}
  \begin{center}
    \includegraphics[width=0.32\textwidth]{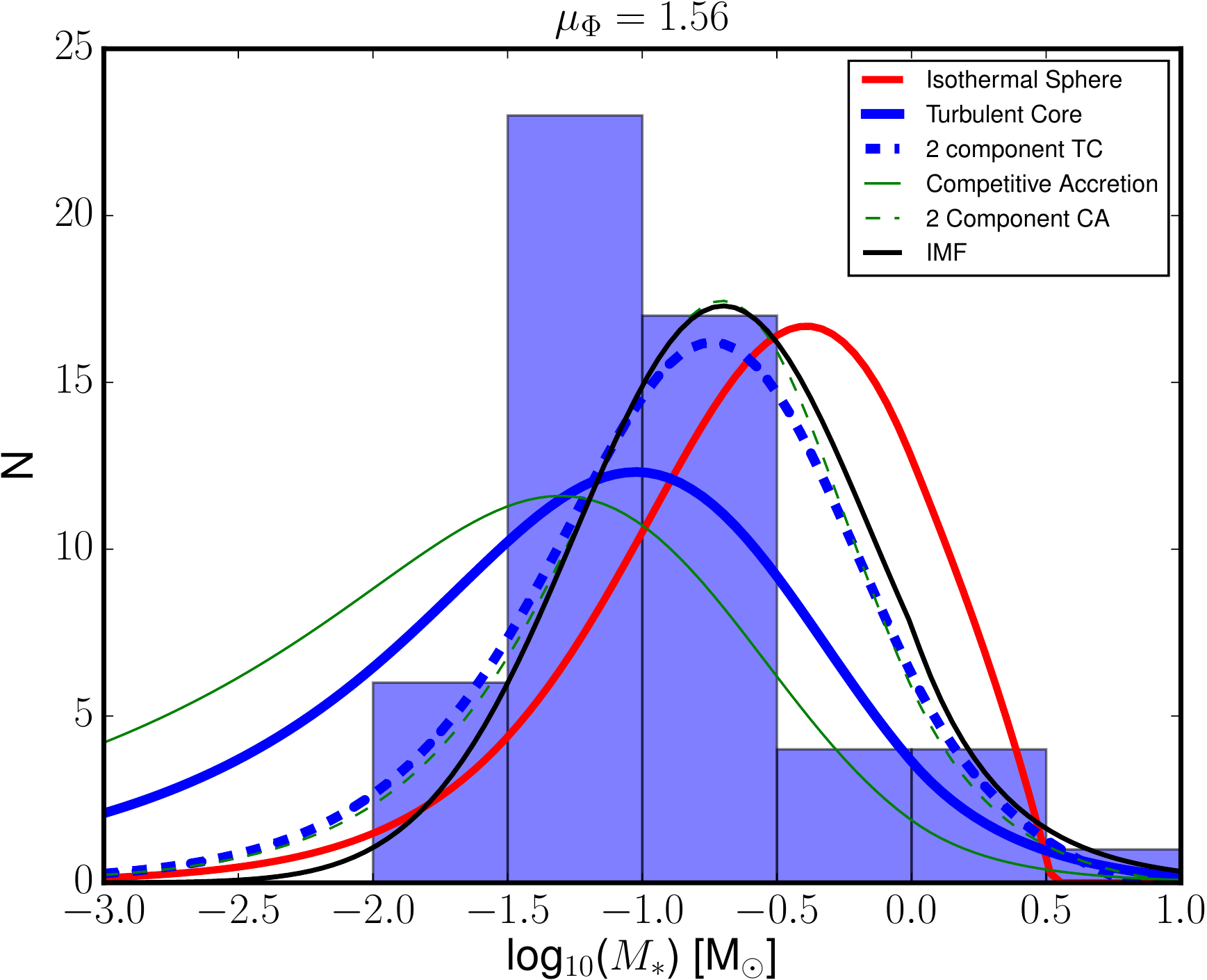}
    \includegraphics[width=0.32\textwidth]{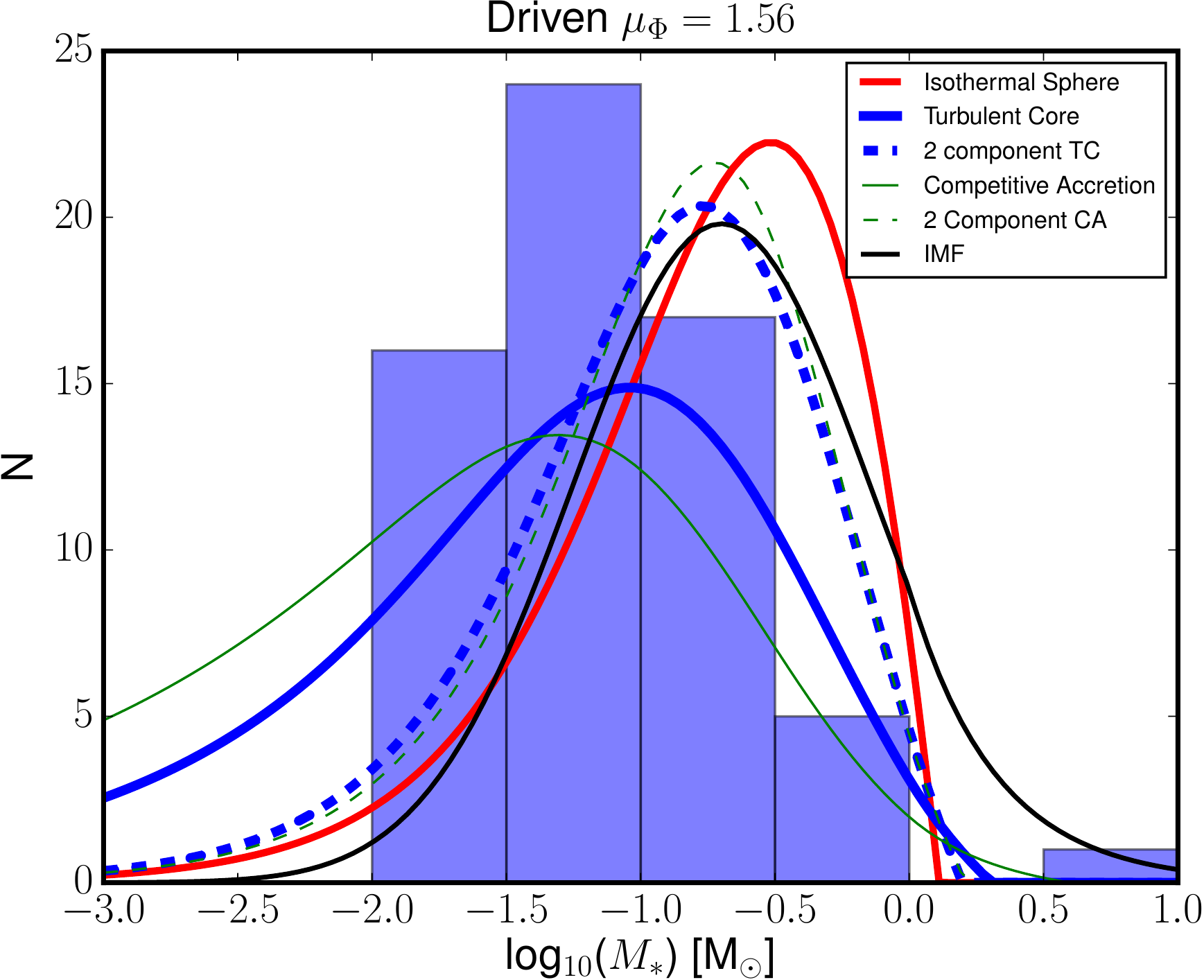} \\
    \includegraphics[width=0.32\textwidth]{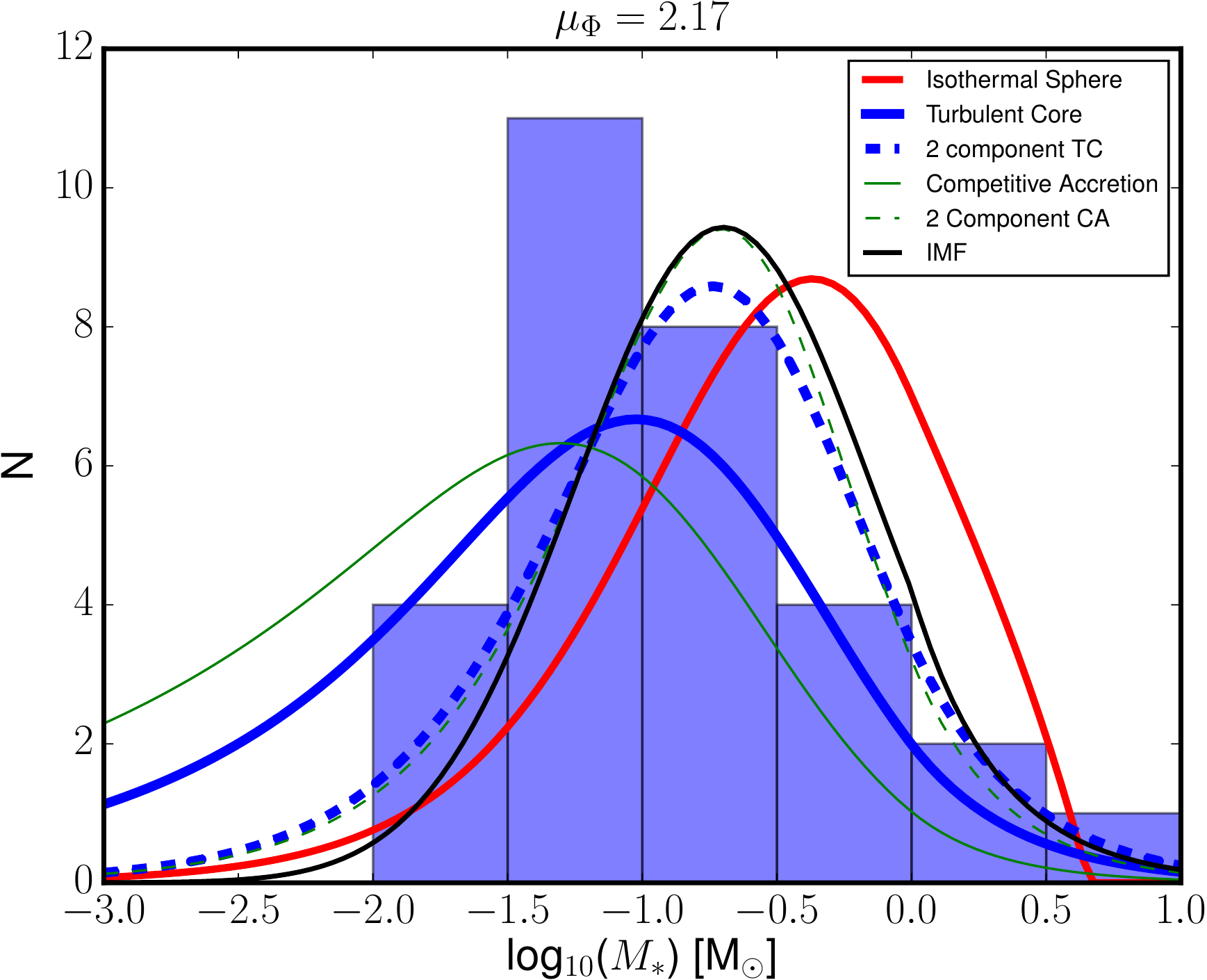}
    \includegraphics[width=0.32\textwidth]{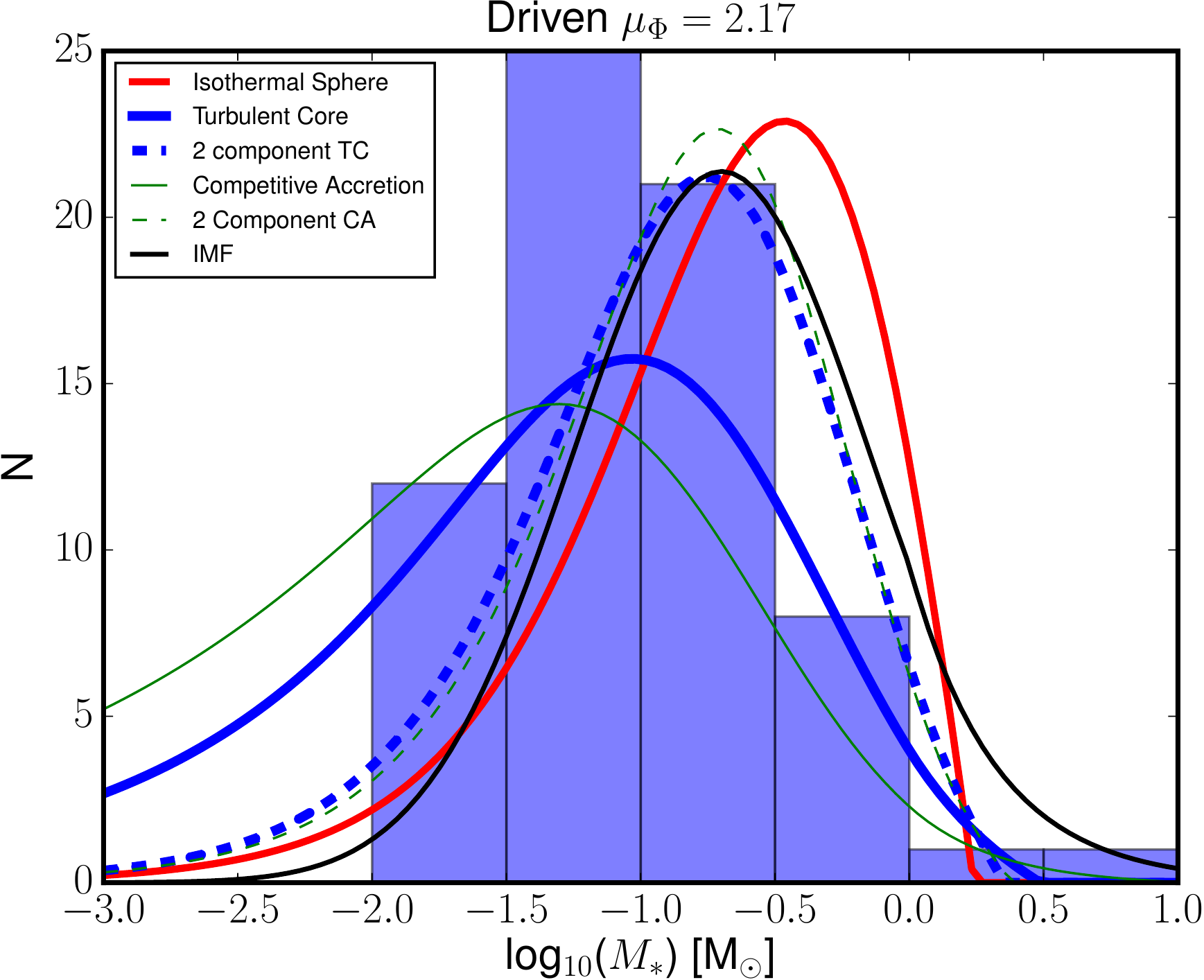} \\
    \includegraphics[width=0.32\textwidth]{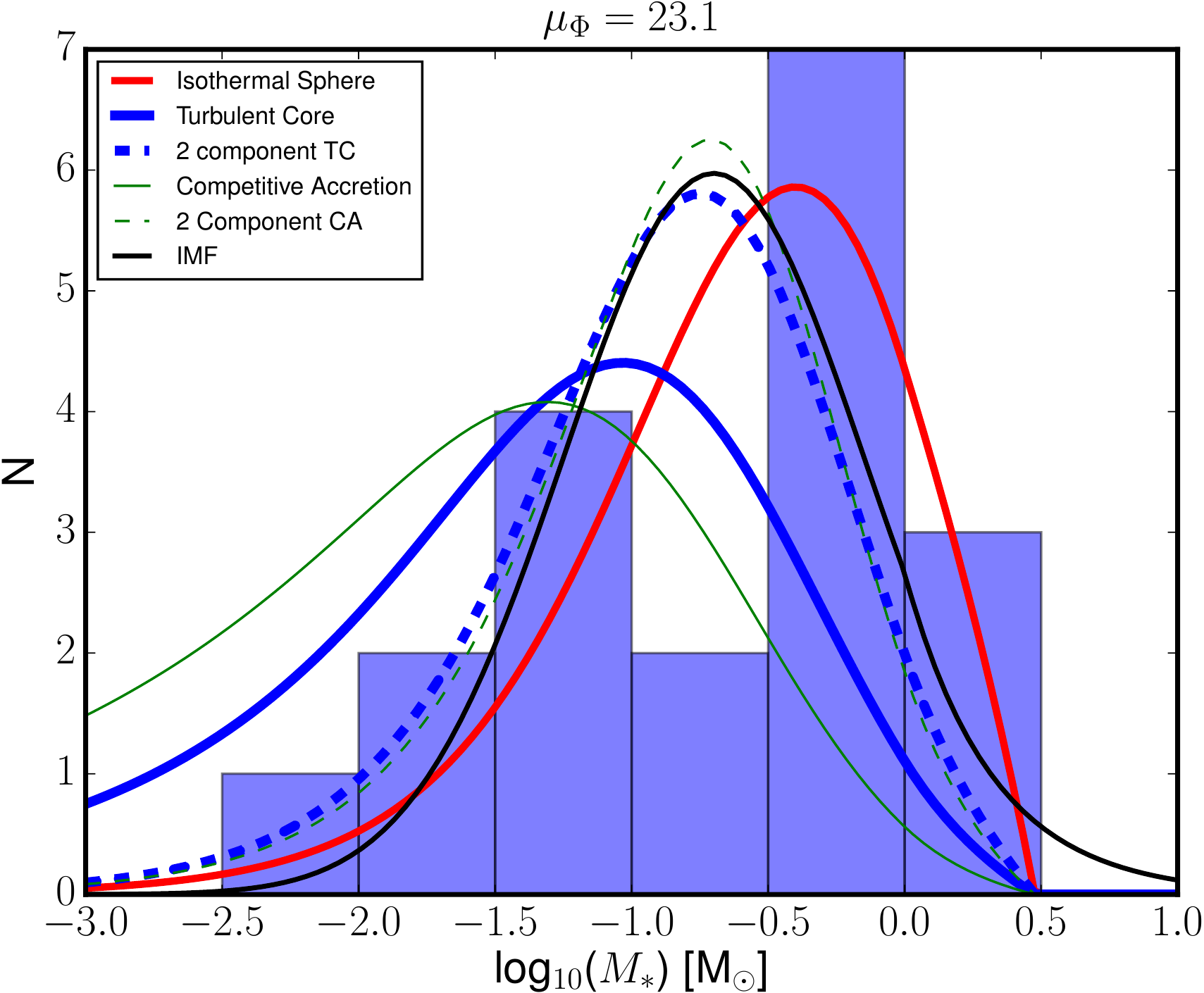}
    \includegraphics[width=0.32\textwidth]{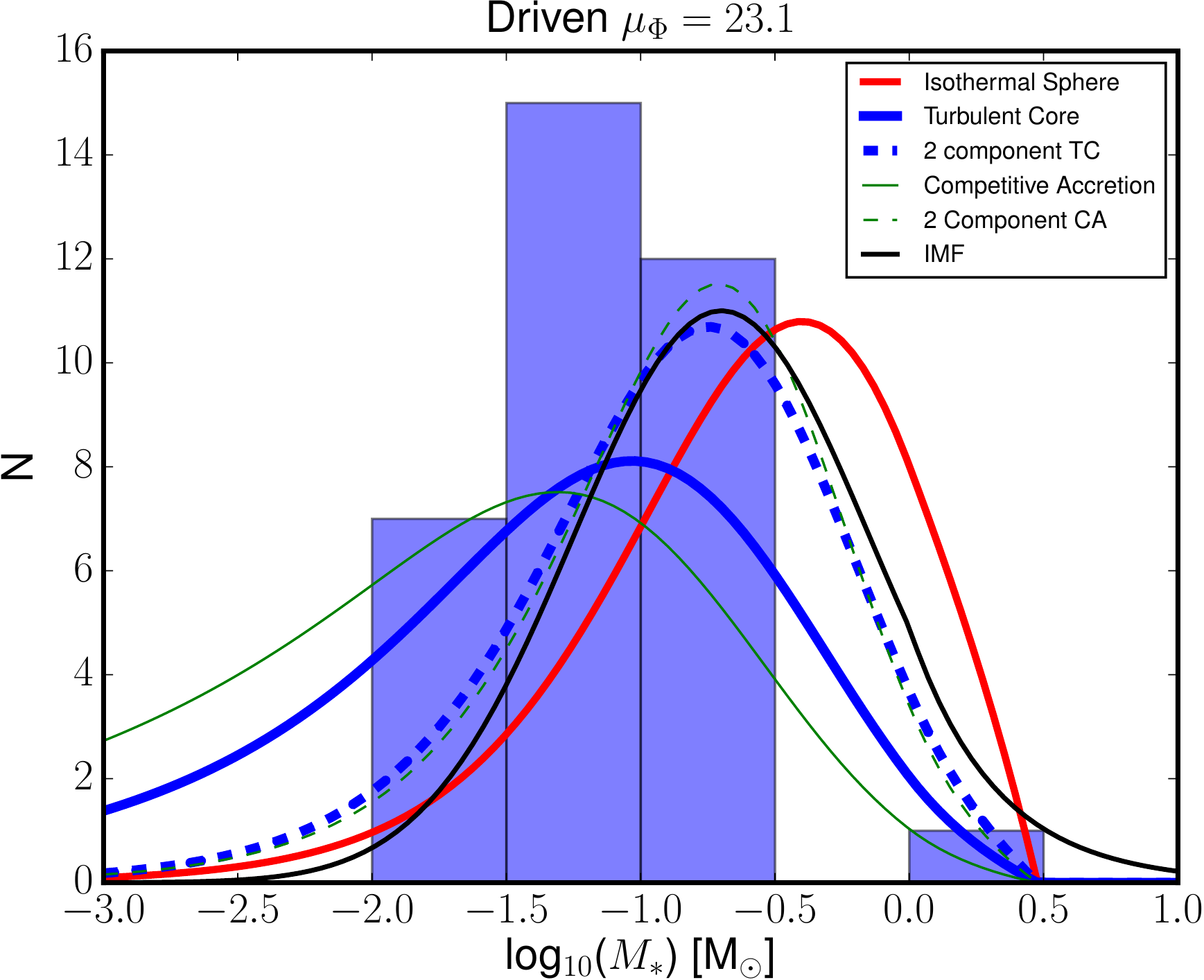} \\
    \includegraphics[width=0.32\textwidth]{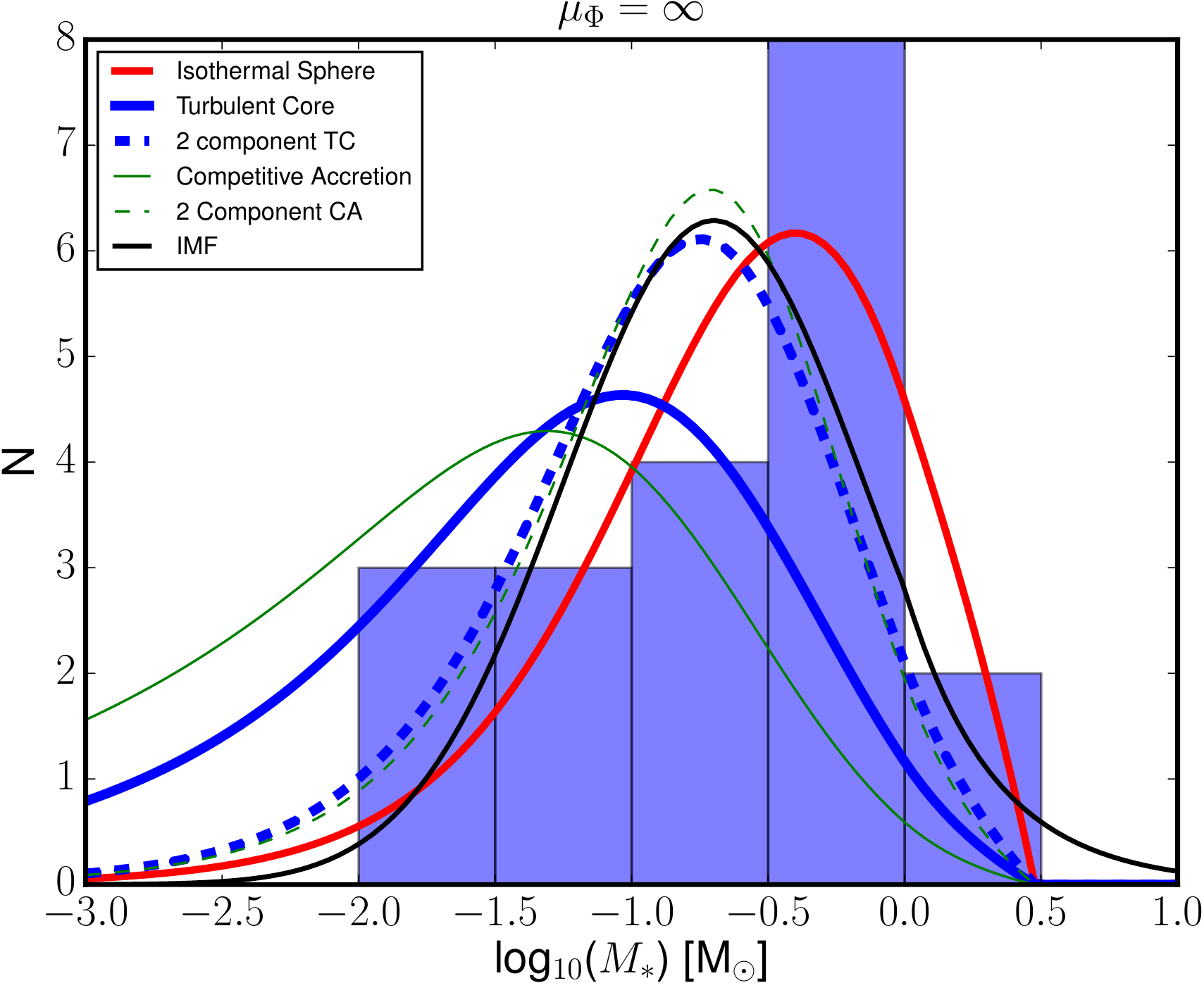}
    \includegraphics[width=0.32\textwidth]{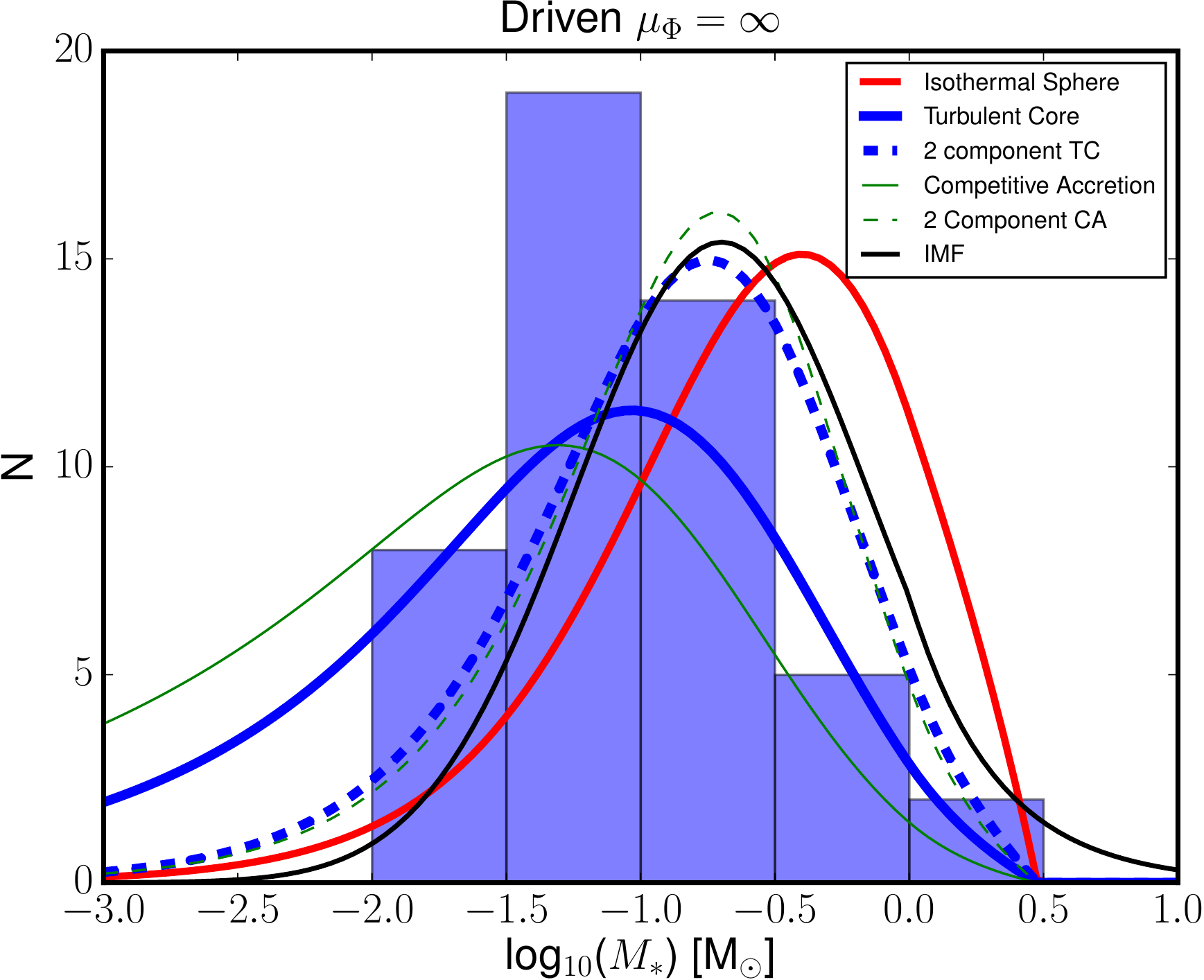} \\
  \end{center}
  \caption{Histograms of protostellar masses at the end of each of the
    models with turbulent forcing (right) and with decaying turbulence
    (left).  The four rows show the $\mu_\Phi=1.56$, $\mu_\Phi=2.17$,
    $\mu_\Phi=23.1$, and $\mu_\Phi=\infty$ from top to bottom.  The
    black line shows the observed initial mass function
    \citep{chabrier2005}.  The other curves denoted in the figure
    legend show the analytic protostellar mass functions of
    \citet{mckeeoffner10} for the single and two component turblent
    core models, the single and two component competetive accretion
    models and the isothermal sphere model. \label{f6}}
\end{figure*}

\autoref{f6} shows the protostellar mass distributions at the
termination of each model.  The black curve in the figure shows the
observed initial mass function of \citep{chabrier2005}.  The mass distributions in our models have stabilized over dynamical timescales ($t_{\rm ff, >med}$), as shown in figure \ref{f5}. However,
because the sources in our models are still accreting, our models are
not exactly comparable to the initial mass function that is
representative of the distribution of final stellar masses.
For this reason, we also compare our models to the 
protostellar mass functions (PMFs) from the analytic theory of \citet{mckeeoffner10}; these
PMFs are the mass distributions that one expects a series of protostars
to possess if their final mass distribution matches the
\citet{chabrier2005} IMF, and if the numbers of stars formed versus
time $N_*(t)$ and the accretion histories of individual stars
follow some specified functional form. We observe that $N_*(t)$
is a roughly linear function in all simulations except those
that display runaway collapse, and thus we adopt a linear
function for $N_*(t)$. For the accretion rate versus time of
individual protostars, \citet{mckeeoffner10} consider a range
of cases, all of which we plot together with the simulation
results in \autoref{f6}: competitive accretion (CA),
turbulent core (TC), 2-component competitive accretion (2CCA),
and 2-component turbulent core (2CTC); the latter two both
approach the constant accretion rate expected for an isothermal
sphere for low mass sources (IS), and we also plot the pure IS
case.\footnote{We do not consider any
of \citet{mckeeoffner10}'s ``tapered accretion" analytic models, since
the median accretion rate in our simulations shows no signs of
tapering (\autoref{f5}).} The only other parameter required by
the \citet{mckeeoffner10} PMF analytic model is the upper limit to the
most massive star that will form in the cluster
$m_{\rm u}$.  We have generated the theoretical PMFs in
\autoref{f6} by setting $m_{\rm u}$ to the larger of $3~\msol$ or the
value of $m_{\rm u}$ necessary so that the theoretical PMF has exactly
one source heavier than the second-most massive particle present in
the at the termination of the simulation.  The choice of $m_{\rm u}$
largely affects the shape of the high-mass tail of the PMF and the
conclusions that follow from comparing our simulation to the
theoretical PMFs are insensitive to this choice.  The $p$-value
of the Kolmogorov-Smirnov (K-S) statistic comparing each simulation
to each mass function is reported in table \ref{t3}.  
The relative disagreement between simulations and mass functions 
pairs with $p$-values less than 0.05 is unlikely due to random 
chance.  These distribution pairs are different with high
statistical confidence.  We denote the $p$-value of these pairs 
in bold text in the table.

\begin{table*}
\begin{tabular}{lcccccc}
\hline
Comparison Analytic Model & IMF                     & TC                       & 2CTC                    & CA                      & 2CCA                     & IS \\
\hline
$\mu_\Phi=1.56$          & {\bf 3.2$\times10^{-4}$} & 0.094                    & {\bf 1.7$\times10^{-3}$} & 0.69                    & {\bf 1.0$\times10^{-3}$} & {\bf 1.1$\times10^{-7}$} \\
$\mu_\Phi=1.56$ No Wind  & {\bf 1.1$\times10^{-3}$} & {\bf 7.7$\times10^{-5}$} & {\bf 1.0$\times10^{-3}$} & {\bf 6.0$\times10^{-6}$} & {\bf 4.4$\times10^{-4}$} & 0.054 \\
$\mu_\Phi=2.17$          & 0.19                    & 0.92                     & 0.27                     & 0.56                    & 0.25                     & {\bf 0.013} \\
$\mu_\Phi=23.1$          & 0.058                   & {\bf 4.2$\times10^{-3}$} & {\bf 0.014}              & {\bf 1.1$\times10^{-3}$} & {\bf 9.9$\times10^{-3}$} & 0.35 \\
$\mu_\Phi=\infty$        & {\bf 0.50}              & {\bf 1.3$\times10^{-3}$} & {\bf 0.011}              & {\bf 1.4$\times10^{-4}$} & {\bf 9.2$\times10^{-3}$} & 0.35 \\
$\mu_\Phi=1.56$ Driven   & {\bf 3.0$\times10^{-3}$} & 0.52                    & 0.13                     & 0.91                     & 0.080                   & {\bf 2.1$\times10^{-3}$}\\
$\mu_\Phi=2.17$ Driven   & {\bf 3.5$\times10^{-3}$} & 0.72                    & 0.086                    & 0.94                     & {\bf 0.050}             & {\bf 1.6$\times10^{-4}$}\\
$\mu_\Phi=23.1$ Driven   & {\bf 0.033}             & 0.19                    & {\bf 0.032}               & 0.63                     & {\bf 0.033}             & {\bf 2.9$\times10^{-5}$}\\
$\mu_\Phi=\infty$ Driven & {\bf 4.4$\times10^{-3}$} & 0.43                    & {\bf 0.036}              & 0.60                     & {\bf 4.4$\times10^{-3}$} & {\bf 1.9$\times10^{-5}$}\\
\end{tabular}
\caption{\label{t3}
Kolmogorov Smirnov statistic $p$-values for the distribution of protostars in each simulation as compared to the observed IMF \citep{chabrier2005} and several analytical protostellar mass functions including the turbulent core (TC), 2 component turbulent core (2CTC), competitive accretion (CA), 2 component competitive accretion (2CCA) and the isothermal sphere models \citep{mckeeoffner10}.  Simulations that include turbulent forcing are denoted as ``Driven'' in the left column.  The other models are the cases of decaying turbulence.  Simulation to mass function pairs with $p$-values less than 0.05 can be confidently interpreted as a sample drawn from a different distribution than the given mass function.  The $p$-value for these ''rejected'' models is denoted in bold-face in the table.}
\end{table*}

We find that our runs with weak initial fields ($\mu_\Phi=23.1$ and
$\infty$) and no driving have peak masses that exceed the observed
mode of the IMF ($m_* = 0.2 ~\msun$ -- \citealp{chabrier2005}), and
that are continuing to rise (\autoref{f5}), at the termination of the
models.  As discussed in \autoref{s3.1}, these models also approach a
state of rapid, near global free-fall collapse, contrary to
observational constraints.  Furthermore, the shape of the mass
distributions is closest to that of the IS case.  The ever-increasing
peak of the PMFs in these simulations appears to be a manifestation of
the overheating problem identified by \citet{krumholz11}, whereby
values of $\epsilon_{\rm ff}$ close to unity produce such high
accretion luminosities that fragmentation is strongly suppressed,
leading to a mass function that shifts to ever-higher masses as
collapse continues but no new stars can be created.  Furthermore,
these simulations have low K-S statistic $p$-values when compared
against all analytic models except for the IS one.

In contrast, the moderate and strong initial field cases all achieve
stable or slowly increasing median masses in the range $0.05 \msol <
m_* < 0.15 \msol$, independent of the presence of external turbulent
forcing.  All of the models with continuously driven turbulence also
show stable peak masses independent of the strength of the initial
magnetic field.  At the termination of these models, the median masses
lie between the analytically predicted peak masses for the single
component competitive accretion (CA) and single component turbulent
core accretion (TC) models of $0.05\msol$ and $0.095\msol$,
respectively.  Because of the median masses for the CA and TC models
are close to those of the numerical simulations, these simulations
also have high K-S $p$-values when compared to the CA and TC
theoretical models.  The shape of the mass functions are not amenable
to reliably differentiating among the moderately magnetized models or
the weakly magnetized models with continuous turbulent forcing to
within the statistical limitations of the number of sources in these
simulations.

\begin{figure}
  \begin{center}
    \includegraphics[width=0.32\textwidth]{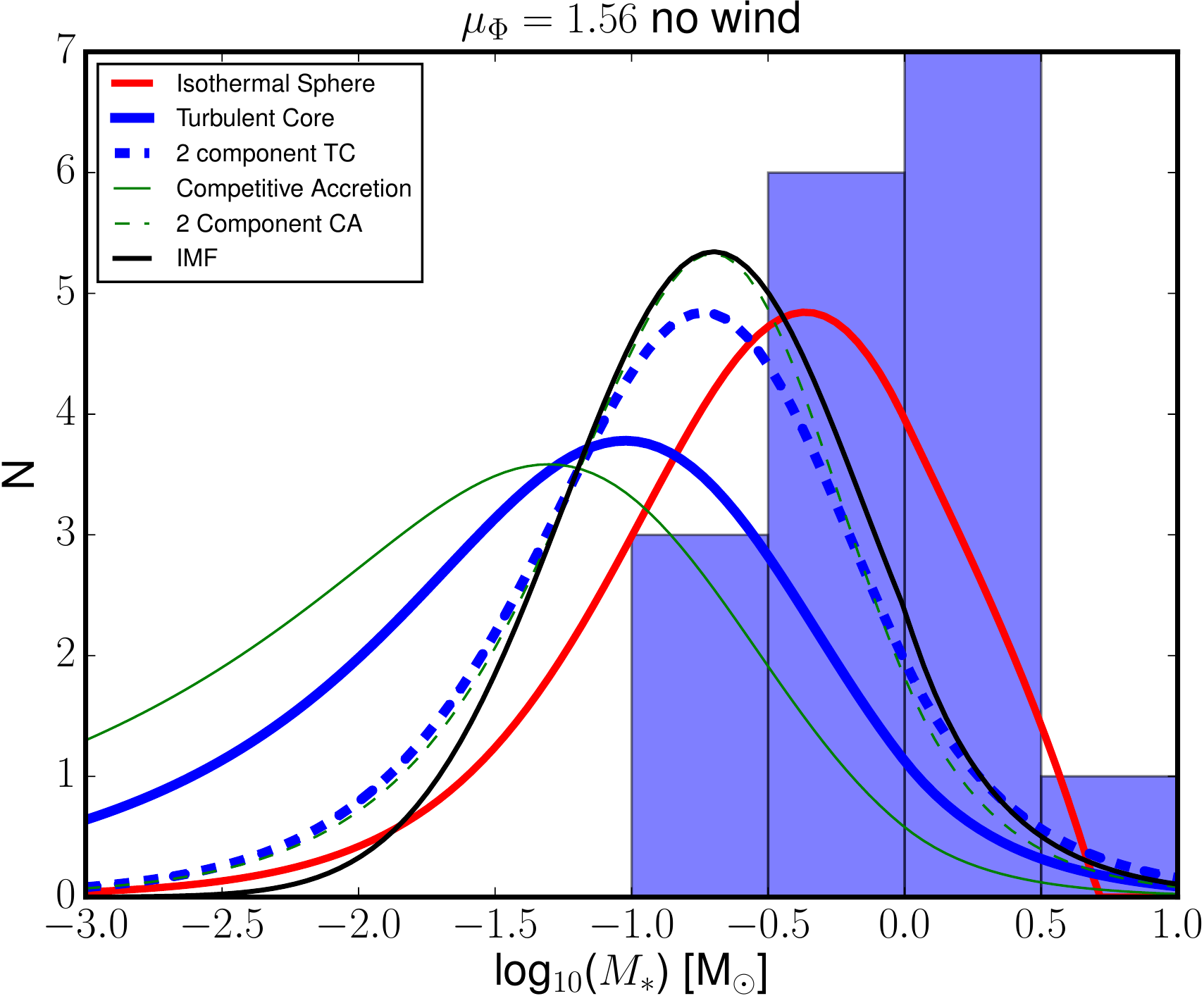}
  \end{center}
  \caption{Histogram of the protostellar masses at the end of the
    $\mu_\Phi=1.56$ model without protostellar feedback. \label{f7}}
\end{figure}

\autoref{f7} shows the protostellar mass distribution at the
termination of the strongly magnetized $\mu_\Phi=1.56$ without
turbulent driving or protostellar outflow ejection.  The effect of
outflows on the protostellar mass distribution can be seen clearly by
comparing this figure to the top left panel of \autoref{f6}, which
shows the undriven $\mu_\Phi=1.56$ model with outflows. Clearly
omission of outflows leads to a significant increase in the typical
stellar mass and very low K-S test $p$-values. 
This difference is partly due to direct removal of mass
by the outflows, but this effect alone cannot explain the difference
in mass distribution, partly because it is too large (a factor of
$\gtrsim 10$ in mass), and partly because there is also a dramatic
drop in the total number of stars when we turn off outflows, something
that cannot easily be explained by mass ejection. Instead, the shift
in the mass function here appears to result from the interaction
between outflows and radiative heating. Outflows indirectly reduce the
luminosity of individual stars by lowering their accretion rates
\citep{krumholz12, hansen, myers14}, and create paths of reduced
optical depth, allowing more efficient escape of radiation from
the gas immediately around the protostars \citep{krumholz05, cunningham11}.
Due to the omission of these two effects in the
no-outflow model, radiative heating is much more effective at
suppressing fragmentation, leading to the production of fewer, much
more massive stars. We showed in \autoref{s3.1} that the strong
initial magnetic support in the no-outflow model was sufficient to
limit the star formation efficiency to values consistent with
observation, but clearly the shape of the mass distribution of this
model is not consistent with observation.

It is also interesting to compare our results to previous simulations
making use of a similar set of physical processes. Our $\mu_\Phi=1.56$
model with continuous driving is comparable to the recent models
by \citet{li17}; the simulations have similar
initial ($\beta_{\rm IC}$ and $\mathcal{M}_{A,\rm IC}$) and 
post-driving ($\beta_{\rm >med}$, $\mathcal{M}_{A,\rm >med}$ 
and $\rho_{\rm >med}$) conditions. However, while in our case this
model produces an IMF peak at slightly less than $0.1\msol$, the
\citet{li17} simulations yield at peak at $0.2\msol$, comparable to
the observed IMF peak. Given the similarities of the
dimensionless conditions of the two cluster models, we attribute this
difference to the outflow ejection parameters discussed
in \ref{ssec:physics}.  The models presented here eject three times
the momentum flux in the early phases of prestellar collapse and
therefore have a more disruptive effects on their parent cores.  While
\citet{cunningham11} point out that observations do not precisely
constrain the outflow model ejection parameters, the fact
that the model in \citet{li17} achieves better agreement with the
well-constrained median IMF suggests that the outflow ejection parameters
used in that model are probably more realistic. This conclusion,
however, does not affect the relative differences between the
various runs that we have identified; indeed, a shift to higher
masses as a result of somewhat weaker outflows would worsen the
agreement between the observed IMF and those produced in the
weakly-magnetised, non-driven models.

The other obvious prior work to which we can compare are the simulations
of \citet{hansen}. Indeed, our non-magnetised, decaying turbulence
model is identical to theirs apart from the random
phases used during the initial turbulent driving phase.  Both models
ran for comparable times as well. However, the model presented in this work
exhibits rapidly increasing accretion rates and commensurately
increasing source masses at late time, while the simulations of
\citeauthor{hansen} do not. While this might at first seem quite
surprising, the result can be understood by examining \autoref{f1}.
This figure shows that the stellar mass at first increases relatively
slowly, but then rapidly increases as the turbulence decays and
free-fall collapse begins. The behaviour in the \citeauthor{hansen}
simulation is consistent with what we see just before the onset of
rapid collapse in our simulation. Since the exact time at which
runaway collapse begins is almost certainly sensitive to the random
driving, the most likely explanation for the difference is that the
\citeauthor{hansen} simulation was simply not run quite long
enough to reach the runaway collapse phase where the collapse
begins to alter the IMF.

\subsection{Stellar Multiplicity}
The statistics of multiple systems provides another observational
constraint against which our models can be tested.  To compute the
stellar system multiplicity we use the algorithm of \cite{bate09}.
The algorithm works recursively: one identifies the most strongly
bound pair of stars in the simulation, and replaces it with a
single object at the centre of mass position of the pair with the
total mass and momentum of its constituent protostars, then repeats
until there are no more bound pairs that can be combined. In cases
where combining the most bound pair of objects would create a system
with a multiplicity greater than four, one skips it and
combines the next-most bound combination instead. This is motivated
by the fact that high-multiple bound systems are dynamically unstable
and such systems are likely to break apart if the model were continued
further in time.  At the end of this procedure we obtain $S$ single
star systems, $B$ binary systems, $T$ triple systems and $Q$
quadruple systems.  For the set of bound systems the multiplicity
fraction is defined as \citep{bate09,krumholz12,myers14}
\begin{equation}
\label{eq:mf}
\textrm{MF} = \frac{B+T+Q}{S+B+T+Q}
\end{equation}

\begin{figure*}
  \begin{center}
    \includegraphics[width=0.32\textwidth]{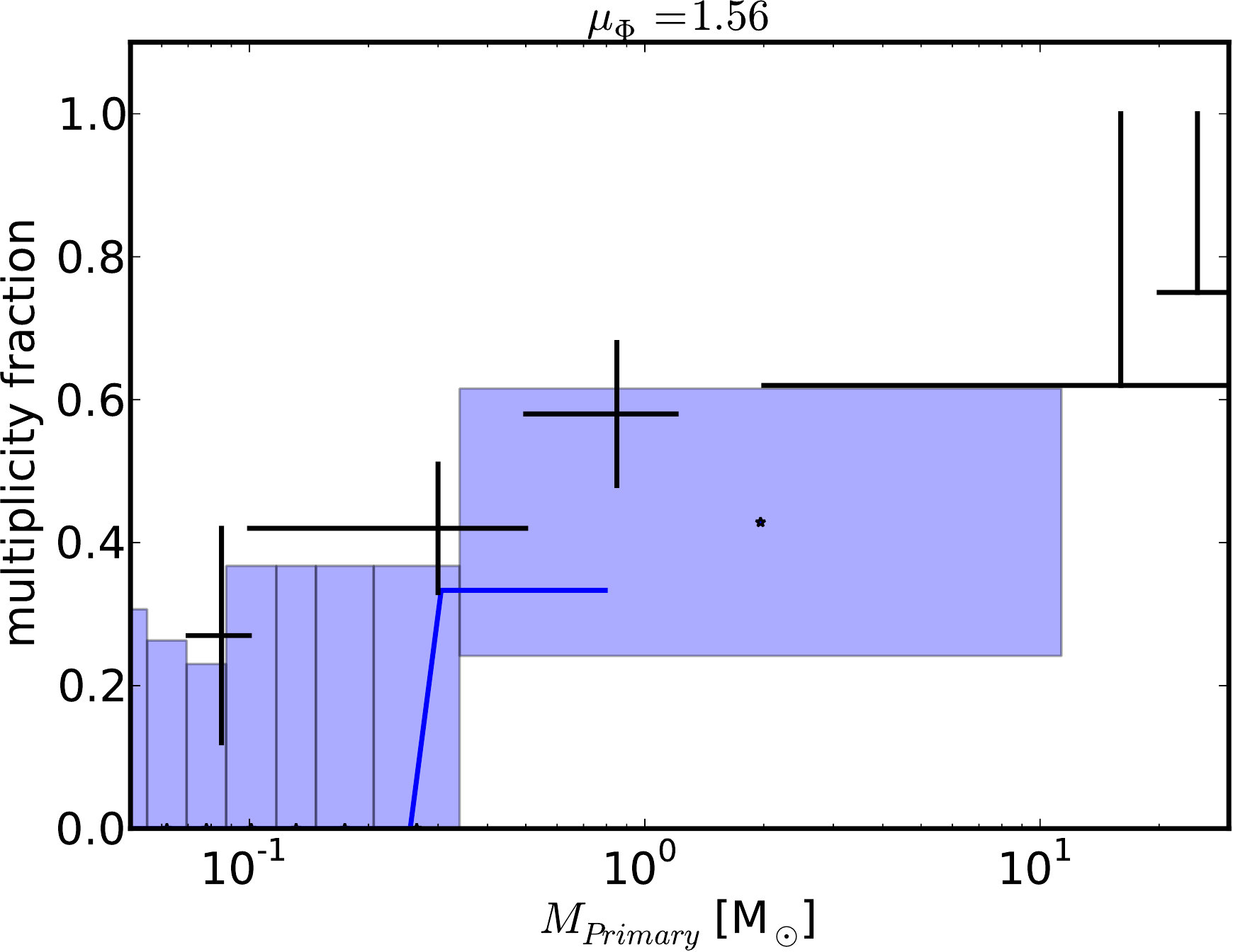}
    \includegraphics[width=0.32\textwidth]{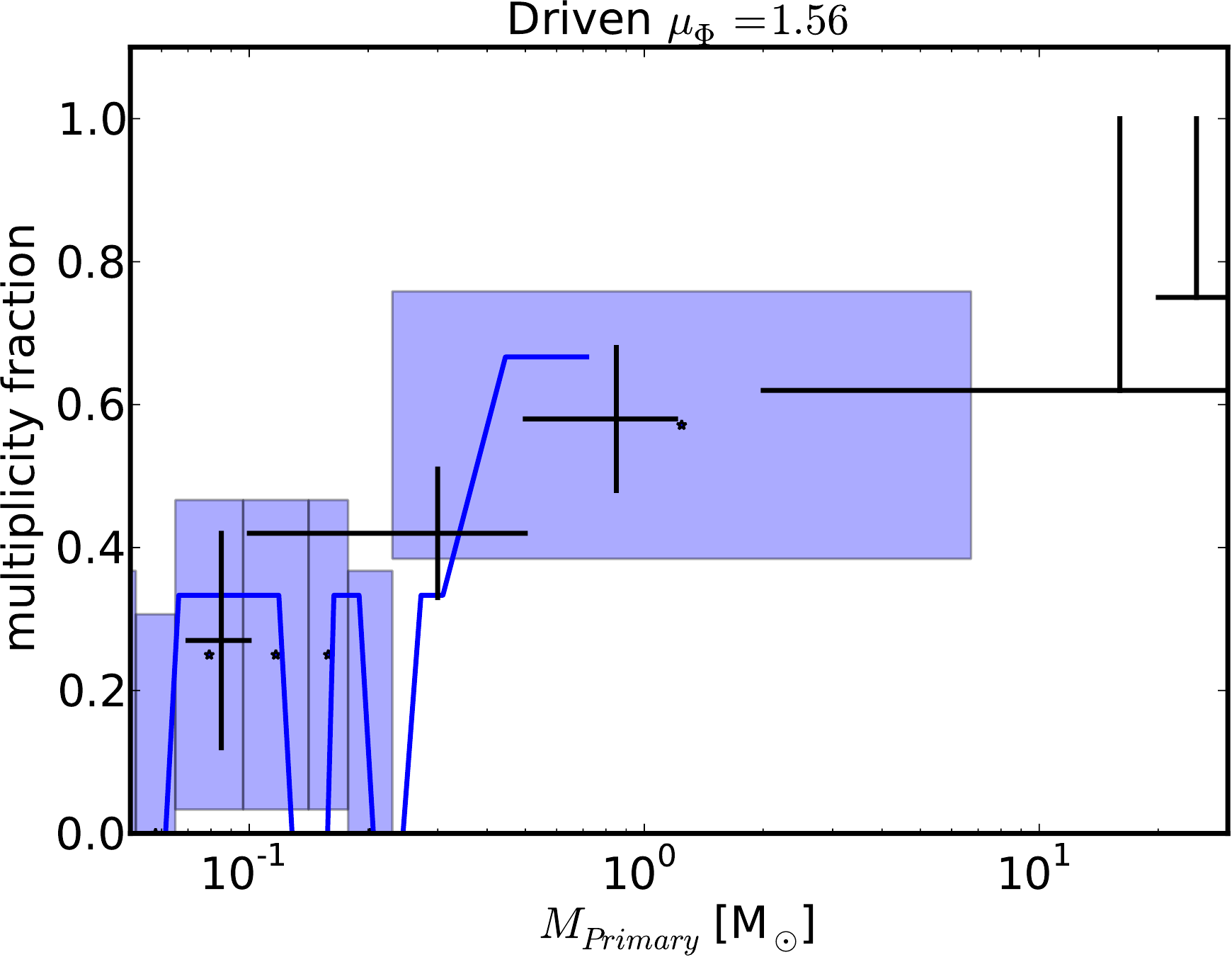} \\
    \includegraphics[width=0.32\textwidth]{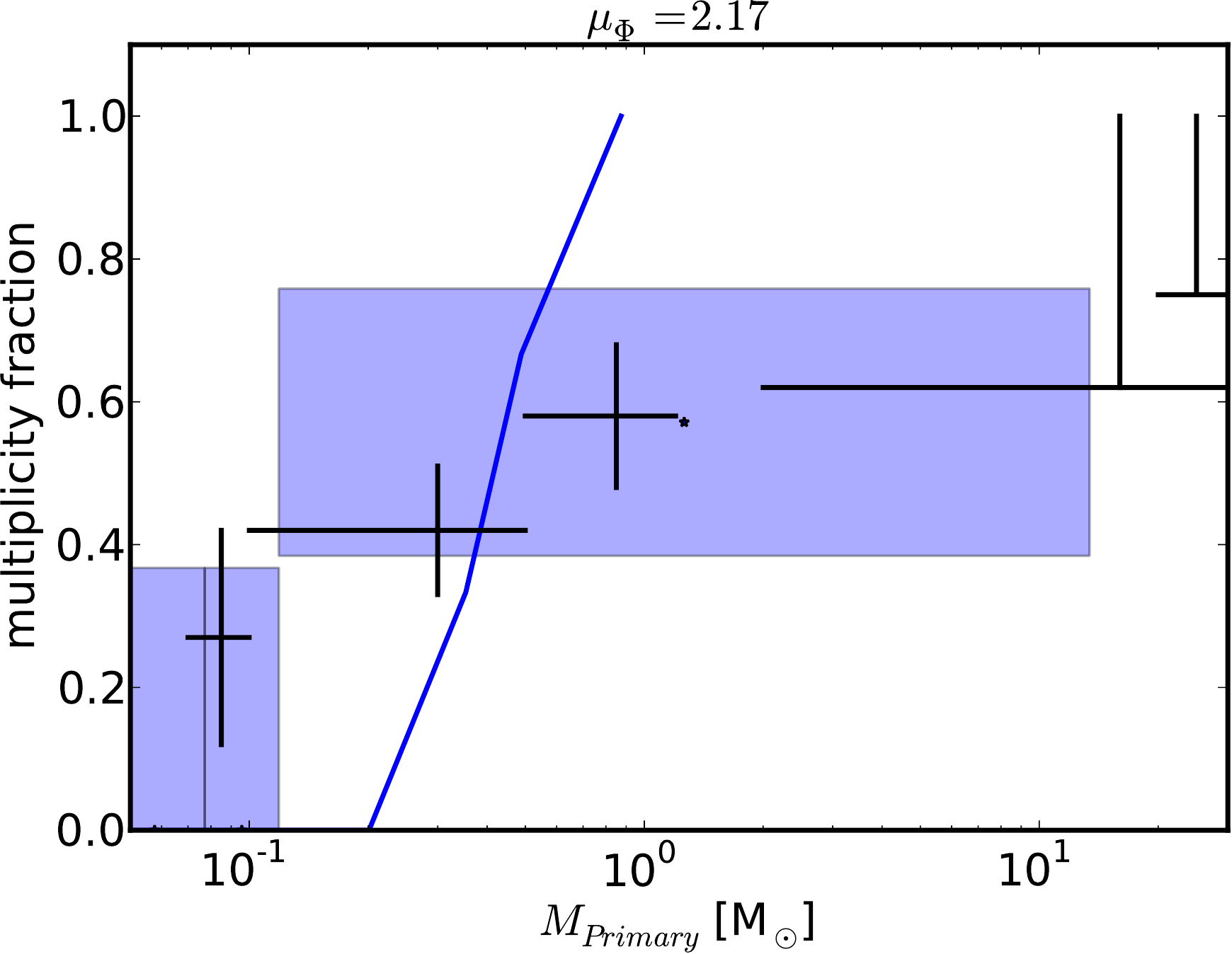}
    \includegraphics[width=0.32\textwidth]{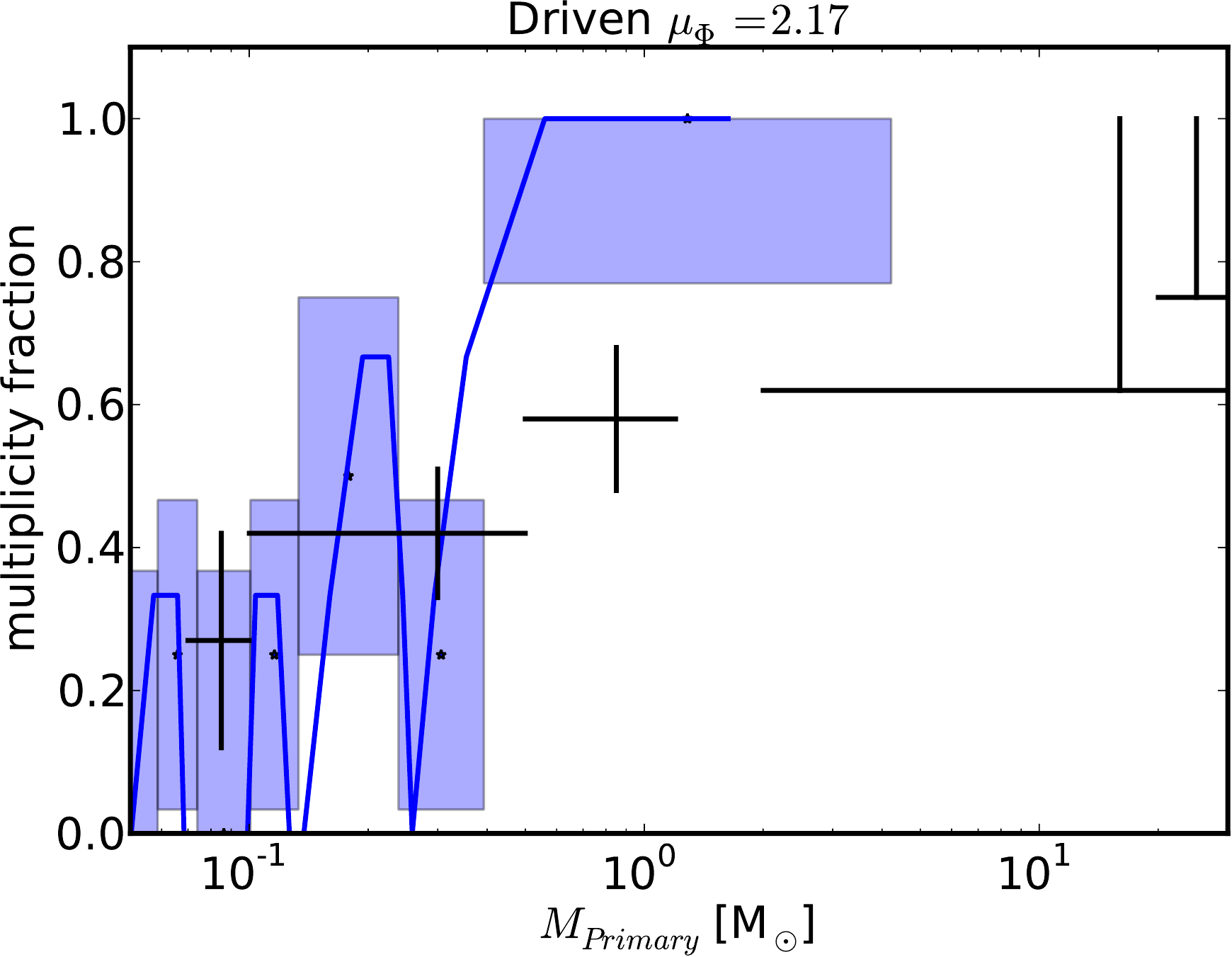} \\
    \includegraphics[width=0.32\textwidth]{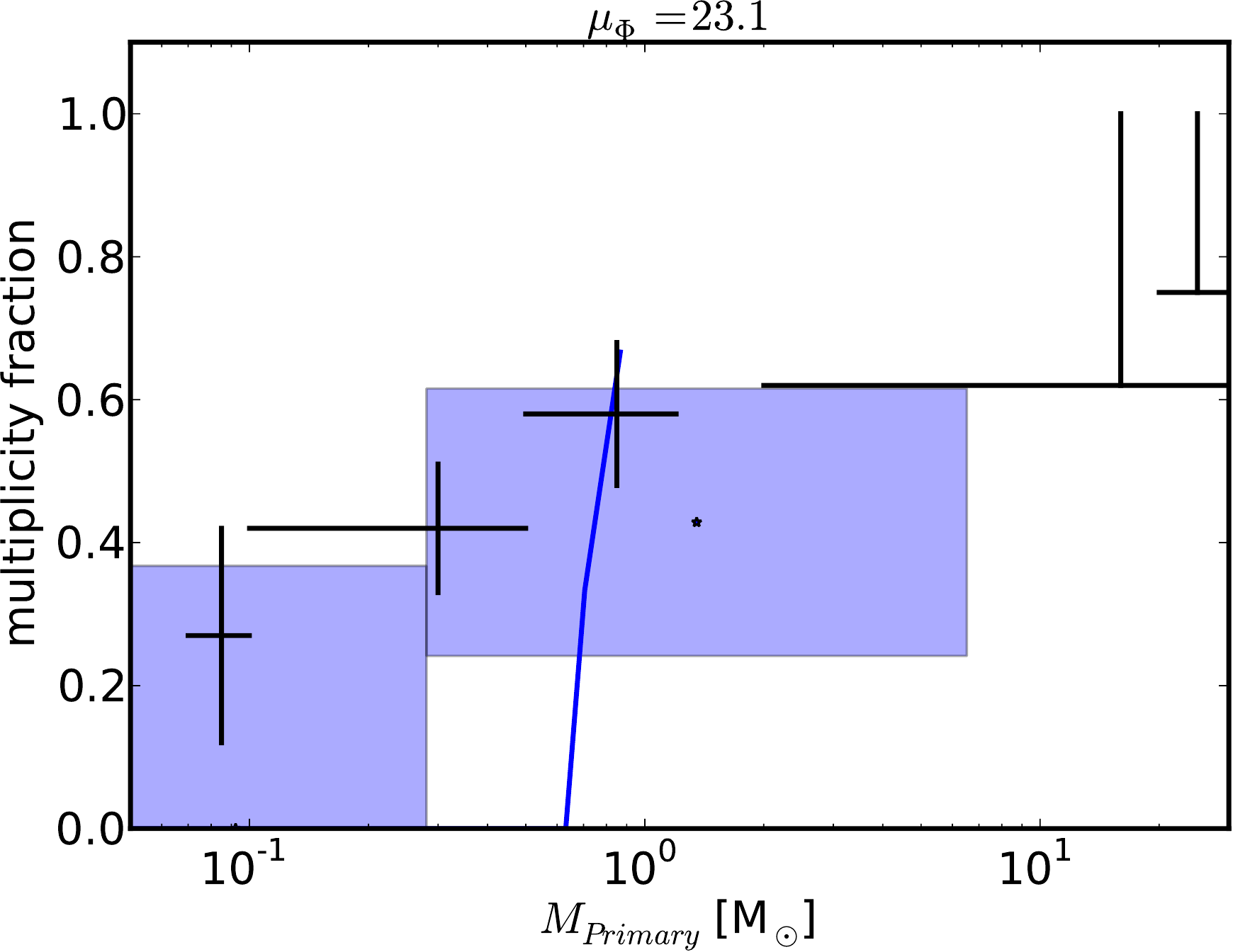}
    \includegraphics[width=0.32\textwidth]{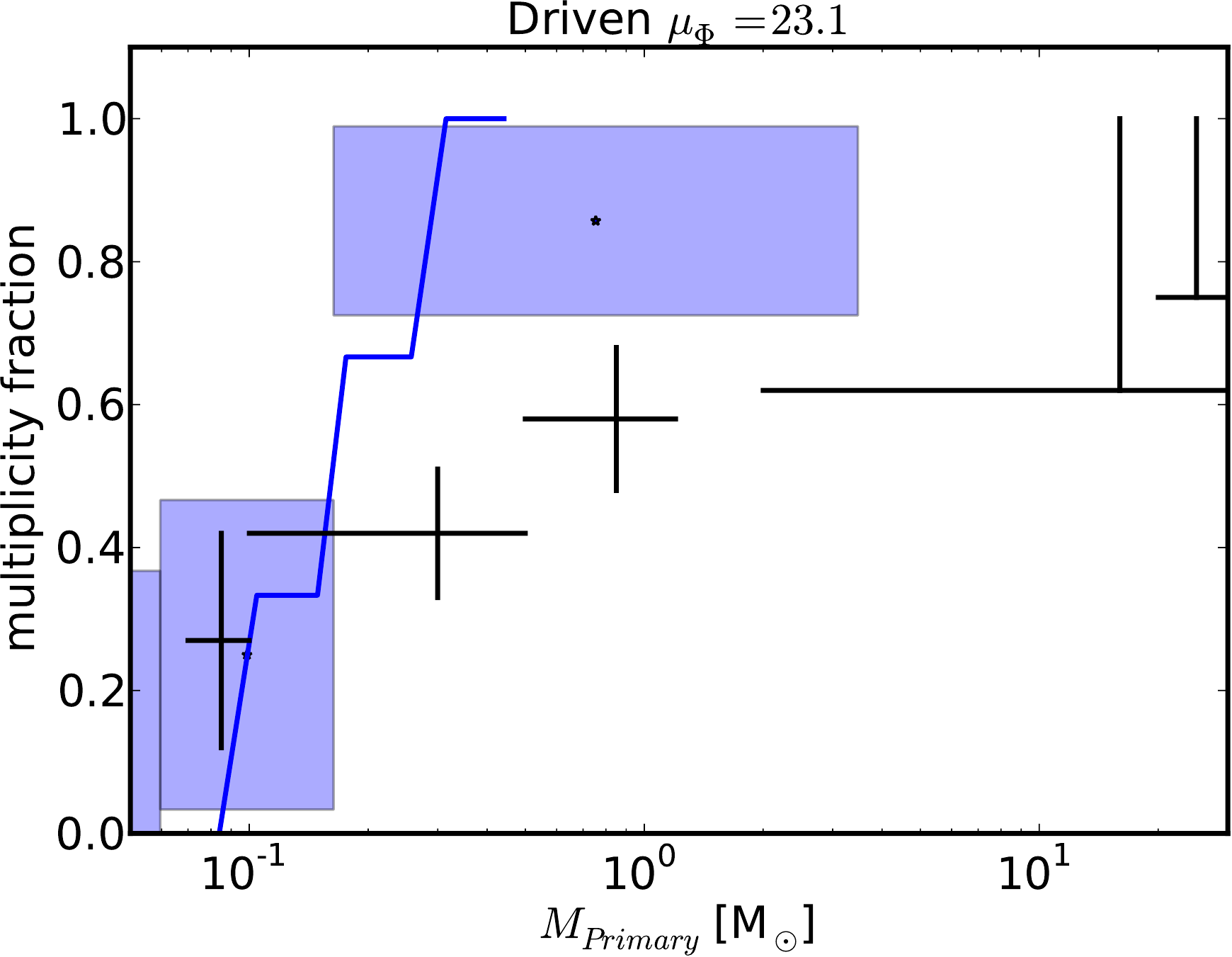} \\
    \includegraphics[width=0.32\textwidth]{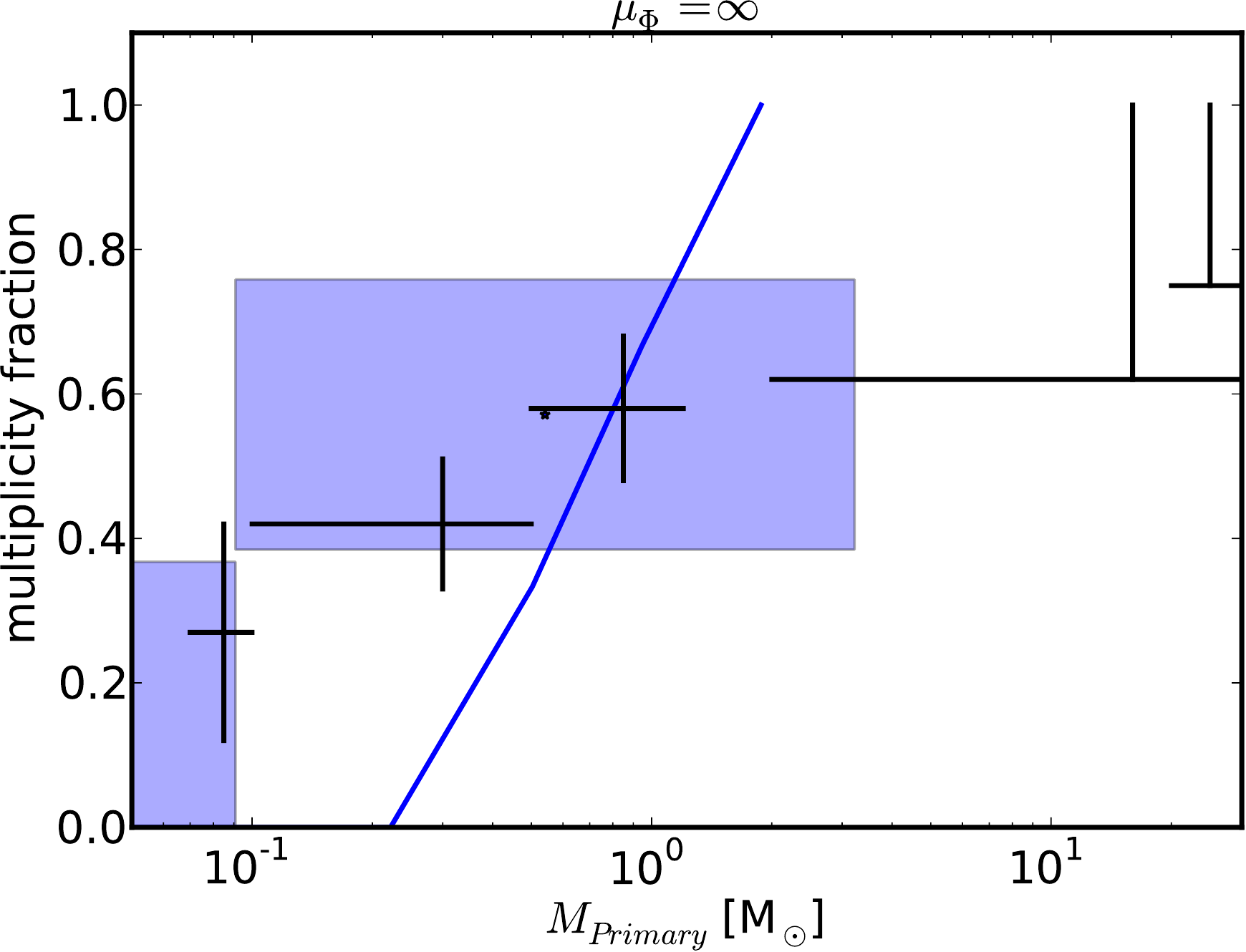}
    \includegraphics[width=0.32\textwidth]{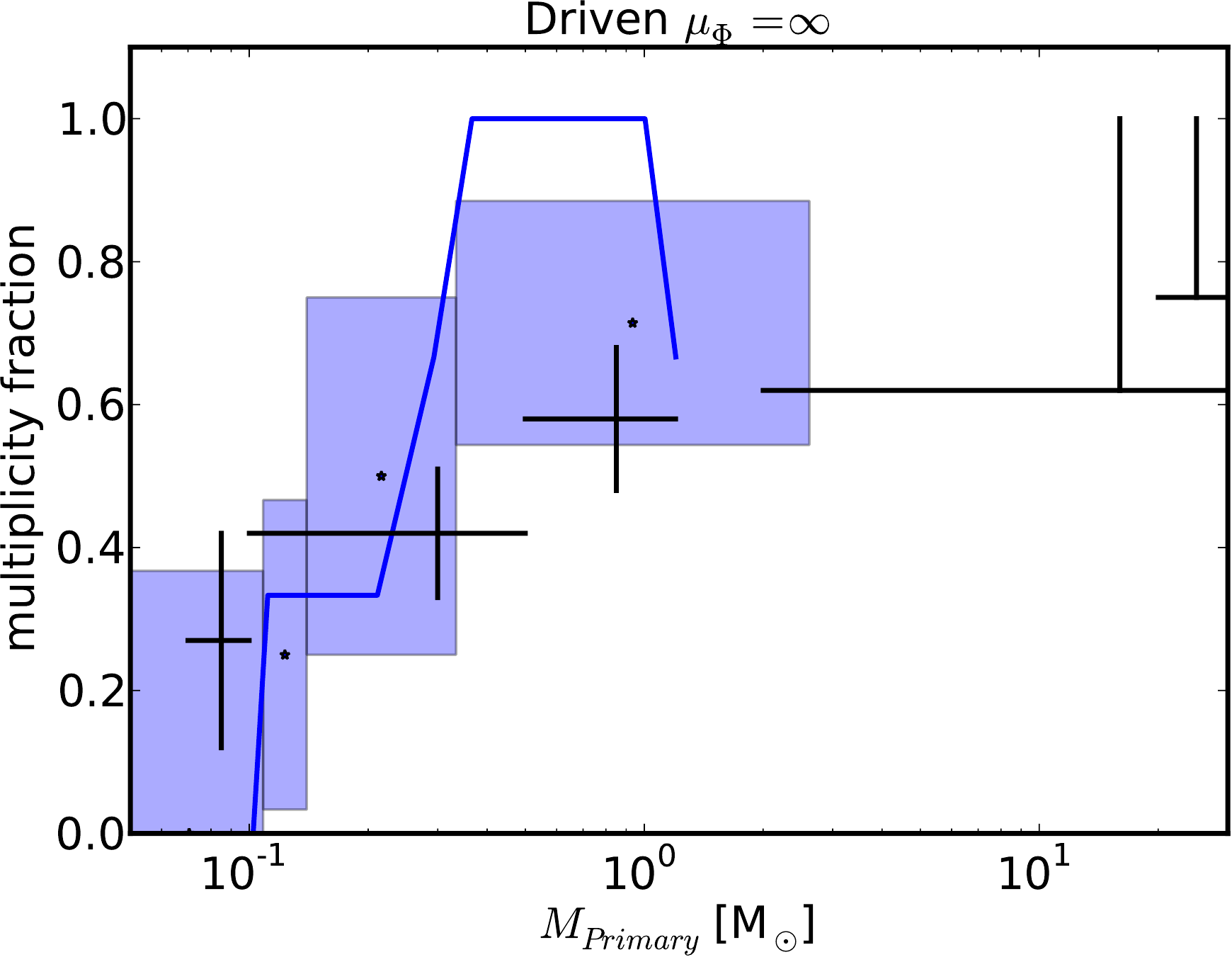} \\
  \end{center}
  \caption{Multiple stellar system fraction at the end of each of the
    models with turbulent driving (right) and with decaying turbulence (left), computed
    as a function of primary star mass.  The
    four rows show the $\mu_\Phi=1.56, 2.17, 23.1$,
    and $\infty$, from top to bottom. 
     For the simulations, the blue curve indicates evolution of a running average over three
     systems as a function of mass, while shaded regions indicate the mean
     multiplicity and central 68\% probability range on the multiplicity in
     bins -- see main text for details on how the running average and binned
     values are computed. Black
    crosses denote the mass range and mean multiplicity found in
    observational surveys.  The two highest mass surveys are
    statistical lower bounds.  The horizantal marker on these
    observations indicate the lower bound.  The observational data,
    from low to high mass, are compiled from \citet{basri2006} and
    \citet{allen2007} (combined into the leftmost point),
    \citet{fischer1992}, \citet{raghavan}, \citet{preibisch},
    \citet{mason2009} and these are the same data used in
    \citet{bate2012} and \citet{krumholz12}.\label{f8}}
\end{figure*}

\autoref{f8} shows the multiplicity fraction as a
function of the primary star mass in each of our simulations.
Given the limited number of systems
in at least some of our simulations, computing the multiplicity
fraction requires some care. We do so in two ways.  First, we form
running averages, meaning that, for each primary star (i.e., single
star or the most massive member of a multiple), we evaluate
\autoref{eq:mf} for the set of systems consisting of that primary and
the next most massive and next least massive primaries. This quantity
is shown as the blue curve in \autoref{f8}. Second, we can compute
multiplicity in bins.  To do so, we start with the least massive
primary, and then place the primaries in bins of mass, with the bin
width taken to be either sufficient to contain four primaries, or 0.2
dex, whichever is greater. For each bin, we regard the systems in it
as samples drawn from a binomial distribution of multiples or singles,
and from those samples we compute a central estimate and a 68\%
confidence interval on the true multiple system fraction in that bin
(see \citealt{krumholz12} for full details on how this computation is
carried out).  We plot the binned 68\% confidence intervals as the
shaded regions in \autoref{f8}.

Comparing either the running averages or the binned results to the
observed multiplicity fractions that we also plot in \autoref{f8}, we
find that all of the models generally agree with observations 
within the range of statistical uncertainties.  It is noteworthy that
the weakly magnetized models without turbulent forcing are consistent
with observed multiplicity statistics even though their
protostellar mass functions (\ref{s3.3}) are strongly inconsistent
with observations. The driven
turbulence $\mu_\Phi=2.17$ and $23.1$ models somewhat over-predict the
multiplicity fraction for systems with primary mass $M_p\sim 1~\msol$,
but this may be the result of poor statistics in this mass range in
those models, and we note that the observational surveys are based on
field stars, while there are strong hints that multiplicity fraction
is higher for still-embedded systems \citep{tobin16}, which is the
more relevant comparison for our simulations.\footnote{We
  unfortunately cannot place the data from \citet{tobin16} on
  \autoref{f8} because the masses of the individual stars are largely
  unconstrained.} We also note that, while the bins at low mass are
generally consistent with the observations individually, taken
together it is clear that we systematically under-produce multiple
systems at low mass relative to the observational data.  This is a
likely a resolution effect.  Our models do not resolve gas-particle
gravity forces below the scale where sink source terms couple to the
gas -- $4 \Delta x$.  Low-mass multiple systems must necessarily be
close and consequently more difficult to resolve to remain bound.
This limitation is also noted in the multiplicity fractions extracted
from the models of \citet{bate09}, \citet{krumholz12}, and
\citet{myers14}.

\subsection{Turbulent Decay and Protostellar Feedback Rates}

\begin{figure*}
  \begin{center}
    \includegraphics[clip=true,width=0.45\textwidth]{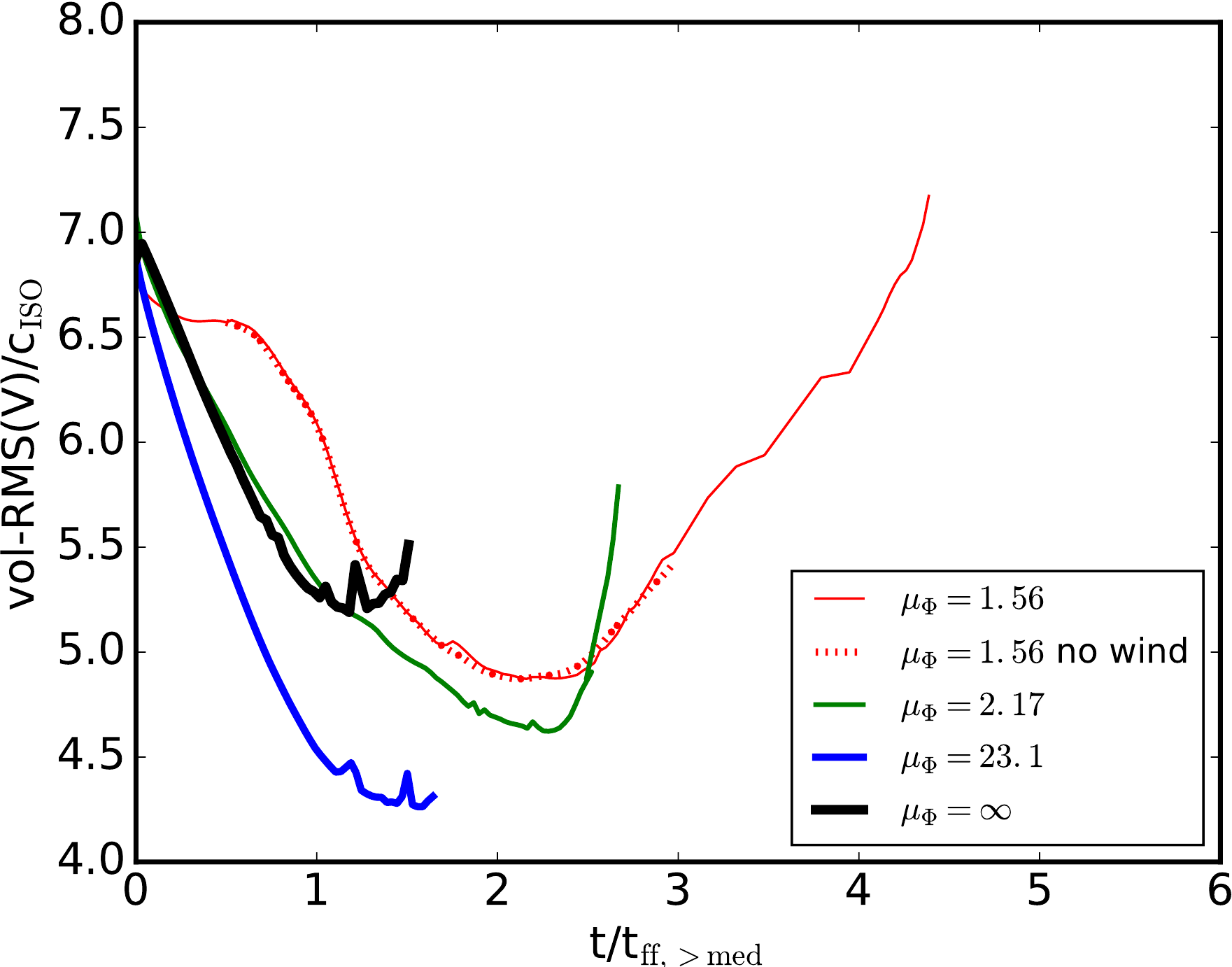}
    \includegraphics[clip=true,width=0.45\textwidth]{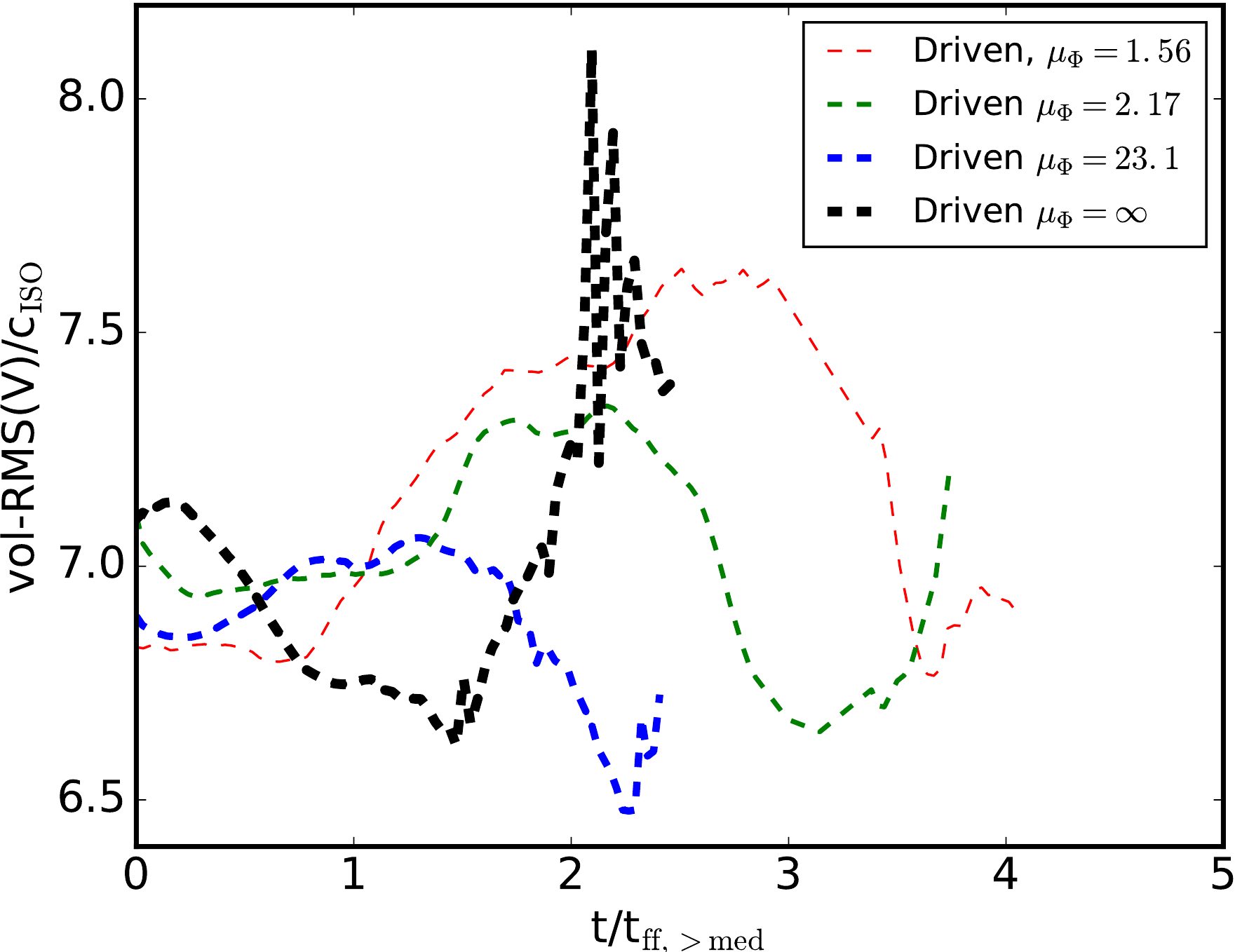} \\
  \end{center}
  \caption{Volume-weighted root mean squared gas velocity, normalized
  by the gas sound speed, as a
    function of time for the models with turbulent forcing (right) and
    with decaying turbulence (left). Note the difference in scales between
    the left and right panels. \label{f9}}
\end{figure*}

\begin{figure*}
  \begin{center}
    \includegraphics[clip=true,width=0.45\textwidth]{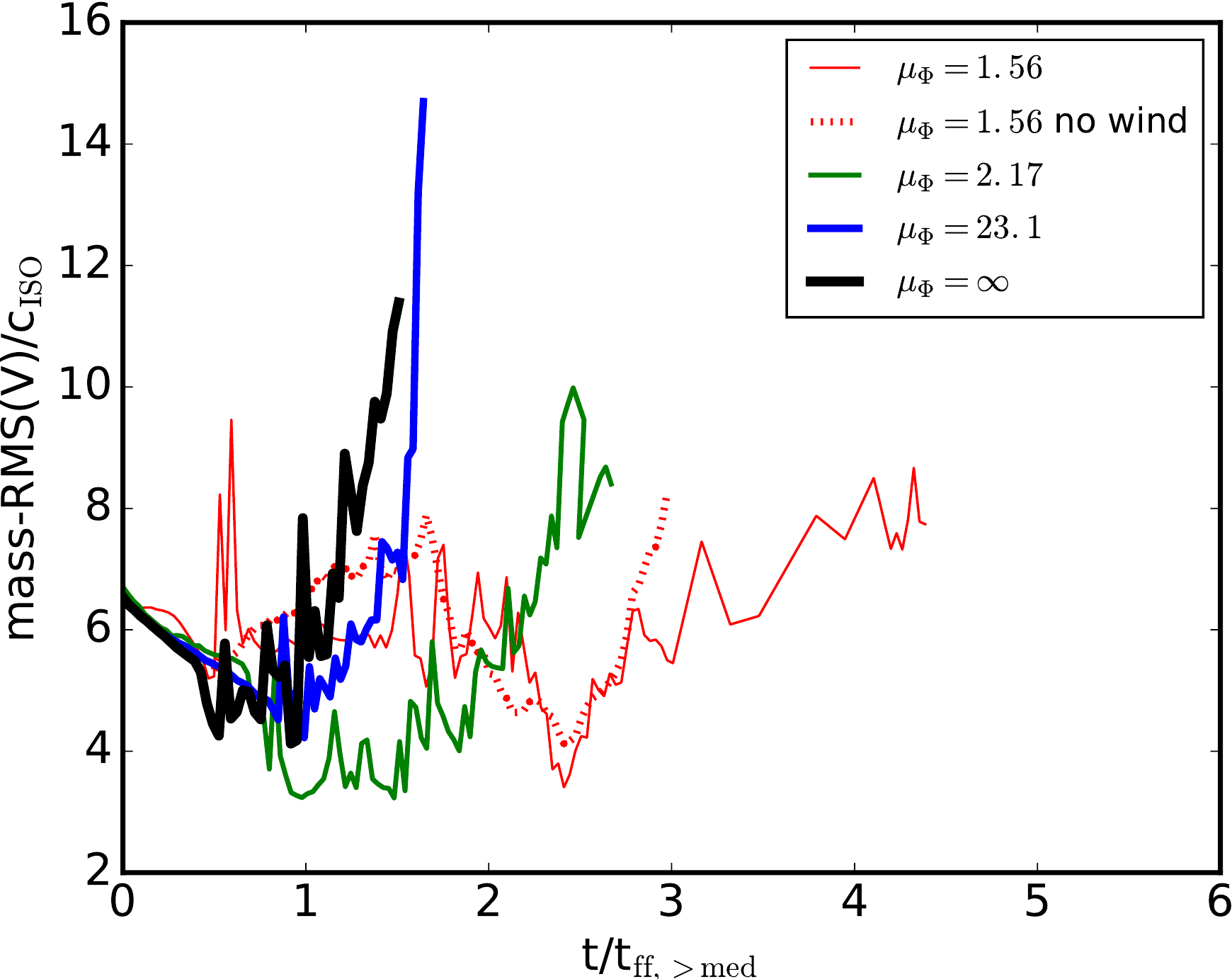}
    \includegraphics[clip=true,width=0.45\textwidth]{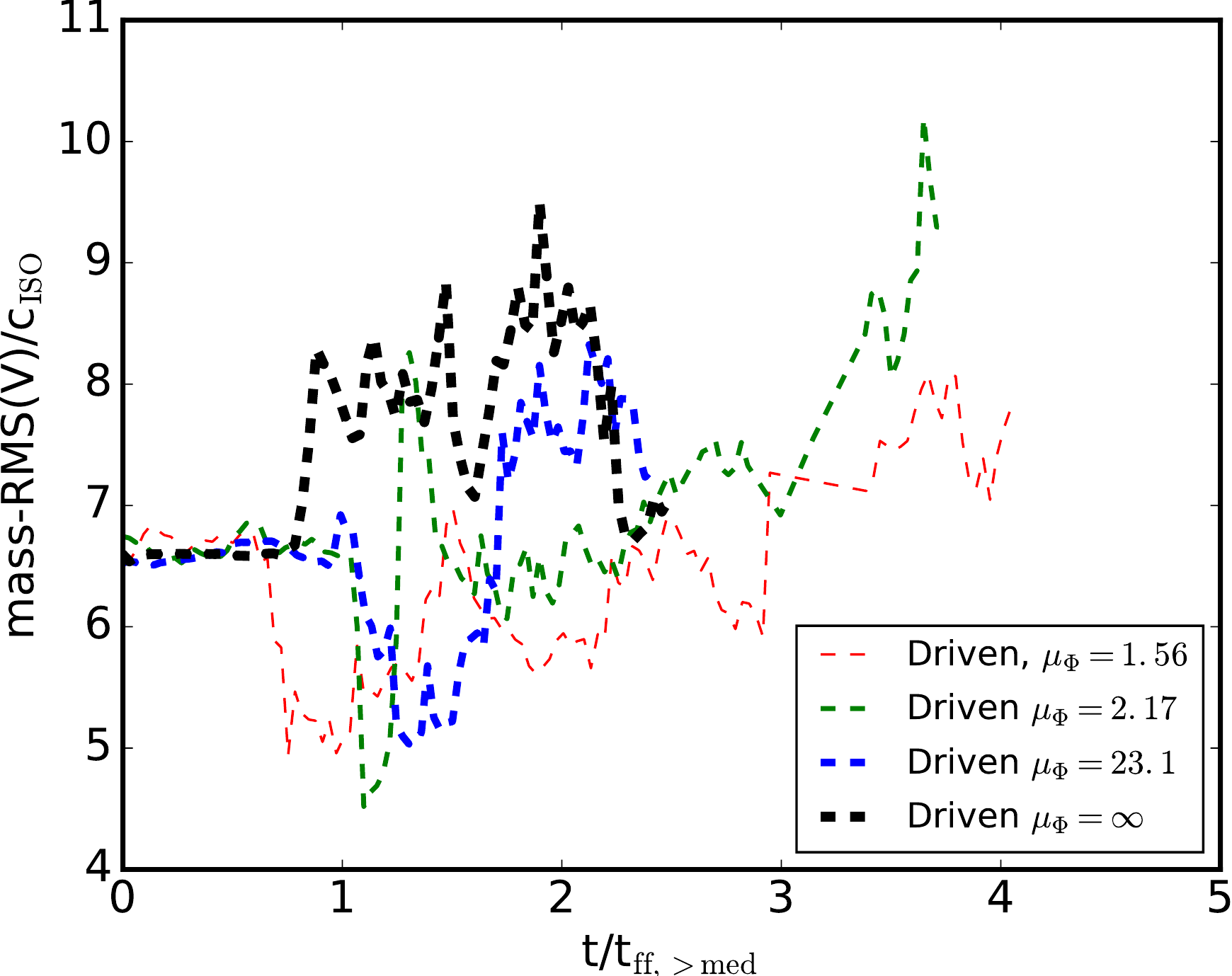} \\
  \end{center}
  \caption{Same as \autoref{f9}, but now showing mass-weighted rather
  than volume weighted velocities. As with \autoref{f9}, note the difference in
  scales between the two panels. \label{f10}}
\end{figure*}

\begin{figure*}
  \begin{center}
    \includegraphics[clip=true,width=0.45\textwidth]{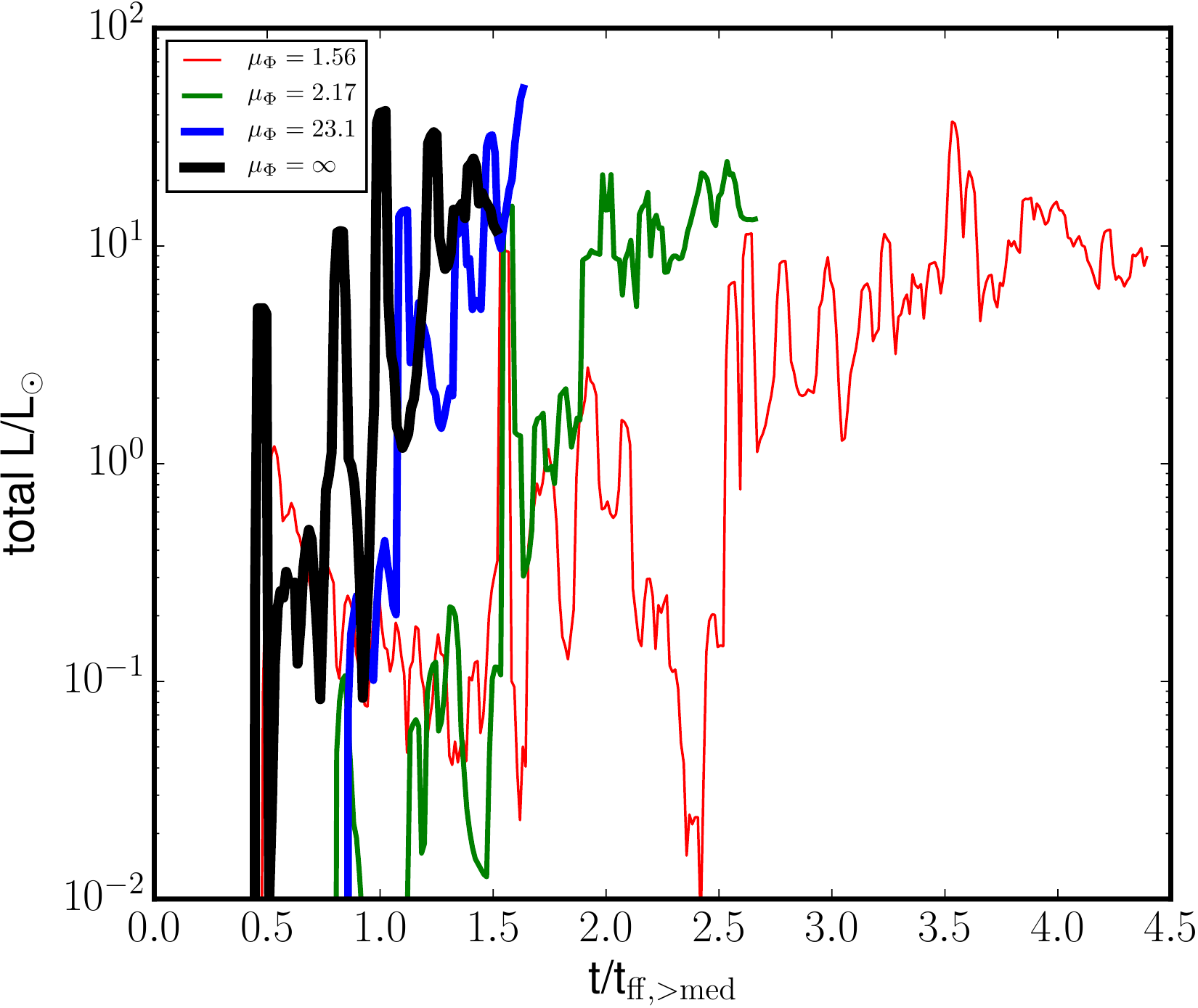}
    \includegraphics[clip=true,width=0.45\textwidth]{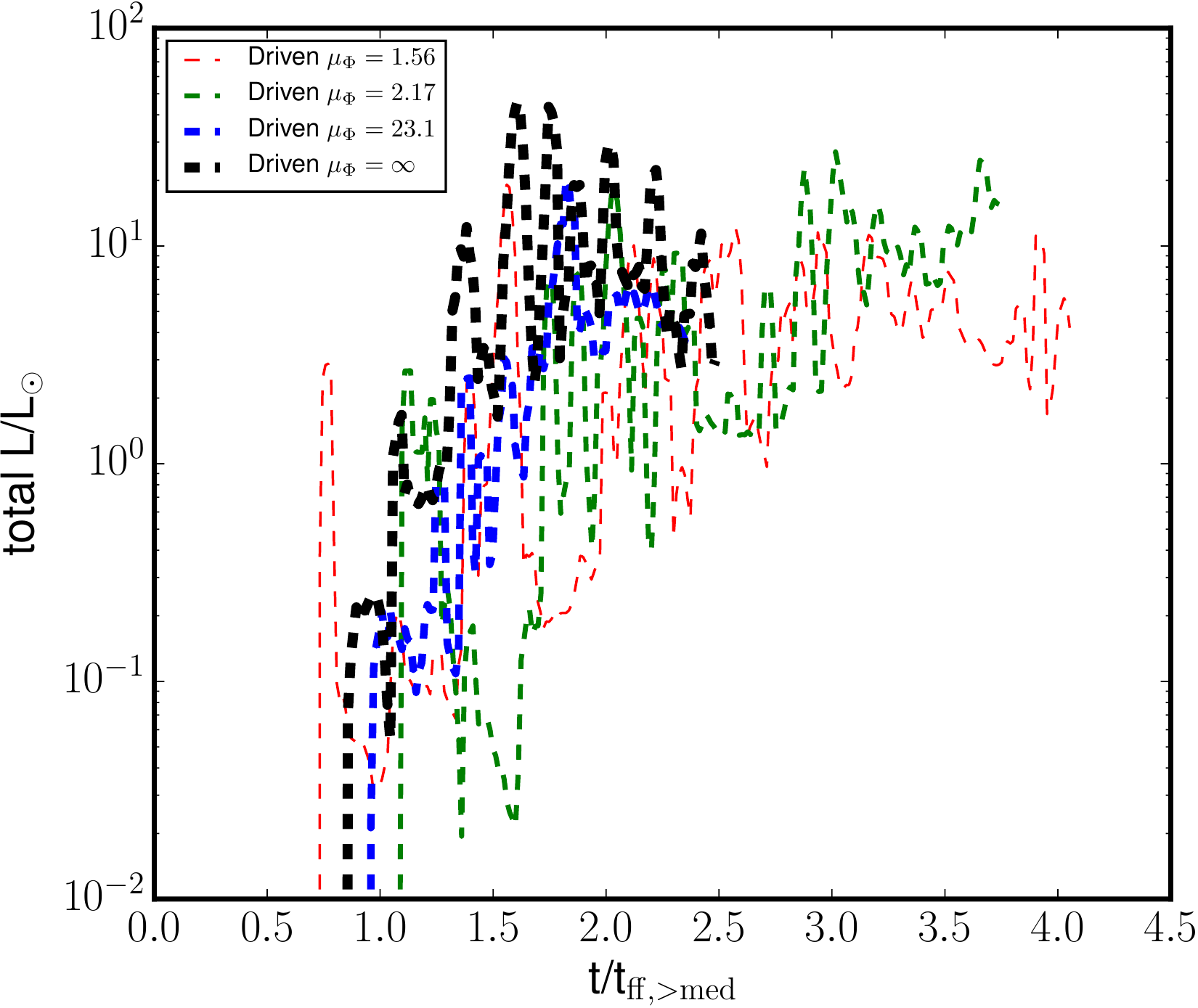} \\
  \end{center}
  \caption{Total mechanical luminosity of the protostellar outflows as
    a function of time for the models with turbulent forcing (right)
    and with decaying turbulence (left). The data have been boxcar-smoothed
    with an interval of $0.05~t_{\rm ff}$.  \label{f11}}
\end{figure*}

\autoref{f9} shows the temporal evolution of the volume-weighted RMS
gas velocity for each model.  As intended, the models that include
steady turbulent forcing remain within 10\% of a constant RMS velocity
with $\alpha_{\rm vir}\sim 1$. For the runs without driving, at early
time, $t\lesssim t_{\rm ff}$, turbulent decay causes the RMS velocity
to drop. However, after $1-2$ free-fall times (depending on the level
of magnetic support) the system begins to undergo a global collapse,
and this causes the velocity dispersion to rise again.

\autoref{f10} shows the temporal evolution of the mass-weighted RMS
velocity dispersion.  This averaging is more analogous to
mass-biased observational line width-size measurements. However, it
is also heavily biased toward dense, locally collapsing regions,
which somewhat mutes the effect of turbulent decay relative to infall. 
Consequently, the clear drop in velocity dispersion seen in \autoref{f9}
for the non-driven runs is absent here. We
infer that mass-weighted line-width diagnostics are surprisingly
insensitive to interruptions in large-scale driving source of
durations up to $\la 1 t_{\rm cross}$.

It is also interesting to compare the $\mu_\Phi=1.56$ cases with and
without outflows. In principle the outflows have more than enough
mechanical power to alter the turbulence. We illustrate this in
\autoref{f11}, which shows the outflow mechanical
luminosity versus time, $L_{\rm outflow} = (1/2) \dot{M}_{\rm wind}
v_{\rm wind}^2$. This is typically $\sim 1-10 L_\odot$ at late times
in our simulations, For comparison, the turbulent decay rate is
$\sim \bar{\rho} v_\rms^2 L^3/t_{\rm cross} \sim 10^{-3}~\lsun$,
roughly 3 orders of magnitude smaller. However, when we examine
the volume-weighted RMS velocity, the $\mu_\Phi=1.56$ runs with
and without outflows are nearly identical, and in the mass-weighted
plot they are only slightly different.  These results indicate that outflow
mechanical power is efficiently dissipated by radiative shocks and does
not quickly couple to the volume-filling turbulence of the surrounding
molecular cloud, consistent with earlier works exploring the coupling of
continuously driven jets through a stratified environments \citep{banerjee}.
What effects the
outflows do have are limited to the dense gas surrounding protostellar
cores, which is why the mass-weighted average shows an effect from
outflows but the volume-weighted one does not. However, we
emphasise that even in a mass-weighted sense the effect is modest.
In our simulations, outflows do not appear to be efficient
drivers of turbulence.

\subsection{Local Magnetic Support}
\label{ssec:mag_support}

The turbulent driving phase introduces stretching and amplification of
the magnetic field, relative to the initial magnetic field strength.
The \Alfven Mach number captures the strength of the turbulence
relative to the initial magnetic field.  As discussed in
\autoref{models}, the magnetized models in this work span the range $1 <
\ma< 15$.  However, the weakly magnetized models endure
more significant stretching of the magnetic field and more small-scale
magnetic field amplification in the turbulent driving phase than the
more strongly magnetized models.  The tangled magnetic field is
further amplified due to collapse after self-gravity is switched on.
Therefore, it is only the mean, large-scale field that is weak in our
``weak-field'' model.  On the scale of the prestellar dense cores, the
magnetic field is significantly more tangled and stretched than in
stronger field models.  Similar field amplification due to turbulence
and the early gravitational collapse phase into starless dark clumps
was noted in \citet[section 8 in particular]{li15}.  In
this section we examine the magnetic field structure in the vicinity
of protostars, with the goal of understanding how the combination
of turbulent and gravitational amplification maps from the initial,
large-scale magnetic field to the final, small-scale one in the
dense gas that is bound to and actively feeding
accretion onto protostellar sources.

\begin{figure*}
  \begin{center}
    \includegraphics[clip=true,width=0.45\textwidth]{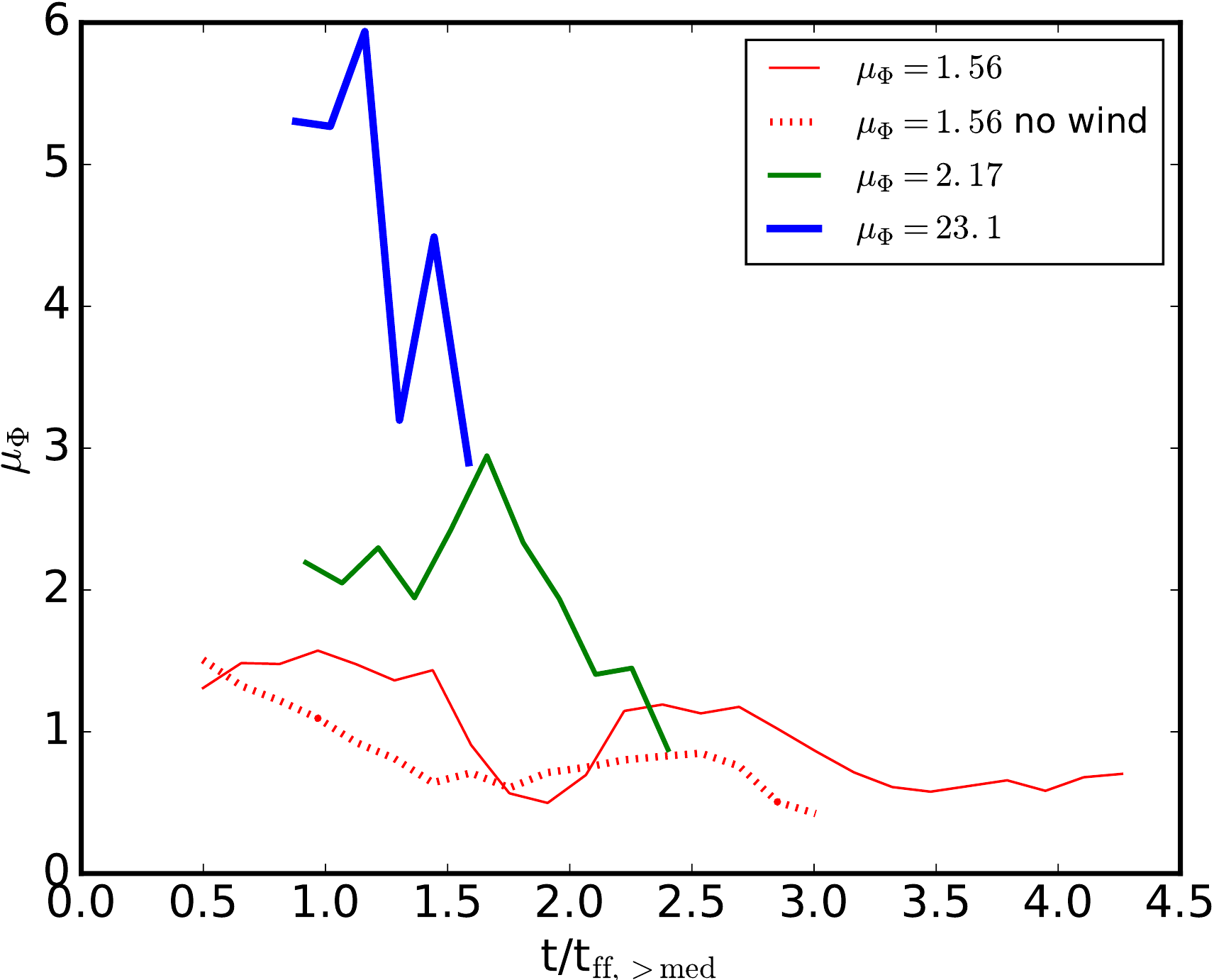}
    \includegraphics[clip=true,width=0.45\textwidth]{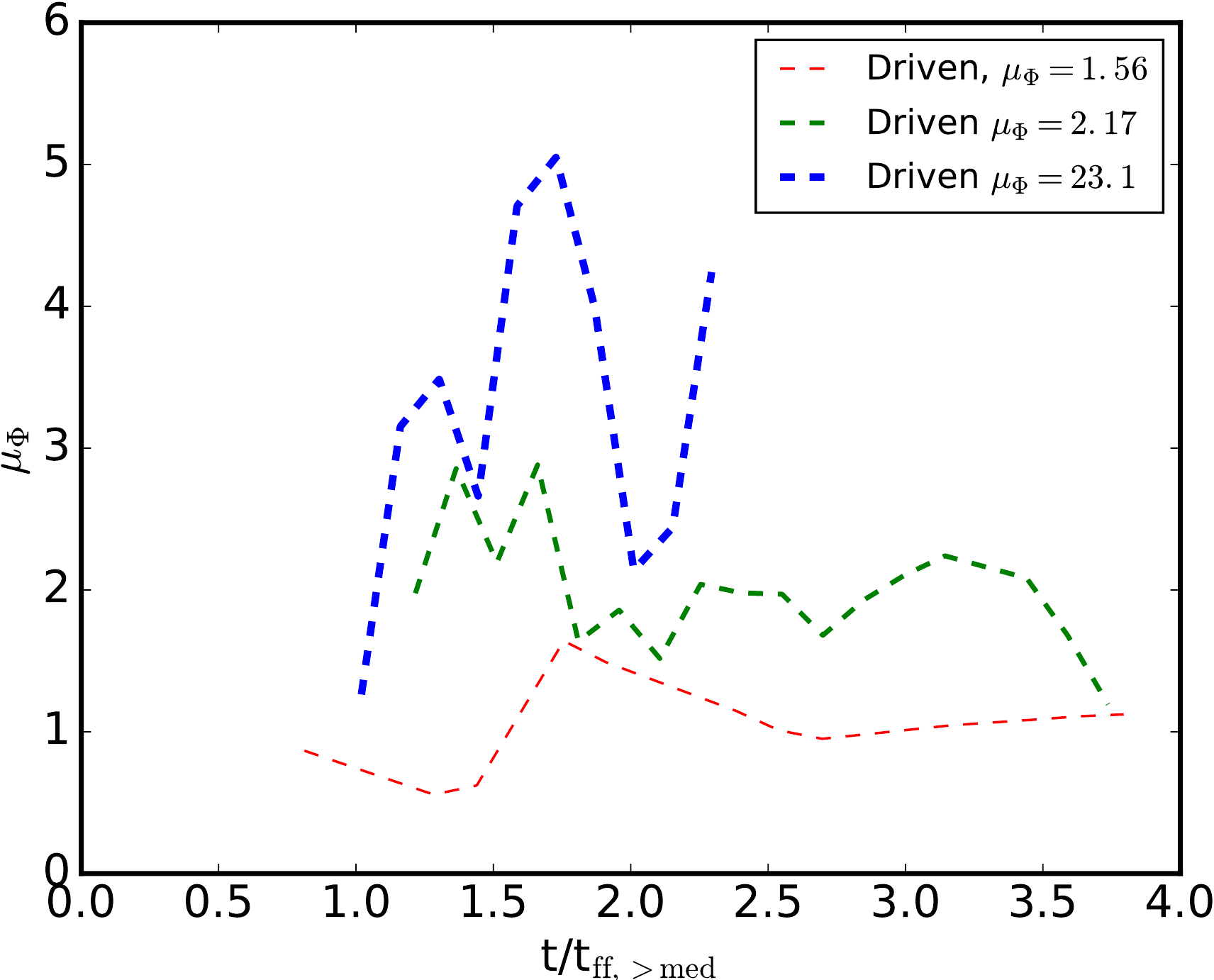} \\
  \end{center}
  \caption{Temporal evolution of the mean mass-to-flux ratio averaged
    over the bound gas around each protostar.  The averaging is
    weighted by protostellar mass.  The
    plot is presented in units of the magnetically critical mass, for
    the models with turbulent forcing (right) and with decaying
    turbulence (left). \label{f12}}
\end{figure*}

We seek to characterize the mean mass-to-flux ratio, averaged over the
bound gas around each protostar that is poised for imminent accretion.  
We take this gas to be that which satisfies the condition
\begin{equation}
\frac{(\vec{v}-\vec{v_*})\cdot(\vec{x}-\vec{x_*})}{|\vec{x}-\vec{x_*}| } + c_s < \sqrt{\frac{Gm_*}{|\vec{x}-\vec{x_*}|}},
\end{equation}
where $\vec{x}$, $\vec{v}$ and $c_s$ are the position, velocity and
sound speed of the gas in the computational domain and $\vec{x_*}$,
$\vec{v_*}$ and $m_*$ are the position, velocity and mass of each
protostar.  This criterion selects gas with insufficient thermal or
kinetic energy to escape eventual accretion to the protostar, but
distinguishes between infalling gas (for which $\vec{v}-\vec{v_*}$
is anti-aligned with $\vec{x} - \vec{x_*}$, and thus the dot product
is negative), which is more bound, and outflowing gas
(vectors aligned), which is less so.  We
assign gas that is bound to more than one star under this criterion 
only to the star to which it is most strongly bound.

For each star $i$, we have now defined a mass $m_{{\rm gas}, i}$ of
``bound gas." We next compute the volume weighted mean magnetic 
field direction in this gas, $\hat{\bf B}_i$, 
and we define the associated
magnetic flux $\Phi_i$ as the net flux through
the intersection of the plane defined by $\hat{\bf B}_i$ and the volume of
assigned gas.  Finally, we characterize the local mass-to-flux ratio
of the gas assigned to each star as $\mu_{\Phi,i} = m_{{\rm gas},i}
/ M_{\Phi,i}$ where $M_{\Phi,i}$ is the magnetic critical mass
computed from \autoref{mcrit}.
It should be noted that the significance of $\mu_\Phi$ changes if a protostar is present in the cloud: The protostellar gravity also acts on the gas, so that accretion can occur normal to the field even for $\mu_\Phi<1$.

\autoref{f12} shows the temporal
evolution of the mean mass-to-flux ratio averaged over the bound gas
around each protostar.  The average is weighted the by the mass of
each protostar. Due to the computational burden
associated with post-processing this result from our models, the plots
in \autoref{f12} are sampled only every $0.2 t_{\mathrm{ff,>med}}$ in
time.  In the figure, the curve for each model begins after the
formation of the first sink particle.

In interpreting \autoref{f12} it is important to bear in mind that
this diagnostic characterizes the net impact of multiple competing
processes.  Under our sink particle prescription \citep{lee14},
infalling gas is released from its associated magnetic flux the
instant that it deposited onto the sink particle. This prescription
is intended to mimic resistive effects near the surface of the
protostar at length scales below that which can be resolved by our
models. Consequently accretion causes the mass-to-flux ratio associated with
locally bound dense gas clumps to decrease. The accreted mass is
removed from the gas but the flux remains. However, the magnetic
flux may subsequently escape, because it is no longer anchored
to the gravitationally bound material -- this is a form of magnetic
interchange instability \citep{zhao,krasnop,cunningham12,libp}.
If the magnetic flux tubes escape the gravitationally bound region,
this drives the mass to flux ratio back up. Similarly, turbulent magnetic
reconnection acts to 
diffuse
magnetic flux and
increases the local mass-to-flux ratio in regions that undergo
reconnection.  In this case, the gas mass remains but the magnetic
flux is displaced outward \citep{santoslima}. Outflow ejection and
turbulent forcing also impart mechanical energy to the system,
and the resulting flows can further amplify, entangle, and ultimately reconnect
magnetic flux tubes \citep{sur,li15}.  In this way accretion, outflow,
field tangling and reconnection all influence the mass to flux ratio.
Since accretion itself is episodic, so too is the evolution of the mean
mass to flux ratio of the bound gas in \autoref{f12}.  We
anticipate the degree of tangling and episodic reconnection to be most
pronounced in the weak-field models as weaker fields are more
readily tangled and amplified by the turbulent flow.  This too is
borne out in \autoref{f12}.

In the undriven models shown in the left panel of \autoref{f12}, the
mass-to-flux ratios in the bound gas systematically decreases with
time for each of the models.  This indicates that buoyant flux tubes
rising away from the protostars as a consequence of accretion act
enhance the degree of magnetic support to the bound gas, and this
enhancement overwhelms the effect of turbulent reconnection. The
$\mu_\Phi=1.56$ model without outflows indicates 
systematically
stronger magnetic support in bound gas than the case
with outflows.  This indicates that outflows contribute to magnetic
reconnection and consequent outward transport of magnetic flux, and
explains why these models achieve comparable star formation efficiency
($\epsilon_{\rm ff,IC}=12\%$ in the case without outflows and
$\epsilon_{\rm ff,IC}=23\%$ in the case with outflows). The enhanced
magnetic support in the case without outflows offsets the loss of
mechanical feedback support when the outflows are turned off.  The
$\mu_\Phi=2.17$ model is also strongly magnetically supported, with
the mass-to-flux ratio in the bound gas approaching $\mu_\Phi \sim 1$
at late time, consistent with its accretion rate $\epsilon_{\rm
  ff,IC}=14\%$, slightly higher than the more strongly magnetized
$\mu_\Phi=1.56$.  The model that began with a field characterized
globally as $\mu_\Phi=23.1$ attains a characteristic mass-to-flux
ratio $\mu_\Phi \approx 5$ via field amplification and gravitational
compression by the time the first protostellar particle forms.  While
the level of magnetic support in the bound gas increases with time,
the degree of magnetic support achieved by the end of the model is
insufficient to significantly reduce the net accretion rate from that
of the purely hydrodynamical case.

The evolution of the mass-to-flux ratio of the bound gas in the models
with driven turbulence models is shown in the right panel of
\autoref{f12}.  In this case the model that began with a global
$\mu_\Phi=2.17$ retains a roughly constant mass-to-flux ratio in the
bound gas of $\mu_\Phi \sim 2$ throughout its evolution.  The model
that began with a global $\mu_\Phi=1.56$ attains stronger magnetic
stabilization in the dense gas with $\mu_\Phi < 1$ by the time the
first protostars have formed.  Subsequent evolution indicates a slowly
decreasing level of magnetic support in the bound gas, stabilizing to
$\mu_\Phi \gtrsim 1$.  The model that began with a global
$\mu_\Phi=23.1$ enhances the degree of magnetic support in the bound
gas with time, reaching to $\mu_\Phi \sim 4$ by the end of the model.

Taken in aggregate, our results indicate that small-scale dissipation
effects near the surface of a protostar provide a feedback loop for
stabilising the star formation efficiency.  Bound gas cores that
are under-supported initially undergo more rapid collapse than cores
that are better supported, independent of whether the initial support
is predominately turbulent or magnetic.  The more rapidly collapsing
regions reconnect and expel magnetic flux to the core from the
newly-formed protostar at a faster rate.  The end result of this 
process over a time scale $>1 t_{\textrm{ff}}$ is that cores with
continuous mechanical support that remain in near virial equilibrium
approach mass-to-flux ratios of $\mu_\Phi < 5$ independent of the
large-scale mean field strength.  Models with decaying or otherwise
sub-virial levels of turbulent support will lead to greater local
enhancement of the level of magnetic support in the bound gas,
approaching $\mu_\Phi < 2$ at late time.

\section{Magnetic and Radiative Effects in Shaping the IMF} \label{s4}

In \autoref{s3.3} we demonstrated that our models have
protostellar mass distributions that are stable or 
evolving slowly toward higher mass only slowly compared to cloud-collapse
timescales. The models with weak initial fields or a lack of driving
all have median masses that are too large in comparison to the observed
IMF, while the strong field and driven cases have smaller median masses
that are at least roughly compatible with these models reproducing the
observed IMF. It is therefore interesting to investigate in more depth
what mechanisms shape this mass distributions. To explore this question, we turn
to the techniques described in \citet{krumholz16}, who analysed
the simulations of \citet{myers14}; these simulations use the same
physics we include here, but focus on a much denser region appropriate
to sites of high mass star formation. Comparing the two sets of simulations
therefore enables us to understand how this question depends on the
star-forming environment.

\subsection{Analysis Methodology}\label{s4.1}

We begin with a brief summary of the \citet{krumholz16} analysis, and
refer readers interested in the details to that paper for a full description.
The central idea of this analysis is to determine what physical process
is responsible for halting fragmentation in the vicinity of each protostar,
thereby leaving the mass available for accretion and the build up of
more massive stars, rather than the production of additional low mass
stars. To this end, we examine spherical regions of gas within a
$4000~{\rm AU}$ radius around each protostellar sink particle at time
intervals of $0.05 t_{\rm ff,>med}$.  We divide the volume of interest
into 128 spherical shells of radius $r$ concentric with the position of the sink
particle.  For each shell, we compute the total gas enclosed $m_{\rm
gas}$, excluding the sink particle mass, the mean density $\rho$
of the enclosed material, and its mass-weighted mean sound speed 
$c_s$.\footnote{As in \citet{krumholz16}, we focus on
cumulative rather than differential quantities because they are less
subject to noise; see Appendix A of \citet{krumholz16} for a demonstration
that this choice does not alter the qualitative results.}
From the resulting profiles in $\rho$ and 
$c_s$ we compute an
effective temperature
\begin{equation}
T_{\rm eff} = \frac{\mu m_H}{k_B} c_s^2 \label{Teff},
\end{equation}
and, more importantly from the standpoint of determining whether
gas will fragment, the Bonnor-Ebert mass \citep{ebert,bonnor},
\begin{equation}
  m_{\rm BE} = 1.86 \sqrt{\frac{c_s^3}{G^3\rho}}. \label{mbe}
\end{equation}
The Bonnor-Ebert mass characterizes the degree of thermal support.
Absent other forces, objects less massive than $m_{\rm BE}$ are stable
against collapse and objects more massive than $m_{\rm BE}$ are
unstable against collapse.  To elucidate the importance of heating by
protostars in stabilizing the core, we repeat the computation with the
sound speed $c_s=0.19~\textrm{cm s}^{-1}$, the sound speed of
molecular gas at $10~\textrm{K}$ used as the background isothermal temperature in
the simulations.  We shall denote this quantity as $m_{\rm BE,10}$.
Comparison of $m_{\rm BE}$ with $m_{\rm BE,10}$ enables us to
determine the importance of radiative heating in limiting fragmentation.

To characterise the importance of magnetic forces, we compute the
magnetic flux $\Phi$ threading each spherical shell, defined so that
$\Phi = \int |\vec{B}\cdot\hat{n}|\, d^2x$ is computed along a
circular cross section that is concentric with the spherical shell
and has normal $\hat{n}$.  The absolute value is taken in the integration
on the assumption that oppositely-directed fluxes do not reconnect and
cancel. We consider 12 possible orientations for $\hat{n}$, uniformly
distributed on the unit sphere, with one value aligned with the star's
angular momentum vector, and use the largest value of $\Phi$ we find
for all the $\hat{n}$ values considered. From this we can compute
an effective mean magnetic field strength inside a radius $r$,
\begin{equation}
B_{\rm eff}(<r) = \frac{\Phi}{\pi r^2}, \label{Beff}
\end{equation}
a magnetic critical mass $m_\Phi$ (\autoref{eq:mphi}), and the
magnetically-supported mass \citep{mou76,mckeeostriker}
\begin{equation}
  m_B=\frac{m^3_\Phi}{m^2_{\rm gas}}.
\end{equation}
We prefer $m_B$ to $m_\Phi$ for this analysis as the former
is an intensive quantity that does not depend on the size of the volume
considered.

It has been well established that magnetic fields influence
the spectrum of stars formed and reduce the star formation efficiency \citep{price08}.  Here, we elucidate the relative contribution or radiative feedback to magnetic support by
comparing the enclosed mass $m_{\rm gas}$ with the thermally-
and magnetically-supported masses $m_{\rm BE}$ and $m_B$.  This
provides direct insight into the role of thermal support (augmented
by radiative heating) and magnetic fields in shaping the IMF. For
sufficiently small shells around each protostar, we always find
$m_{\rm gas} \ll m_{\rm BE}$ and $m_{\rm gas} \ll m_B$, i.e.,
the enclosed mass is too small to fragment and form another
star given the level of thermal and magnetic support. This gas
will almost certainly be accreted onto the star that it surrounds,
augmenting this star's mass. At sufficiently large radii,
$m_{\rm gas} \gg m_{\rm BE}$ and $m_{\rm gas} \gg m_B$, i.e.,
the mass enclosed is large enough that is able to fragment and
produce additional stars, rather than being accreted onto the
star it surrounds. Thus the shell at which
$m_{\rm gas} \approx \max(m_{\rm BE}, m_{B})$ marks a
rough dividing line between gas that will end up in the existing
star and gas that will end up elsewhere, and $m_* + m_{\rm gas}$
provides a rough estimate of the mass to which the star is likely
to grow. We define the thermal critical mass
$m_{\rm BE, crit}$ as the smallest (and almost always sole) mass
where $m_{\rm gas} = m_{\rm BE}$, and analogously for
$m_{B, {\rm crit}}$ and $m_{\rm BE, 10, crit}$. Thus these
critical masses provide a rough estimate of the future mass
supply available for each star.

\citet{krumholz16} show that, in the simulations of \citet{myers14},
stars with mass $m_* \ll 0.2$ $M_\odot$ almost invariably have
stabilised gaseous envelopes such that $m_{\rm BE,crit} \gg m_*$,
explaining the relative rarity of brown dwarfs as compared
to stars: brown dwarf mass fragments can form, but when they do,
they stabilise enough gas around themselves that they usually
continuing accreting well out of the brown dwarf mass regime. Moreover,
\citet{krumholz16} find that their stabilised regions have
$m_{\rm BE, crit} \gg m_{B, {\rm crit}} \gg m_{\rm BE,10,crit}$,
indicating that, while magnetic support is more important than
thermal support in the absence of radiative heating, once one
considers the effects of stellar radiation, it is the dominant factor
in limiting fragmentation and thus establishing the location of the
IMF peak.

\subsection{Profiles of Mean Density, Temperature and Magnetic Field}\label{s4.2}

\begin{figure*}
  \begin{center}
    \includegraphics[clip=true,width=0.8\textwidth]{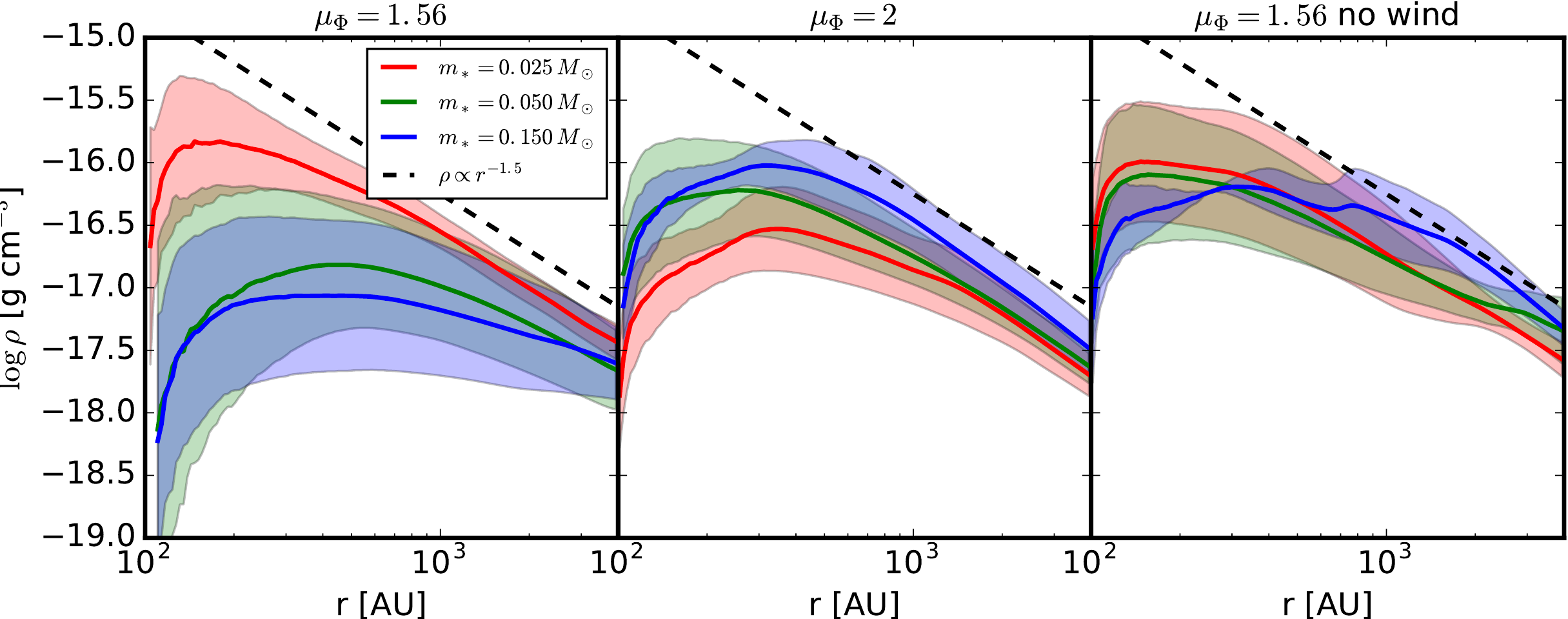} \\
    \includegraphics[clip=true,width=0.8\textwidth]{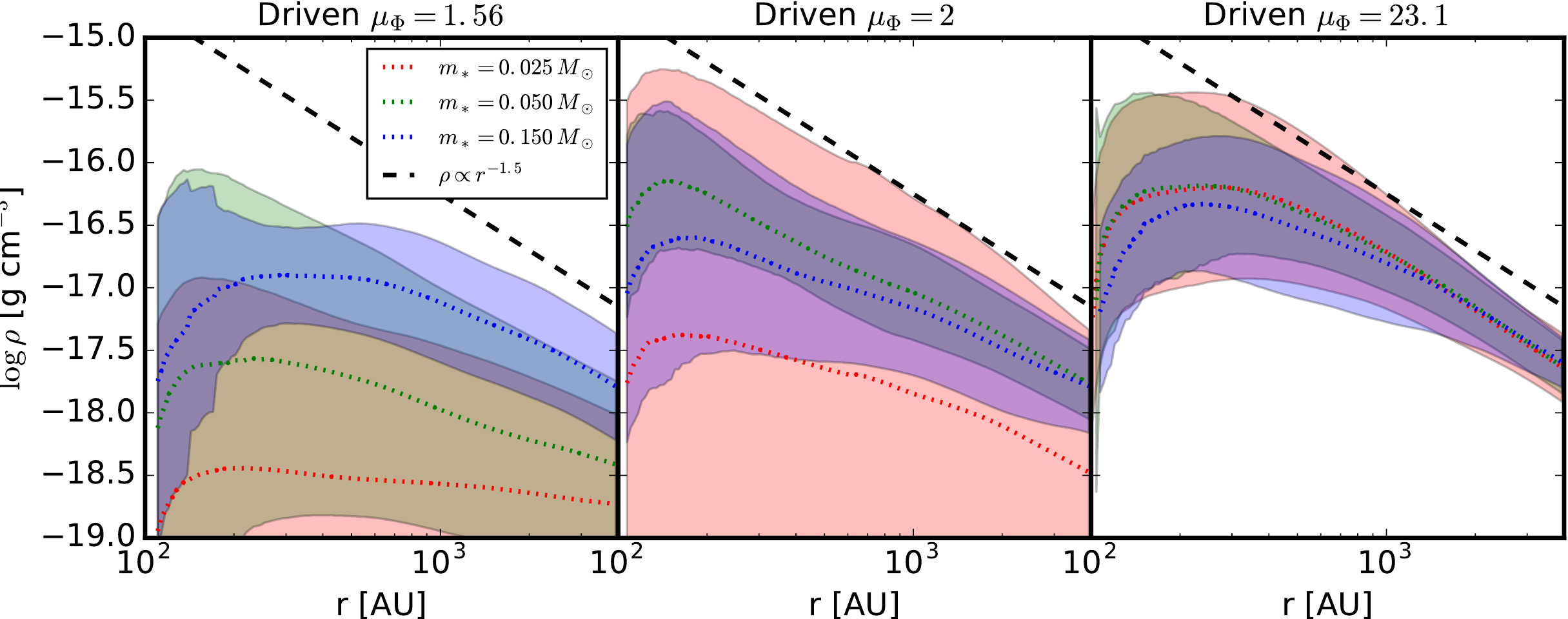} \\
  \end{center}
  \caption{Mean density profiles $\rho(r)$ averaged in spherical
    shells around protostars.  The panels shows the results for the
    strong (left), moderate (centre) and weak (right) magnetic field
    models.  The top row depicts results for the models without
    turbulence driving after the onset of self-gravity at $t=0$.  The
    bottom row depicts results for the model with continued turbulence
    driving after the onset of self-gravity.  Different colours show
    means for protostars of mass $0.025~\msol$, $0.05~\msol$ and
    $0.15~\msol$.  For each mass bin, the central line shows the mean
    while the shaded band shows the $1~\sigma$ dispersion for protostars in
    that mass bin. The dashed black line shows the density scaling $\rho
    \propto r^{-3/2}$, expected for free-fall collapse. \label{f13}}
\end{figure*}

\begin{figure*}
  \begin{center}
    \includegraphics[clip=true,width=0.7\textwidth]{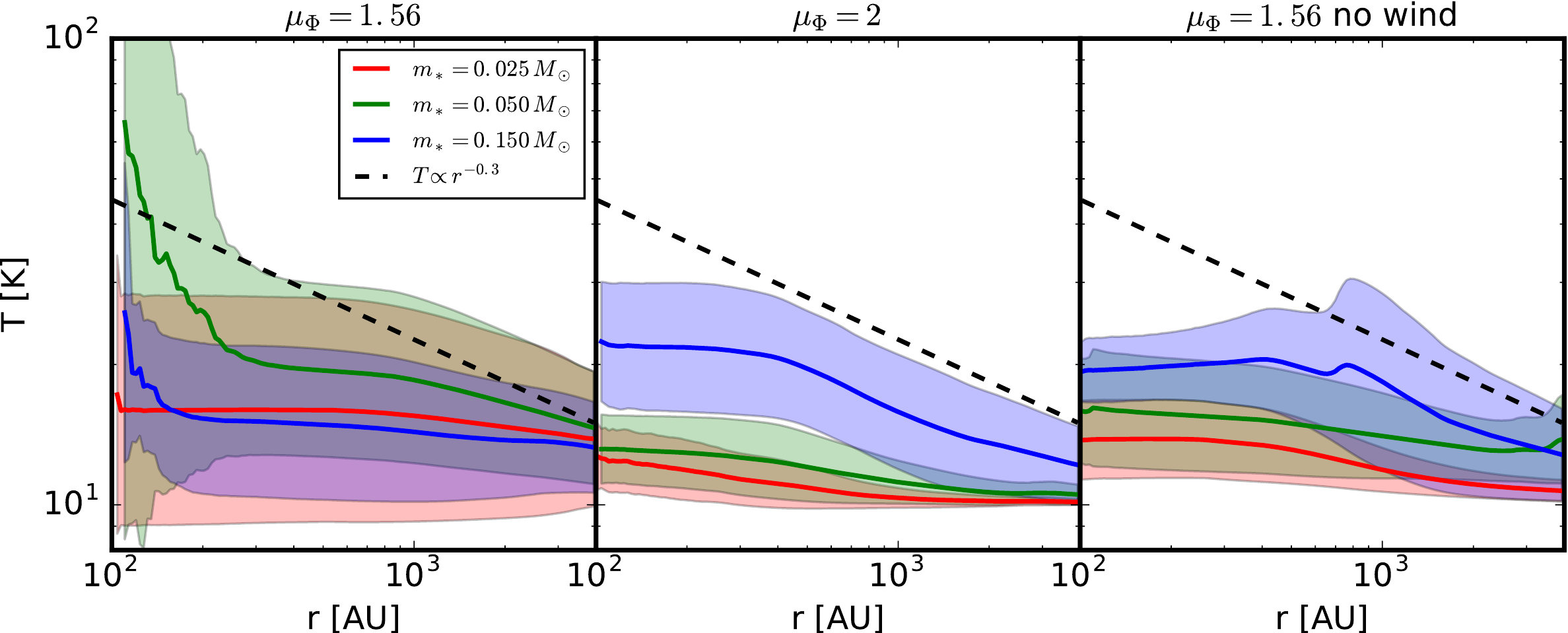} \\
    \includegraphics[clip=true,width=0.7\textwidth]{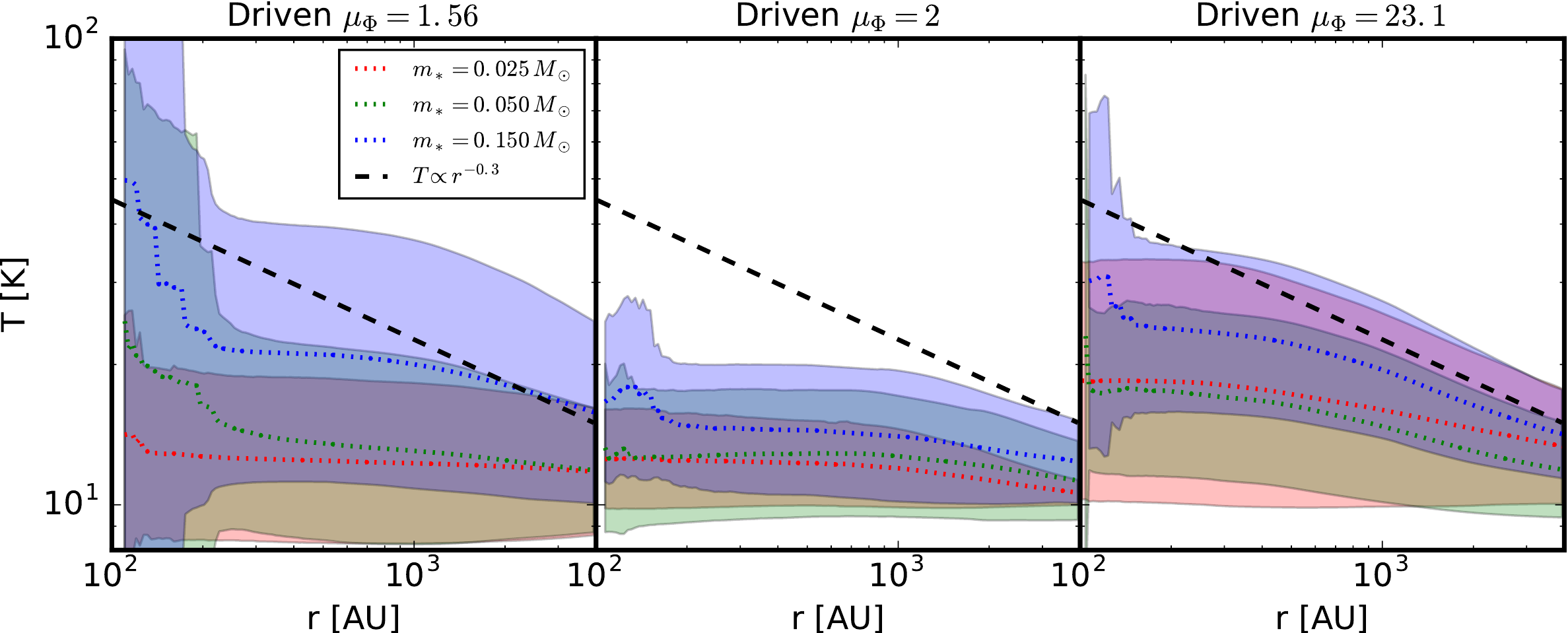} \\
  \end{center}
  \caption{Same as \autoref{f13} but showing the mean effective
    temperature $T_{\rm eff}(r)$ (equation \ref{Teff}).  The black
    dashed line shows a $T_{\rm eff} \propto r^{-0.3}$.  The initial
    temperature of the models is $T=10~{\rm K}$. \label{f14}}
\end{figure*}

\begin{figure*}
  \begin{center}
    \includegraphics[clip=true,width=0.7\textwidth]{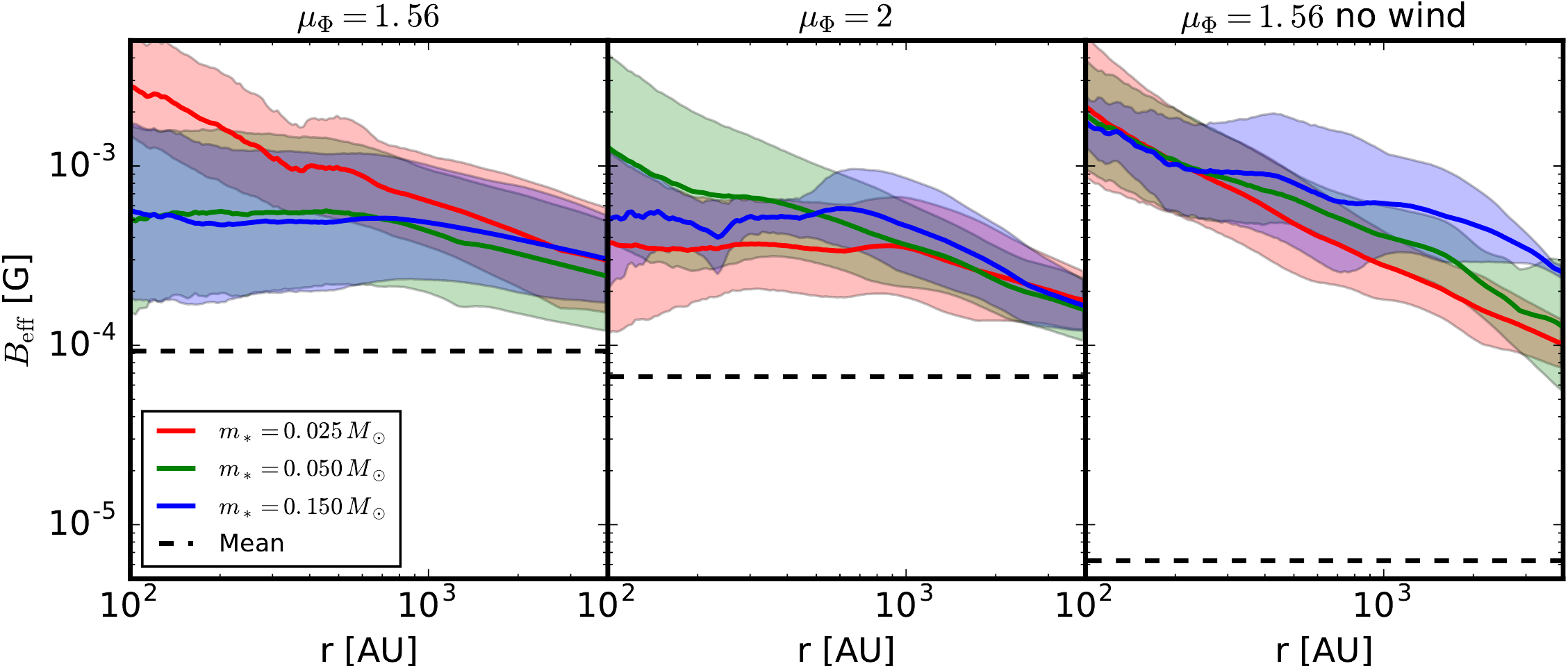} \\
    \includegraphics[clip=true,width=0.7\textwidth]{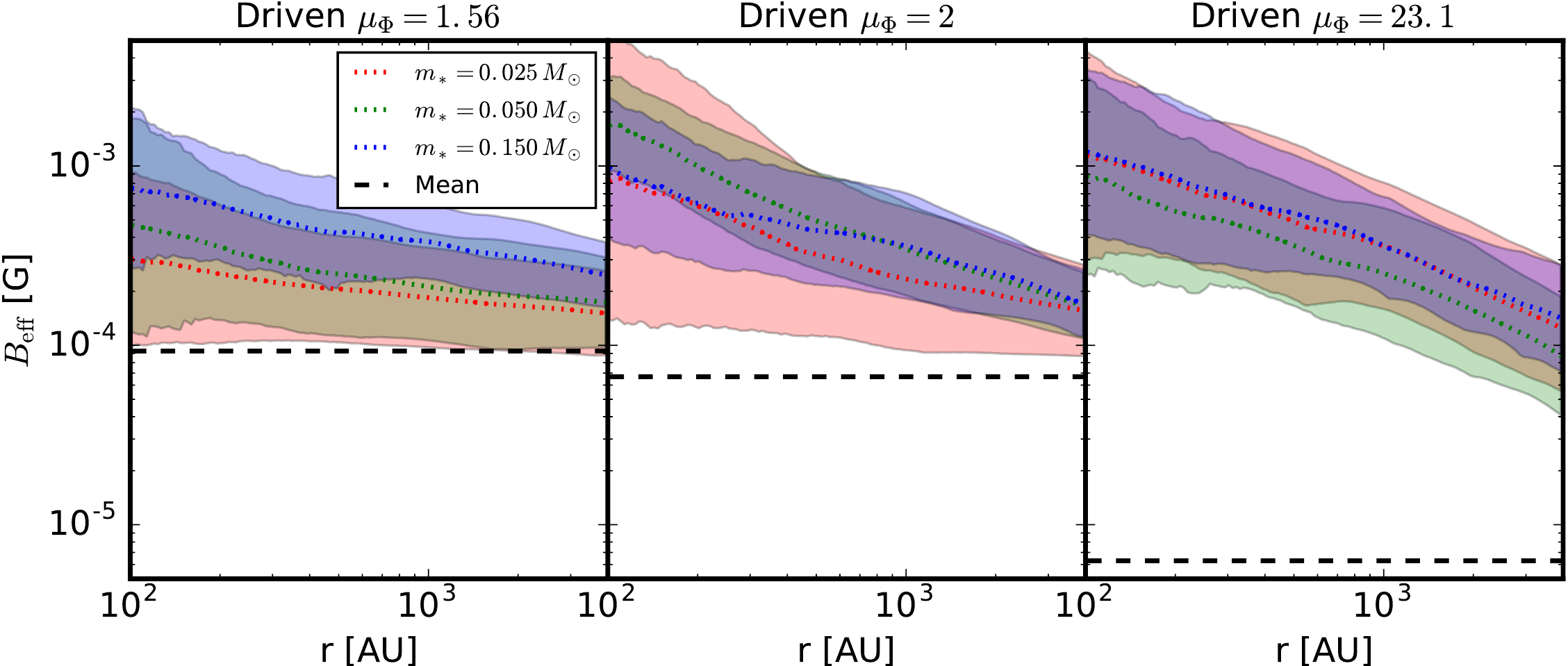} \\
  \end{center}
  \caption{Same as \autoref{f13} but showing the mean effective
    magnetic field $B_{\rm eff}(<r)$ (equation \ref{Beff}).  The black
    dashed horizontal line indicates the initial magnetic field
    strength before driving to the initial turbulent
    state. \label{f15}}
\end{figure*}

Figure \ref{f13} shows the average density profile around central
stars with three different masses; the averages plotted are taken
over all protostars at all times. The density profiles always
show a central dip at $130~\textrm{AU} < r < 260~\textrm{AU}$, which
is a result of our sink particle accretion zones; we therefore
ignore this region in our analysis. At larger radii, we find that
the moderately and weakly magnetized models ($\mu_\Phi \geq$ 2)
without driving have density profiles $\rho \propto r^{-3/2}$. 
This profile is consistent with that expected for free-fall
collapse onto a point mass.  However, for Bondi-type flow the
density at a fixed distance from the central object
should increase with mass. The fact that it does not is
inconsistent with pure Bondi flow, but is
consistent with the turbulent core model of \citet{mckeetan} and with
the model of \citet{murray15} for the for collapsing, self-gravitating,
turbulent cores. In contrast, the strongly magnetized ($\mu_\Phi = 1.56$) undriven
cases, and all driven cases for a central star mass of $m_* = 0.15~\msun$,
show slightly shallower density profiles. This is consistent with
a buildup of magnetic flux resulting in a slower than free-fall
collapse.

Figure \ref{f14} shows averaged profiles of $T_{\rm eff}$. Except for the $\mu_\Phi = 1.56$ undriven
model, these profiles indicate increasing $T_{\rm eff}$ with central star mass.
As \citet{krumholz16}
point out, to maintain a constant density profile, infall velocities
and accretion rate must increase as the progenitor mass increases.
Since accretion luminosity is the dominant source of radiative heating
in low mass sources, the temperature around them rises with their
mass. The weakly magnetized models show the most rapid increase in
heating with mass. \citet{krumholz16} noted a similar trend in models
of denser high mass star forming regions, and also found that
the temperature profiles were always close to $T_{\rm eff}\propto r^{-0.3}$. 
Here we find substantially shallower profiles, likely because the
lower optical depth environment we are simulating is less effective
at trapping the radiation \citep{krumholz08}.

In Figure \ref{f15} we show the corresponding average profiles of
$B_{\rm eff}$. Consistent with our findings in 
\autoref{ssec:mag_support}, despite the large differences in
large-scale magnetic flux, there is very little difference
between the small-scale magnetic field strengths around
protostars in the different simulations. Nor does the
effective field strength increase with star mass, as one
might expect from a naive picture in which accretion drags
magnetic flux toward a star.

The non-dependence of the small scale magnetic fields on either
the large-scale flux or the amount of gas that has already been
accreted strongly suggests that the magnetic field profiles we
find are the result of a self-regulation mechanism operating on
small scales. We can think of this self-regulation as due to two
pairs of competing processes,
although given our limited resolution
and the complex nature of the flows in our simulations, we cannot
easily identify which mechanisms dominate.
First, there is a competition between
dynamo amplification and reconnection. In 
a turbulent dynamo, magnetic fields amplify until these two
effects balance \citep{sur,li15}. 
Second, there is a
competition between advection and magnetic interchange instabilities. As gas accretes, magnetic flux tubes are advected into 
the cores around the protostars, which tends to drive up
$B_{\rm eff}$. The countervailing process is escape of flux through
magnetic interchange instabilities. In reality, the interchange instabilities are triggered by resistive effects close to the
surface of the protostar, which decouple flux tubes from the mass that
anchors them and allows them to drift outward 
\citep{zhao,krasnop,cunningham12,libp}; we model this by not accreting any magnetic flux into the star particles. If the flux liberated from the protostar diffused
out to a radius $R$, then it would contribute a mean field 
\begin{equation}
    B_{\rm eff,*}=2G^{1/2}\left(\frac{L}{Z}\right)\frac{m_*}{\mu_\Phi R^2},
\end{equation}
where we have assumed that the protostar has formed from a cylinder of gas with a height $Z$ less than the box height $L$.
For $m_*=0.15\,M_\odot$ and $\mu_\Phi=1.56$, the results in Figure \ref{f15} place a lower limit $R> 1000$ au on the distance that the field has diffused outward.

\subsection{Critical Mass}\label{s4.3}

\begin{figure*}
  \begin{center}
    \includegraphics[clip=true,width=0.7\textwidth]{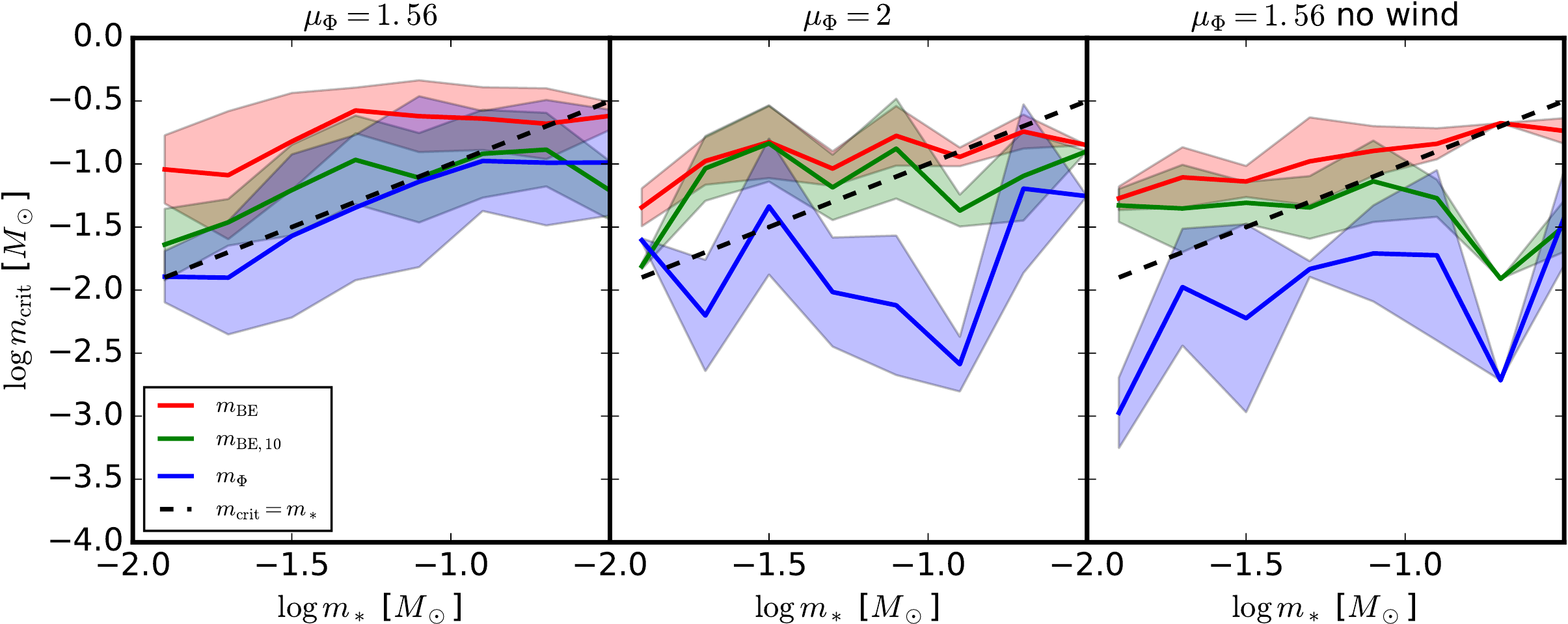} \\
    \includegraphics[clip=true,width=0.7\textwidth]{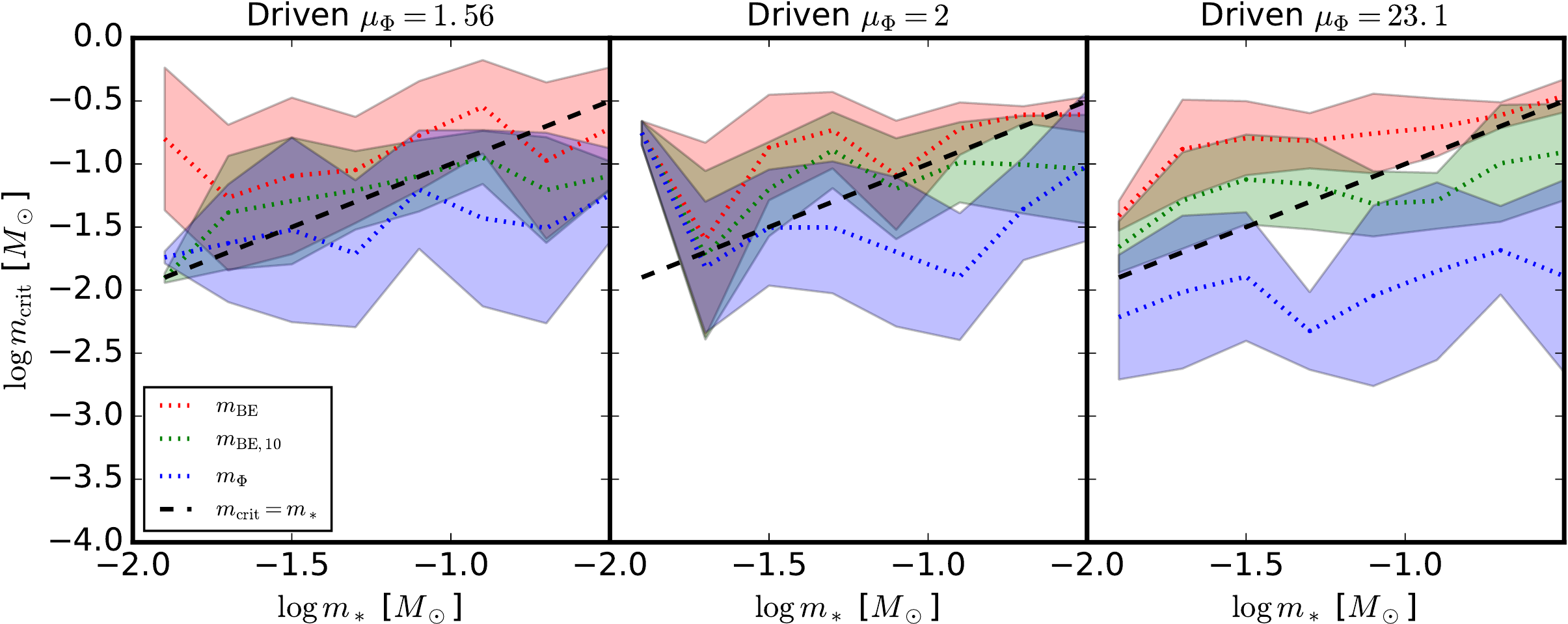} \\
  \end{center}
  \caption{Critical mass of gas supported by thermal pressure $m_{\rm
      BE}$, thermal pressure fixed to the initial temperature $10~{\rm
      K}$ $m_{\rm BE,10}$ and magnetic fields $m_B$ as a function of
    central stellar mass $m_*$.  Shaded bands indicate $1~\sigma$
    dispersion over the protostars.  The dashed black line indicates
    where the mass supported against collapse by pressure or magnetic
    forces is sufficient to double the present stellar mass, $m_* =
    m_{\rm crit}$.  The panels shows the results for the strong (left),
    moderate (centre) and weak (right) magnetic field models.  The top
    row depicts results for the models without turbulence driving
    after the onset of self-gravity at $t=0$.\label{f16}}
\end{figure*}

In figure \ref{f16}, we show the critical mass of gas supported by
thermal pressure, $m_{\rm BE}$, thermal pressure fixed to the initial
temperature, $10~{\rm K}$ $m_{\rm BE,10}$, and magnetic fields, $m_B$, as
a function of central stellar mass $m_*$ as described in \S\ref{s4.2}.
When $m_{\rm crit} > m_*$, 
the mass of the stabilised envelope around
a protostar is much greater than the mass of the star itself, and thus
is likely to greatly increase its mass.

First consider thermal support. \cite{krumholz16} found that
$m_{\rm BE}$ was typically a factor of $\sim 10$ larger than
$m_{\rm BE,10}$, indicating that radiative heating increased the
amount of mass that could be thermally supported by a factor of
$\sim 10$. In our lower mass cluster models the difference between
$m_{\rm BE}$ and $m_{\rm BE,10}$, is less pronouned, closer to
$0.5$ dex than 1 dex. However, the difference is smaller not
because $m_{\rm BE}$ is smaller, but rather because $m_{\rm BE,10}$
is larger. That is to say, in our lower density simulation the
characteristic density is lower, and thus the Bonnor-Ebert mass
in unheated gas is larger. In comparison, the mass that can
be supported after heating, $m_{\rm BE}$, is almost unchanged
and independent of environment---the increase in the temperature of the 
gas has been compensated by an increase in its density.

Next consider magnetic support. The results in \autoref{f16}
are consistent with the findings of \citet{krumholz16}
that the magnetic fields are relatively unimportant in
supporting the cores compared to thermal pressure. Again,
this points to the importance of local processes in regulating
the magnetic field structure in the vicnity of protostars. 
The 
magnetic flux
in the $\mu_\Phi=1.56$ case is sufficient to
support a mass $m_\Phi=119~\msun$ against collapse, but the field
threading the local region around the protostars is strong enough
to support only $0.01~\msun$ to $0.1~\msun$. This result explains why the 
mode of the IMFs (\S\ref{s3.3}) in our models are relatively
insensitive to the large-scale magnetic field.  The critical mass
supported against fragmentation is determined primarily by thermal
effects enhanced by radiation. 
Magnetic fields thus have a significant effect on the rate of star formation, but not on the IMF.

\section{Conclusions} \label{conclusion}

We present a numerical study of low-mass star cluster formation, and
how this process is affected by the strength of the large-scale
magnetic field, the presence or absence of turbulent energy input, and
the effects of protostellar outflows. Our simulations include MHD,
radiative transfer and protostellar radiation feedback, and
protostellar outflows, a combination of physical processes that has
received little attention in the literature to date.

Our primary finding is that our models are able to reproduce the
observed (low) efficiency of star formation and the observed location
of the IMF peak (to within a factor 2) only in models that include
outflows and that do not undergo global collapse, either because they
are supported by external energy input that drives the turbulence, or
because they are close enough to the line of magnetic criticality for
magnetic fields to inhibit collapse. In contrast, multiplicity
statistics are insensitive to all variations in the initial
conditions.

\citet{krumholz11, krumholz12} had previously conjectured that there
might be a link between the low efficiency of star formation and the
location of the IMF peak, but this was based on a single set of
non-magnetised simulations. Our much more extensive suite of
radiation-MHD simulations puts this conclusion on a much firmer
footing, and suggests a general principle: the IMF peak location is
directly linked to the inefficiency of star formation.

The primary mechanism controlling the IMF peak appears to be
accretion-powered radiation from forming protostars, which produces
enhanced thermal pressure on small scales in the vicinity of forming
stars, as noted earlier by \cite{bate09} as well.  In comparison, the
small-scale magnetic field structure around forming protostars is both
insensitive to the large-scale magnetic field strength, and
dynamically sub-dominant compared to radiation when it comes to
regulating gas fragmentation. These results strongly favour a picture
in which the small-scale magnetic field structure around protostars is
regulated by local competitions between accretion of magnetized
material and dynamo amplification, which tend to enhanced the field
strength, and magnetic interchange instability and turbulent
reconnection, which tend to suppress it. The competition between these
processes tends to force the local mass to flux ratio in gas bound to
protostars to $\sim 2$ (normalized to the critical value), regardless
of the large-scale field.

Our models lack resolution to capture the effects of thermal support
against gravitational contraction deep within very low mass cores and
we lack non-ideal MHD effects that are most strongly coupled to
turbulent density profiles on small scales.  However these effects
should influence our modeled protostellar mass distributions only on
the low mass tail $m_* \lesssim 0.05 \msol$ and our conclusions are not
sensitive to these effects.  We acknowledge that higher resolution
models with non-ideal effects and multiple realizations to capture
better statistics at protostellar mass extrema could improve the
comparisons with observation made here by capturing a wider range of
source masses with higher fidelity.

\section*{Acknowledgements}
Support for this research was provided by NASA through NASA ATP grants
NNX13AB84G (RIK \& CFM), NNX15AT06G (MRK), the US Department of Energy
at the Lawrence Livermore National Laboratory under contract
DE-AC52-07NA27344 (AJC \& RIK), the NSF through grant AST-1211729 (RIK
\& CFM), and the Australian Research Council's \textit{Discovery
  Projects} funding scheme (project DP160100695, MRK). This research
was also supported by grants of high performance computing resources
from the National Center of Supercomputing Application through grant
TGMCA00N020, under the Extreme Science and Engineering Discovery
Environment (XSEDE), which is supported by National Science Foundation
grant number OCI-1053575, the computing resources provided by the NASA
High-End Computing (HEC) Program through the NASA Advanced
Supercomputing (NAS) Division at Ames Research Center and resources
and services from the National Computational Infrastructure (NCI),
which is supported by the Australian Government. LLNL-JRNL-737724.

\label{lastpage}
\end{document}